\documentclass[a4paper,11pt]{article}
\usepackage{amssymb}
\usepackage{graphicx}
\usepackage{amsbsy}
\usepackage{amsmath}
\usepackage{cite}
\usepackage{slashed}
\usepackage{amsfonts}
\usepackage{hyperref}

\hypersetup{
colorlinks   = true,
linkcolor    = blue,
citecolor    = blue,
}
\newcommand*{\doi}
[1]{\href{http://dx.doi.org/#1}{doi: #1}}

\begin{document}

\thispagestyle{empty} \setcounter{page}{0} 
\begin{flushright} February 2023\\
\end{flushright}

\vskip3.4 true cm

\begin{center}
{\huge On the fermionic couplings of axionic dark matter}\\[1.9cm]

\textsc{Christopher Smith}$^{\ast}$\vspace{0.5cm}\\[9pt]
\smallskip {\small \textsl{\textit{Laboratoire de Physique Subatomique et de
Cosmologie, }}}\linebreak{\small \textsl{\textit{Universit\'{e}
Grenoble-Alpes, CNRS/IN2P3, Grenoble INP, 38000 Grenoble, France}.}} \\[1.9cm%
]
\textbf{Abstract}\smallskip
\end{center}

\begin{quote}
\noindent In the non-relativistic limit, two types of dark matter axion
interactions with fermions are thought to dominate: one is induced by the
spatial gradient of the axion field and called the axion wind, and the other
by the time-derivative of the axion field, generating axioelectric effects.
By generalizing Schiff theorem, it is demonstrated that this latter operator
is actually strongly screened. For a neutral fermion, it can be entirely
rotated away and is unobservable. For charged fermions, the only effect that
can peek through the screening is an axion-induced electric dipole moment
(EDM). These EDMs are not related to the axion coupling to gluons, represent
a prediction of the Dirac theory analogous to the $g=2$ magnetic moments,
are not further screened by the original Schiff theorem, and are ultimately
responsible for inducing the usual axioelectric ionization. The two main
phenomenological consequences are then that first the axion-induced nucleon
EDM could be significantly larger than expected from the axion gluonic
coupling, and second, that the electron EDM could also become available, and
could actually be highly sensitive to relic axions.

\let\thefootnote\relax\footnotetext{$^{\ast}\;$chsmith@lpsc.in2p3.fr}
\end{quote}

\newpage

\setcounter{tocdepth}{2}
\tableofcontents

\section{Introduction}

The axion mechanism~\cite{Peccei:1977hh,Peccei:1977ur} is currently our best
solution to the strong CP puzzle. The non-observation of a neutron electric
dipole moment (EDM) constrains the QCD theta term, $\theta G_{\mu\nu}%
\tilde{G}^{\mu\nu}$ to be tiny, $\theta\lesssim10^{-10}$~\cite%
{Abel:2020pzs}. As this coupling receives contributions from two unrelated
sectors of the Standard Model (SM), a topological QCD contribution and an
electroweak contribution from the quark Yukawa couplings, both a priori of $%
\mathcal{O}(1)$, such a tiny value requires an unacceptable fine-tuning. The
axion solution\cite{Weinberg:1977ma,Wilczek:1977pj} relies on the axion $a$
being the Goldstone boson associated to the spontaneous breaking of an
anomalous $U(1)$ symmetry, the PQ symmetry\cite{Peccei:1977hh,Peccei:1977ur}%
. This ensures a coupling of the axion to gluons, $aG_{\mu\nu}\tilde{G}%
^{\mu\nu}$, which develops into a potential for the axion in the low-energy
limit. At the minimum of this potential, the axion field absorbs the $\theta$
term, making it unobservable.

The initial implementation of the axion mechanism relied on the axion
emerging at the electroweak scale, and was quickly ruled out as this would
imply way to large couplings to matter particles.\ Invisible axion scenarios
were then developed, most notably the DFSZ\cite%
{Dine:1981rt,Zhitnitsky:1980tq} and KSVZ\cite{Shifman:1979if,Kim:1979if}
models, in which the axion is very light and very weakly coupled. In most
cases, being in addition very long-lived, the axion emerged as a viable dark
matter (DM) candidate (for a review of the axion in a cosmological context,
see e.g. Ref.~\cite{Marsh:2015xka}). Interestingly, experimental strategies
could then take advantage of the rather high flux of such dark matter axions
(see e.g. Ref.~\cite{Graham:2015ouw,Irastorza:2018dyq}). In practice, dark
matter axion production mechanisms ensure the axion is rather cold, and
being in addition very light, it can be represented by a classical coherent
pseudoscalar field, typically $a(\mathbf{r},t)=a_{0}\cos(Et-\mathbf{p}\cdot%
\mathbf{r})$, $E^{2}=\mathbf{p}^{2}+m_{a}^{2}$, $m_{a}$ the axion (or
axion-like particle) mass, and $a_{0}$ set by the local DM density, $%
m_{a}a_{0}=\sqrt{2\rho_{DM}}$ with $\rho_{DM}\approx 0.4\ $GeV$/cm^{3}~$\cite%
{Catena:2009mf}.

The goal of the present paper is to analyze the couplings to SM fermions of
such a dark matter axion background, in the non-relativistic limit. The
usual starting point is the axion Lagrangian (to simplify the notation, a
coupling constant $g=m/\Lambda$ with $\Lambda$ the PQ breaking scale is
understood to be absorbed into $a$ throughout this paper)%
\begin{equation}
\mathcal{L}_{D}=\bar{\psi}\left( \slashed \partial-m+\frac{%
\gamma^{\mu}\gamma_{5}\partial_{\mu}a}{m}\right) \psi\ .  \label{Intro1}
\end{equation}
Such a derivative interaction to the axion is reminiscent of its Goldstone
boson nature. The corresponding Dirac equation is $i\partial_{t}\left\vert
\psi\right\rangle =\mathcal{H}_{D}\left\vert \psi\right\rangle $ with 
\begin{equation}
\mathcal{H}_{D}=\gamma^{0}\left( \boldsymbol{\gamma}\cdot\mathbf{p}+m-\frac{%
\gamma^{0}\gamma_{5}\dot{a}}{m}+\frac{\gamma_{5}\boldsymbol{\gamma }\cdot%
\mathbf{\nabla}a}{m}\right) \ ,  \label{Intro2}
\end{equation}
where $\dot{a}=\partial_{t}a$. In the Dirac representation, where $%
\gamma^{0} $ is diagonal, $\gamma^{5}$ directly couples the fermion and
antifermion degrees of freedom of $\left\vert \psi\right\rangle $.
Consequently, in the non-relativistic limit, the $\dot{a}$ term receives a
dependence on $\mathbf{p}=-i\mathbf{\nabla}$:%
\begin{equation}
\mathcal{H}_{D}^{\mathrm{NR}}=\gamma^{0}\left( m+\frac{\mathbf{p}^{2}}{2m}+%
\frac{i\gamma^{5}\boldsymbol{\gamma}\cdot\mathbf{\nabla}a}{m}\right) +\frac{%
\gamma^{5}\{\boldsymbol{\gamma}\cdot\mathbf{p},\dot{a}\}}{2m^{2}}+\mathcal{O}%
(1/m^{3})\ .  \label{Intro3}
\end{equation}
These two leading interactions have been extensively studied in the
literature~\cite{Stadnik:2013raa,Sikivie:2020zpn}. The so-called axion wind
term, $\gamma^{5}\boldsymbol{\gamma}\cdot\mathbf{\nabla}a$, leads to a
coupling of the gradient of the axion field to the spin of the fermion. It
can be searched for experimentally e.g. using NMR techniques~ \cite%
{Budker:2013hfa,Abel:2022vfg,Graham:2017ivz,Garcon:2019inh} or magnons~\cite%
{Barbieri:2016vwg}.

The second term is dubbed the axioelectric effect~\cite%
{Derevianko:2010kz,Pospelov:2008jk,Avignone:1986vm,Dimopoulos:1985tm}. It
translates as a coupling of $\dot{a}$ to the combination $\boldsymbol{p}\cdot%
\mathbf{S}$ of the momentum $\mathbf{p}$ and spin $\mathbf{S}$ of the
fermion. As a result, sufficiently energetic axions could kick bound
electrons out, in analogy with the photoelectric effect. The sun could
produce a consequent flux of such axions, whose possible detection via these
ionization processes, or more generally electron recoil effects, gave rise
to a rather intense experimental activity~\cite%
{Bernabei:2005ca,CoGeNT:2008yoi,PandaX:2017ock,Armengaud:2013rta,CDEX:2016rpr,CDMS:2009fba,Abe:2012ut,Hochberg:2016sqx}%
. The corresponding constraints on the axion are reviewed e.g. in Ref.~\cite%
{Bloch:2016sjj}, as well as more recently in Refs.~\cite%
{Bloch:2020uzh,Takahashi:2020bpq} in the context of the excess events
observed at XENON1T~\cite{XENON:2020rca}. Note, though, that these
experiments also probe different mechanisms and/or the coupling of the axion
to photons.

A peculiar feature of Goldstone bosons is that there are different ways to
parametrize them. For the axion, an equally valid Lagrangian uses the
so-called polar or exponential parametrization: 
\begin{equation}
\mathcal{L}_{E}=\bar{\psi}\left( \slashed \partial-m\exp\left( 2i\gamma ^{5}%
\frac{a}{m}\right) \right) \psi\ .  \label{Intro4}
\end{equation}
The derivative interaction is replaced by an infinite tower of interactions,
starting by the pseudoscalar coupling $a\bar{\psi}i\gamma^{5}\psi$. The
corresponding Hamiltonian is then 
\begin{equation}
\mathcal{H}_{E}=\gamma^{0}\left( \boldsymbol{\gamma}\cdot\mathbf{p}%
+m\exp\left( 2i\gamma^{5}\frac{a}{m}\right) \right) =\gamma^{0}\left( 
\boldsymbol{\gamma}\cdot\mathbf{p}+m+2i\gamma^{5}a\right) +\mathcal{O}%
(a^{2})\ ,
\end{equation}
and its non-relativistic limit can be worked out to be%
\begin{equation}
\mathcal{H}_{E}^{\mathrm{NR}}=\gamma^{0}\left( m+\frac{\mathbf{p}^{2}}{2m}+%
\frac{i\gamma^{5}\boldsymbol{\gamma}\cdot\mathbf{\nabla}a}{m}\right) +\frac{%
\gamma^{5}\{\boldsymbol{\gamma}\cdot\mathbf{p},\dot{a}\}}{4m^{2}}+\mathcal{O}%
(1/m^{3})\ .  \label{Intro5}
\end{equation}
The same axion wind and axioelectric interactions emerge, but the
coefficient of the latter differs by a factor two. Historically, this fact
was well known in the context of nucleon-pion interactions. The equivalence
of the pseudoscalar and axial interaction was first discussed by Dyson in
1948~\cite{Dyson} (see also Ref.~\cite{Case:1949zza,Berger:1952zz}), on the
basis of the axion wind term being the same. Later, this ambiguity in the
time-dependent term, as well as in some higher order terms in the
non-relativistic expansion, generated a lot of attention~\cite%
{Bolsterli:1974ct,Barnhill:1969ygg,Nieto:1976ca,Friar:1974dk,Lee:1976xi,Friar:1977xh}%
. As we will see, part of the issue was related to the truncation of the
exponential parametrization. After all, many of these works date back to
before Goldstone theorem was formulated, let alone the pion identified as a
pseudo-Goldstone boson of the chiral symmetry breaking. Nowadays, the
equivalence between the derivative and exponential representation is an
established fact, but surprisingly, a non-relativistic expansion truly
reflecting this has not been worked out yet. This is the purpose of the
present paper.

In particular, adopting a modern language, we will see that the $\gamma
^{5}\{\boldsymbol{\gamma }\cdot \mathbf{p},\dot{a}\}$ coupling can be
systematically rotated away for a neutral fermion. The demonstration is
actually quite simple and can readily be given. First, remember that a
non-relativistic expansion is not unique\footnote{%
In this respect, the non-relativistic expansion do depend on the method
chosen to construct it. We have used the standard Foldy-Wouthuysen procedure~%
\cite{Foldy:1949wa} to derive Eq.~(\ref{Intro3}) and (\ref{Intro5}), in
which the $\gamma ^{5}\{\boldsymbol{\gamma }\cdot \mathbf{p},\dot{a}\}$
operator does immediately have different coefficients. This may not be
apparent in all methods, in particular using the elimination method~\cite%
{Pospelov:2008jk}.}. As customary in quantum mechanics, unitary
transformations cannot change the physics. So, performing such a
transformation, and provided the block-diagonal nature of the Hamiltonian is
maintained, an equally valid non-relativistic expansion is found. Now, as
proposed a long time ago in Ref.~\cite{Barnhill:1969ygg,Friar:1974dk},
consider%
\begin{equation}
\left\vert \psi \right\rangle \rightarrow \left\vert \psi ^{\prime
}\right\rangle =\exp (iS)\left\vert \psi \right\rangle ,\ \ S=\frac{\mu }{%
4m^{2}}\gamma ^{5}\{\boldsymbol{\gamma }\cdot \mathbf{p},a\}\ .
\label{Intro6}
\end{equation}%
If $i\partial _{t}\left\vert \psi \right\rangle =\mathcal{H}\left\vert \psi
\right\rangle $, then $i\partial _{t}\left\vert \psi ^{\prime }\right\rangle
=\mathcal{H}^{\prime }\left\vert \psi ^{\prime }\right\rangle $ with $%
\mathcal{H}^{\prime }=\mathcal{H}-\dot{S}$ to $\mathcal{O}(1/m^{2})$ since $[%
\mathcal{H},S]$ starts at $\mathcal{O}(1/m^{3})$. Thus, acting on $\mathcal{H%
}_{D}^{\mathrm{NR}}$ with $\mu =2$, or on $\mathcal{H}_{E}^{\mathrm{NR}}$
with $\mu =1$, the $\gamma ^{5}\{\boldsymbol{\gamma }\cdot \mathbf{p},\dot{a}%
\}$ coupling is replaced by $\mathcal{O}(1/m^{3})$ and higher terms.

At the time, this was interpreted as an ambiguity that should cancel out in
physical observables. Here, we will go one step further and argue that $%
\gamma ^{5}\{\boldsymbol{\gamma }\cdot \mathbf{p},\dot{a}\}$ does not encode
any true physical effects. In other words, for neutral fermions, the
operator $\gamma ^{5}\{\boldsymbol{\gamma }\cdot \mathbf{p},\dot{a}\}$ is
totally screened at $\mathcal{O}(1/m^{2})$ in the non-relativistic
expansion. One reason for this interpretation has to do with Schiff's
theorem~\cite{Schiff:1963zz}, which states that charged fermion EDMs are
screened. The transformation $S$ of Eq.~(\ref{Intro6}) is closely related to
Schiff's transformation, and even \textit{becomes} the Schiff's
transformation for a charged fermion. The consequence in that case is that
the covariant $\gamma ^{5}\{\boldsymbol{\gamma }\cdot \mathbf{P},\dot{a}\}$
coupling becomes equivalent to an axion-induced EDM operator, $a\boldsymbol{%
\sigma }\cdot \mathbf{E}$ at $\mathcal{O}(1/m^{2})$. Phenomenologically,
these two operators are indistinguishable, ensuring that physics is
independent of the choice of parametrization. Yet, this equivalence makes it
manifest that unexplored but promising avenues do exist to search for dark
matter axions.

The paper is organized as follow. To set the stage, we start in the next section by a brief overview of the construction of the non-relativistic expansions via the Foldy-Wouthuysen method~\cite{Foldy:1949wa}. This also gives us the opportunity to introduce Schiff's theorem~\cite{Schiff:1963zz} and its generalizations. Then, in Sec.~\ref{SecAxion}, we enter the core of the subject, and perform the non-relativistic expansion of the axionic Hamiltonian up to and including $\mathcal{O}(1/m^{-3})$ terms, firstly in the absence of electromagnetic (EM) fields, secondly for a charged fermion minimally coupled to EM fields, and thirdly for a neutral fermion having electric and magnetic dipole interactions with the EM fields. These results are then put to use in Sec.~\ref{AxEDMobs} to analyze axion-induced lepton and nucleon EDMs, showing how and when some new effects could be expected. Finally, in Sec.~\ref{Ccl}, our results are summarized along with their phenomenological consequences.

\section{Brief overview of the non-relativistic expansion}

The techniques used in the present paper are covered in most textbooks on
relativistic quantum mechanics. In particular, recovering the Pauli equation
by a non-relativistic expansion of the Dirac Hamiltonian for a spin 1/2
field minimally coupled to EM fields,%
\begin{equation}
i\partial_{t}\left\vert \psi\right\rangle =\mathcal{H}_{EM}\left\vert
\psi\right\rangle \ ,\ \ \mathcal{H}_{EM}=\gamma^{0}(\boldsymbol{\gamma}\cdot%
\mathbf{P}+m)+e\phi\ ,  \label{DiracEM}
\end{equation}
where $\mathbf{P}=\mathbf{p}-e\mathbf{A}$, $\mathbf{p}=-i\mathbf{\nabla}$,
and the EM potential is $A^{\mu}=(\phi,\mathbf{A})$, is a standard exercise.
Though well-known, we think it is nevertheless useful to briefly review this
so as to fix our notations, and because it forms the backbone on which we
will add axions later on. Further, once the magnetic moment and electric
moment operators $\sigma_{\mu\nu}F^{\mu\nu}$ and $\sigma_{\mu\nu}\tilde{F}%
^{\mu\nu}$ are added, it permits to introduce the Schiff's theorem that will
be central to the axion discussion.

\subsection{Foldy-Wouthuysen transformation}

The Dirac equation involves four-dimensional spinors, and thus includes both
particles and antiparticles simultaneously. In the non-relativistic limit
though, the energy is not sufficient for pair creation, and the antiparticle
degrees of freedom are not dynamical. In practice, the Dirac equation must
reduce to a decoupled pair of two-dimensional Pauli equations, describing
the dynamics of spin 1/2 particles only. Several procedures exist to perform
this reduction, starting historically by Pauli's elimination method~\cite%
{Pauli:1927qhd}. To set the stage, let us briefly describe the main idea. We
first adopt the Dirac representation for the gamma matrices, that is,%
\begin{equation}
\gamma^{0}=\left( 
\begin{array}{cc}
\mathbf{1} & 0 \\ 
0 & -\mathbf{1}%
\end{array}
\right) \ ,\ \boldsymbol{\gamma}=\left( 
\begin{array}{cc}
0 & \boldsymbol{\sigma} \\ 
-\boldsymbol{\sigma} & 0%
\end{array}
\right) \ ,\ \ \gamma^{5}=\left( 
\begin{array}{cc}
0 & \mathbf{1} \\ 
\mathbf{1} & 0%
\end{array}
\right) \ ,
\end{equation}
where $\boldsymbol{\sigma}$ are the usual Pauli matrices. Note also the
identities $\gamma^{i}\gamma^{j}=(-\delta^{ij}-i\varepsilon^{ijk}\sigma ^{k})%
\mathbf{1}$ and $\boldsymbol{\sigma}\otimes\mathbf{1=}-\gamma^{0}\gamma^{5}%
\boldsymbol{\gamma}$, as well as the fact that $\boldsymbol{\gamma }%
^{\dagger}=\gamma^{0}\boldsymbol{\gamma}\gamma^{0}=-\boldsymbol{\gamma}$,
but $\gamma^{0\dagger}=\gamma^{0}$ and $\gamma^{5\dagger}=\gamma^{5}$. The
diagonal form of $\gamma^{0}$ is instrumental for performing the
non-relativistic expansion. Indeed, if the Dirac spinor $\left\vert
\psi\right\rangle $ is split into a pair of two-component spinors, the Dirac
equation Eq.~(\ref{DiracEM}) takes the matrix form (after $\chi\rightarrow
-\chi$)%
\begin{equation}
\left( 
\begin{array}{cc}
m-E+e\phi & \boldsymbol{\sigma}\cdot\mathbf{P} \\ 
-\boldsymbol{\sigma}\cdot\mathbf{P} & m+E-e\phi%
\end{array}
\right) \left( 
\begin{array}{c}
\varphi \\ 
\chi%
\end{array}
\right) =\left( 
\begin{array}{c}
0 \\ 
0%
\end{array}
\right) \ .
\end{equation}
Factoring out the large time evolution due to the rest mass and defining $%
E^{\prime}=E-mc^{2}$, this becomes%
\begin{equation}
\left( 
\begin{array}{cc}
e\phi & \boldsymbol{\sigma}\cdot\mathbf{P} \\ 
\boldsymbol{\sigma}\cdot\mathbf{P} & -2m+e\phi%
\end{array}
\right) \left( 
\begin{array}{c}
\varphi \\ 
\chi%
\end{array}
\right) =E^{\prime}\left( 
\begin{array}{c}
\varphi \\ 
\chi%
\end{array}
\right) \ .
\end{equation}
Thus, because of the $2m$ term, $\chi$ is essentially determined by $\varphi$%
. It corresponds to a small $\mathcal{O}(v/c)$ component relative to the
large $\varphi$ component. Plugging $\chi\approx\boldsymbol{\sigma}\cdot 
\mathbf{P}\varphi/2m$ back into the equation of $\varphi$ permits to reduce
the Dirac equation to a Pauli equation for $\varphi$,%
\begin{equation}
i\partial_{t}\varphi=\left[ \frac{(\boldsymbol{\sigma}\cdot\mathbf{P})^{2}}{%
2m}+e\phi\right] \varphi=\left[ \frac{(\mathbf{p}-e\mathbf{A})^{2}}{2m}-%
\frac{e}{2m}\boldsymbol{\sigma}\cdot\mathbf{B}+e\phi\right] \varphi\ .
\end{equation}
This is the essence of Pauli elimination method that can be generalized to
the presence of other interactions and to higher orders. In those cases
though, the method becomes very cumbersome because hermiticity of the
Hamiltonian is not guaranteed, and additional renormalizations of the $%
\varphi$ field are in general required~\cite{DeVries:1970pbg}.

The Foldy-Wouthuysen (FW) procedure is designed to systematize the
block-diagonalization of the Dirac Hamiltonian~\cite{Foldy:1949wa}. Starting
from $i\partial_{t}\left\vert \psi\right\rangle =\mathcal{H}\left\vert
\psi\right\rangle $, the idea is to construct a unitary rotation $%
\psi\rightarrow\psi^{\prime}=e^{iS}\psi$ such that $i\partial_{t}\left\vert
\psi^{\prime}\right\rangle =\mathcal{H}^{\prime}\left\vert \psi^{\prime
}\right\rangle $ with 
\begin{equation}
\mathcal{H}^{\prime}=e^{iS}\left( \mathcal{H}-i\partial_{t}\right) e^{-iS}\ ,
\end{equation}
with $\mathcal{H}^{\prime}$ now block-diagonal. This decouples the large
two-component spinor from the small one, and should be valid as long as the
energy involved does not allow for pair creation. Since we started by
performing a unitary transformation, there is no hermiticity issue with $%
\mathcal{H}^{\prime}$. However, in the presence of interactions, an exact
solution for $S$ cannot be found in general, and one relies instead on a
perturbative expansion in $c^{-1}$. That is, instead of a single unitary
transformation $S$, a sequence of unitary transformations is performed to
bring $\mathcal{H}^{\prime}$ to a block diagonal form, up to some order $%
c^{-n}$. For dimensional reasons, an expansion in $1/c$ is essentially
identical to an expansion in $1/m$, so we will rather concentrate on the
latter and keep $c=1$.

Details of this construction are in Appendix~\ref{AppFW}. In summary, one
first uses the diagonal $\gamma^{0}$ to write the Hamiltonian as 
\begin{equation}
\mathcal{H}=\gamma^{0}(m+\mathcal{O})+\mathcal{E}\ ,\ 
\end{equation}
where $\mathcal{O}$ stand for odd terms, $\mathcal{O}\gamma^{0}=-\gamma ^{0}%
\mathcal{O}$, and $\mathcal{E}$ for even terms, $\mathcal{E}\gamma
^{0}=\gamma^{0}\mathcal{E}$. In general, $\mathcal{O}$ and $\mathcal{E}$ are
differential operators that do not commute. The term $\mathcal{O}$ is the
offending one that couples small and large components. So, in the first
step, we must remove it by some unitary transformation $S$. Since to leading
order, $\mathcal{H}^{\prime}=\mathcal{H}+[iS,\mathcal{H}]-\dot{S}+...$, this
cancellation must come from $[iS,\gamma^{0}m]=-\gamma^{0}\mathcal{O}$, that
is, $iS=\mathcal{O}/(2m)$. Performing that transformation cancels the $%
\mathcal{O}$ term in $\mathcal{H}$, but brings back odd terms at higher
orders (proportional to $[\mathcal{O},\mathcal{E}]$, $\mathcal{O}^{3}$,
etc), so the procedure must be iterated up to some given order in $1/m$.
After three steps, the Hamiltonian becomes%
\begin{equation}
\mathcal{H}^{\mathrm{NR}}=\gamma^{0}\left( m-\frac{\mathcal{O}^{2}}{2m}-%
\frac{\mathcal{O}^{4}}{8m^{3}}+\frac{\mathcal{V}_{1}^{2}}{8m^{3}}\right) +%
\mathcal{E}+\frac{[\mathcal{O},\mathcal{V}_{1}]}{8m^{2}}+\mathcal{O}%
(1/m^{4})\ ,  \label{FW}
\end{equation}
where $\mathcal{V}_{1}\equiv\lbrack\mathcal{O},\mathcal{E}]+i\mathcal{\dot{O}%
}$. When applied on a four-component spinor, the upper two and lower two
components are decoupled. Given the choice of $\gamma^{0}$, only the large
upper component needs to be kept, as the lower small component dynamics is
dampened by the rest mass, i.e., by a $\mathbf{P}/m$ factor.

The FW transformation will be the first step in all our developments. Yet,
it is important to stress that it is not the end of the story. As was
realized comparing various block-diagonalization methods, including the
elimination method, there are some ambiguities in the final form of $%
\mathcal{H}^{\mathrm{NR}}$. This simply reflects the fact that additional
unitary transformations $\psi\rightarrow\psi^{\prime}=e^{iS}\psi$ are still
allowed as long as $S$ is even (for a review, see e.g. Ref.~\cite%
{DeVries:1970pbg}). This feature, at the root of Schiff's theorem, will be
used extensively in the following.

\subsection{Application to electromagnetic interactions}

Taking $\mathcal{O}=\boldsymbol{\gamma}\cdot(\mathbf{p}-e\mathbf{A})\equiv%
\boldsymbol{\gamma}\cdot\mathbf{P}$ and $\mathcal{E}=e\phi$, and keeping
only terms up to $\mathcal{O}(1/m^{3})$, the standard result is recovered:%
\begin{align}
\mathcal{H}_{EM}^{\mathrm{NR}} & =\gamma^{0}\left( m+\frac{\mathbf{P}^{2}}{2m%
}-\frac{e\boldsymbol{\sigma}\cdot\mathbf{B}}{2m}-\dfrac{\mathbf{P}^{4}-e\{%
\mathbf{P}^{2},\boldsymbol{\sigma}\cdot\mathbf{B}\}-e^{2}(\mathbf{E}^{2}-%
\mathbf{B}^{2})}{8m^{3}}\right)  \notag \\
& \ \ \ \ +e\phi-\frac{e\left( (\mathbf{\nabla}\cdot\mathbf{E})+i\boldsymbol{%
\sigma}\cdot(\mathbf{\nabla}\times\mathbf{E})+2\boldsymbol{\sigma}\cdot(%
\mathbf{E}\times\mathbf{P})\right) }{8m^{2}}+\mathcal{O}(1/m^{4})\ .
\label{PauliEM}
\end{align}
By convention, $\mathbf{\nabla}$ acts on the quantity immediately to its
right, but $\mathbf{P}$ acts on everything. Note that the $\boldsymbol{%
\sigma }$ matrices occurring are to be interpreted as $\mathbf{1}\otimes 
\boldsymbol{\sigma}$, since this Hamiltonian still acts on four-dimensional
spinors. Yet, being diagonal, the reduction to the Pauli equation is now
trivial. As is well-known, one can identify the Zeeman magnetic coupling $%
\boldsymbol{\sigma}\cdot\mathbf{B}=2\boldsymbol{S}\cdot\mathbf{B}$ with $%
\mathbf{S}$ the spin operator and $g=2$ the magnetic moment, the spin orbit
coupling $\boldsymbol{\sigma}\cdot(\mathbf{E}\times\mathbf{P})$, and the
Darwin term $\mathbf{\nabla}\cdot\mathbf{E}$.

A more interesting application starts by including the higher order magnetic
moment and electric moment operators%
\begin{align}
\mathcal{H}_{EM}& =\gamma ^{0}\left( \boldsymbol{\gamma }\cdot \mathbf{P}+m+%
\frac{\delta _{\mu }}{2}\sigma ^{\mu \nu }F_{\mu \nu }-i\frac{d}{2}\sigma
^{\mu \nu }\gamma ^{5}F_{\mu \nu }\right) +e\phi   \notag \\
& =\gamma ^{0}\left( \boldsymbol{\gamma }\cdot \mathbf{P}+m+i\gamma ^{0}%
\boldsymbol{\gamma }\cdot ((\delta _{\mu }\mathbf{E}+d\mathbf{B})+i\gamma
^{5}(\delta _{\mu }\mathbf{B}-d\mathbf{E}))\right) +e\phi \ ,  \label{HEMad}
\end{align}%
where electromagnetic fields satisfy $F^{0i}=-E^{i}$, $B^{i}=-1/2\varepsilon
^{ijk}F_{jk}$ and $\delta _{\mu }\equiv ea/2m$. Plugging the odd term $%
\mathcal{O}=\boldsymbol{\gamma }\cdot \mathbf{P}+i\gamma ^{0}\boldsymbol{%
\gamma }\cdot (\delta _{\mu }\mathbf{E}+d\mathbf{B})$ and the even term $%
\mathcal{E}=e\phi +\gamma ^{5}\boldsymbol{\gamma }\cdot (\delta _{\mu }%
\mathbf{B}-d\mathbf{E})$ in Eq.~(\ref{FW}), keeping in mind that $\delta
_{\mu }$ and $d$ are $\mathcal{O}(m^{-1})$, and discarding terms of $%
\mathcal{O}(m^{-4})$ and higher, the block-diagonal Hamiltonian is now%
\begin{align}
\mathcal{H}_{EM}^{\mathrm{NR}}& =\gamma ^{0}\left( m+\frac{\mathbf{P}^{2}}{2m%
}-\frac{e\left( 1+a\right) \boldsymbol{\sigma }\cdot \mathbf{B}}{2m}+d%
\boldsymbol{\sigma }\cdot \mathbf{E}-\frac{\mathbf{P}^{4}-e\{\mathbf{P}^{2},%
\boldsymbol{\sigma }\cdot \mathbf{B}\}-e^{2}(\mathbf{E}^{2}-\mathbf{B}^{2})}{%
8m^{3}}\right)   \notag \\
& \ \ \ \ +e\phi +\frac{ie(1+2a)[\boldsymbol{\gamma }\cdot \mathbf{P},%
\boldsymbol{\gamma }\cdot \mathbf{E}]}{8m^{2}}+\frac{id[\boldsymbol{\gamma }%
\cdot \mathbf{P},\boldsymbol{\gamma }\cdot \mathbf{B}]}{2m}  \notag \\
& \ \ \ \ +\gamma ^{0}\left( \frac{a(a+1)e^{2}}{8m^{3}}\mathbf{E}^{2}+\frac{%
d^{2}}{2m}\mathbf{B}^{2}-\frac{e(1+2a)d}{8m^{2}}\{\boldsymbol{\gamma }\cdot 
\mathbf{B},\boldsymbol{\gamma }\cdot \mathbf{E}\}\right)   \notag \\
& \ \ \ \ +\frac{\gamma ^{5}\{\boldsymbol{\gamma }\cdot \mathbf{P},\{%
\boldsymbol{\gamma }\cdot \mathbf{P},\boldsymbol{\gamma }\cdot (\delta _{\mu
}\mathbf{B}-d\mathbf{E})\}\}+\gamma ^{0}\{\boldsymbol{\gamma }\cdot \mathbf{P%
},\boldsymbol{\gamma }\cdot (\delta _{\mu }\mathbf{\dot{E}}+d\mathbf{\dot{B}}%
)\}}{8m^{2}}+\mathcal{O}(1/m^{4})\ .  \label{HEM}
\end{align}%
The magnetic operator thus describes the deviation of the magnetic moment
from its Dirac value, $a=(g-2)/2$. The $\boldsymbol{\sigma }\cdot \mathbf{E}$
term describes the electric dipole interaction, with $d$ the EDM. If we
remember the identities 
\begin{subequations}
\label{IdComm}
\begin{align}
\lbrack \boldsymbol{\gamma }\cdot \mathbf{P},\boldsymbol{\gamma }\cdot 
\mathbf{X}]& =-(\mathbf{p}\cdot \mathbf{X})-i\boldsymbol{\sigma }\cdot (%
\mathbf{p}\times \mathbf{X})+2i\boldsymbol{\sigma }\cdot (\mathbf{X}\times 
\mathbf{P})\ , \\
\{\boldsymbol{\gamma }\cdot \mathbf{P},\boldsymbol{\gamma }\cdot \mathbf{X}%
\}& =-(\mathbf{p}\cdot \mathbf{X})-i\boldsymbol{\sigma }\cdot (\mathbf{p}%
\times \mathbf{X})-2(\mathbf{X}\cdot \mathbf{P})\ ,
\end{align}%
the Darwin and spin-orbit couplings are identified inside $[\boldsymbol{%
\gamma }\cdot \mathbf{P},\boldsymbol{\gamma }\cdot \mathbf{E}]$, now
modified by a magnetic moment contribution and accompanied by magnetic
interactions induced by $d$.

\subsection{Schiff's theorem and beyond}
\label{SecSchiff}

As stated before, the FW transformed Hamiltonian can still be unitarily
rotated without breaking its block-diagonal character. The simplest such
transformation is 
\end{subequations}
\begin{equation}
iS_{1}=-\frac{i\alpha}{m}\gamma^{5}\boldsymbol{\gamma}\cdot\mathbf{P}\ .
\label{SchiffTf1}
\end{equation}
The transformation $\exp(iS_{1})$ is unitary, and importantly, it commutes
with the mass term $\gamma^{0}m$. One should not be put off by the fact that
this transformation involves the external fields via $\mathbf{P}$. Actually,
we already did many such transformations to block-diagonalize the
Hamiltonian, since the first FW transformation is $\exp(iS)$ with $iS=%
\mathcal{O}/(2m)$ and $\mathcal{O}=\boldsymbol{\gamma}\cdot\mathbf{P}%
+i\gamma^{0}\boldsymbol{\gamma }\cdot(\delta_{\mu}\mathbf{E}+d\mathbf{B})$.
All that differs here is the $\gamma^{5}$ factor, making $S_{1}$ even with
respect to $\gamma^{0}$.

The new Hamiltonian $\mathcal{H}^{\prime }=e^{iS_{1}}\left( \mathcal{H}%
-i\partial _{t}\right) e^{-iS_{1}}$ can be expanded as before, and since $%
iS_{1}\sim \mathcal{O}(m^{-1})$, we need to compute:%
\begin{equation}
\mathcal{H}^{\prime }=\mathcal{H}+[iS_{1},\mathcal{H}]-\dot{S}_{1}+\frac{1}{2%
}[iS_{1},[iS_{1},\mathcal{H}]-\dot{S}_{1}]+\frac{1}{3!}%
[iS_{1},[iS_{1},[iS_{1},\mathcal{H}]-\dot{S}_{1}]]+\mathcal{O}(m^{-4})\ .
\label{SchiffTf2}
\end{equation}%
Now, the key in Schiff's theorem~\cite{Schiff:1963zz} is to note that the $%
\mathcal{O}(1/m)$ terms miraculously combine as%
\begin{equation}
\lbrack iS_{1},e\phi ]-\dot{S}_{1}=-\frac{e\alpha }{m}\gamma ^{5}\boldsymbol{%
\gamma }\cdot (\mathbf{\nabla }\phi +\mathbf{\dot{A}})=\frac{e\alpha }{m}%
\gamma ^{5}\boldsymbol{\gamma }\cdot \mathbf{E}=-\frac{e\alpha }{m}\gamma
^{0}\boldsymbol{\sigma }\cdot \mathbf{E\ }.  \label{SchiffTf3}
\end{equation}%
Thus, with $\alpha =md/e$, the EDM term in $\mathcal{H}_{EM}^{\mathrm{NR}}$
is rotated away! More accurately, we should say that it is transformed into
higher order corrections coming from the rest of Eq.~(\ref{SchiffTf2}).
After some algebra, the transformed Hamiltonian is found to be, keeping only
terms at most linear in either $a$ or $d$, since these quantities are
experimentally small,%
\begin{align}
\mathcal{H}_{EM}^{\mathrm{NR}}& =\gamma ^{0}\left( m+\frac{\mathbf{P}^{2}}{2m%
}-\frac{e\left( 1+a\right) \boldsymbol{\sigma }\cdot \mathbf{B}}{2m}-\frac{%
\mathbf{P}^{4}-e\{\mathbf{P}^{2},\boldsymbol{\sigma }\cdot \mathbf{B}%
\}-e^{2}(\mathbf{E}^{2}-\mathbf{B}^{2})}{8m^{3}}\right) +e\phi   \notag \\
& \ \ \ \ +ie\frac{1+2a}{8m^{2}}[\boldsymbol{\gamma }\cdot \mathbf{P},%
\boldsymbol{\gamma }\cdot \mathbf{E}]+\frac{id}{2m}[\boldsymbol{\gamma }%
\cdot \mathbf{P},\boldsymbol{\gamma }\cdot \mathbf{B}]  \notag \\
& \ \ \ \ +\gamma ^{0}\left( \frac{ae^{2}}{8m^{3}}\mathbf{E}^{2}-\frac{ed}{%
8m^{2}}\{\boldsymbol{\gamma }\cdot \mathbf{B},\boldsymbol{\gamma }\cdot 
\mathbf{E}\}\right)   \notag \\
& \ \ \ \ +\frac{\gamma ^{5}\{\boldsymbol{\gamma }\cdot \mathbf{P},\{%
\boldsymbol{\gamma }\cdot \mathbf{P},\boldsymbol{\gamma }\cdot (\delta _{\mu
}\mathbf{B}-d\mathbf{E})\}\}+\gamma ^{0}\{\boldsymbol{\gamma }\cdot \mathbf{P%
},\boldsymbol{\gamma }\cdot (\delta _{\mu }\mathbf{\dot{E}}+d\mathbf{\dot{B}}%
)\}}{8m^{2}}  \notag \\
& \ \ \ \ \ +\frac{d}{8m^{2}}\gamma ^{5}[\boldsymbol{\gamma }\cdot \mathbf{P}%
,[\boldsymbol{\gamma }\cdot \mathbf{P},\boldsymbol{\gamma }\cdot \mathbf{E}%
]]+\mathcal{O}(1/m^{4})\ ,  \label{HEMSch}
\end{align}%
where the only non-trivial reduction is $[\boldsymbol{\gamma }\cdot \mathbf{P%
},\mathbf{P}^{2}]=-e\gamma ^{0}\gamma ^{5}[\boldsymbol{\gamma }\cdot \mathbf{%
P},\boldsymbol{\gamma }\cdot \mathbf{B}]$, the rest being straightforward
algebraic manipulations.

\subsubsection{Higher order Schiff transformations and operator redundancies}

Central to Schiff's theorem is the presence of the $-\dot{S}_{1}$ piece that
directly enters in the transformed Hamiltonian in Eq.~(\ref{SchiffTf2}), and
can thus directly interfere with the other terms. When applied on $\gamma
^{5}\boldsymbol{\gamma}\cdot\mathbf{P}$, it generates a $\gamma^{5}%
\boldsymbol{\gamma}\cdot\mathbf{\dot{A}}$ term out of which $\gamma ^{5}%
\boldsymbol{\gamma}\cdot\mathbf{E}$ emerges without an additional $m^{-1}$
factor. This same trick can be used for any term that involves time
derivatives of external fields. For instance, consider now%
\begin{equation}
iS_{2}=\frac{i\beta}{8m^{2}}\gamma^{0}\{\boldsymbol{\gamma}\cdot \mathbf{P},%
\boldsymbol{\gamma}\cdot(\delta_{\mu}\mathbf{E}+d\mathbf{B})\}\ .
\label{QEDS2}
\end{equation}
Since it is already of $\mathcal{O}(m^{-3})$, only the leading commutator
with $e\phi$ needs to be computed. Again, $[iS_{2},e\phi]$ combine with the $%
\mathbf{\dot{P}}$ in $-\dot{S}_{2}$ to give a $\mathbf{\nabla}\phi +\mathbf{%
\dot{A}}=-\mathbf{E}$ factor:%
\begin{equation}
\lbrack iS_{2},e\phi]-\dot{S}_{2}=-\frac{e\beta}{8m^{2}}\gamma^{0}\{%
\boldsymbol{\gamma}\cdot\mathbf{E},\boldsymbol{\gamma}\cdot(\delta_{\mu }%
\mathbf{E}+d\mathbf{B})\}-\frac{\beta}{8m^{2}}\gamma^{0}\{\boldsymbol{\gamma 
}\cdot\mathbf{P},\boldsymbol{\gamma}\cdot(\delta_{\mu}\mathbf{\dot{E}}+d%
\mathbf{\dot{B}})\}\ .
\end{equation}
This time though, we find a redundancy among $\mathcal{O}(1/m^{3})$
operators, up to higher order corrections. Our preferred choice is to take $%
\beta=1$ to get rid of the $\mathbf{\dot{E}}$ and $\mathbf{\dot{B}}$
operators, but one could equally well decide to keep the $\mathbf{\dot{E}}$
operator and eliminate the $\mathbf{E}^{2}$ term, or keep the $\mathbf{\dot{B%
}}$ term and eliminate the $\mathbf{E}\cdot\mathbf{B}$ couplings. A third
possible transformation is%
\begin{equation}
iS_{3}=\frac{i\varepsilon}{m^{3}}\gamma^{5}\{\{\boldsymbol{\gamma}\cdot%
\mathbf{P},\boldsymbol{\gamma}\cdot\mathbf{P}\},\boldsymbol{\gamma}\cdot%
\mathbf{P}\}\ ,  \label{QEDS3}
\end{equation}
which also introduces a redundancy among $\mathcal{O}(1/m^{3})$ operators,
up to higher order corrections,%
\begin{equation}
\lbrack iS_{3},e\phi]-\dot{S}_{3}=-\frac{e\varepsilon}{m^{3}}%
\gamma^{5}\left( 2\{\boldsymbol{\gamma}\cdot\mathbf{P},\{\boldsymbol{\gamma}%
\cdot \mathbf{P},\boldsymbol{\gamma}\cdot\mathbf{E}\}\}+\{\boldsymbol{\gamma}%
\cdot\mathbf{E},\{\boldsymbol{\gamma}\cdot\mathbf{P},\boldsymbol{\gamma}\cdot%
\mathbf{P}\}\}\right) \ .  \label{RedunEM}
\end{equation}
These redundancies can be used to reduce the number of relevant operators.
In Appendix~\ref{AppEM}, we present one possible choice of $S_{2}$ and $%
S_{3} $ that bring $\mathcal{H}_{EM}^{\mathrm{NR}}$ to a somewhat optimal
form. It should be stressed though that the final coefficients for the
higher order operators in $\mathcal{H}_{EM}^{\mathrm{NR}}$ should not be
taken too literally. Indeed, once adopting an effective description with the 
$\mathcal{O}(1/m)$ couplings $\sigma^{\mu\nu}F_{\mu\nu}$ and $\sigma^{\mu\nu
}\tilde{F}_{\mu\nu}$, one could in principle also include $\mathcal{O}%
(1/m^{2})$ or $\mathcal{O}(1/m^{3})$ operators. For example, if one adds the 
$F_{\mu\nu}F^{\mu\nu}$ or $F_{\mu\nu}\tilde{F}^{\mu\nu}$ operators in $%
\mathcal{H}_{EM}$, their coefficients will directly correct those of $%
\mathbf{E}^{2}-\mathbf{B}^{2}$ and $\mathbf{E}\cdot\mathbf{B}$ in $\mathcal{H%
}_{EM}^{\mathrm{NR}}$.

Finally, it is worth to stress that this list certainly does not exhaust
possible unitary transformations, and that not all such transformations
encode useful information. For example, consider%
\begin{equation}
iS_{4}=\frac{i\eta }{m^{2}}\{\boldsymbol{\gamma }\cdot \mathbf{P},%
\boldsymbol{\gamma }\cdot \mathbf{P}\}\ ,
\end{equation}%
which is even and hermitian for $\eta $ real. The change in the Hamiltonian
is 
\begin{equation}
\lbrack iS_{4},\mathcal{H}]-\dot{S}_{4}=-\frac{2e\eta }{m^{2}}\{\boldsymbol{%
\gamma }\cdot \mathbf{P},\boldsymbol{\gamma }\cdot \mathbf{E}\}+\mathcal{O}%
(1/m^{4})\ .
\end{equation}%
This transformation just adds the $\{\boldsymbol{\gamma }\cdot \mathbf{P},%
\boldsymbol{\gamma }\cdot \mathbf{E}\}$ operator to the Hamiltonian, up to
higher order terms. To understand why this has no impact on the physics, let
us first expand it using Eq.~(\ref{IdComm}),%
\begin{equation}
-\frac{2e\eta }{m^{2}}\{\boldsymbol{\gamma }\cdot \mathbf{P},\boldsymbol{%
\gamma }\cdot \mathbf{E}\}=-i\frac{2e\eta }{m^{2}}\left( \mathbf{\nabla }%
\cdot \mathbf{E}+i\boldsymbol{\sigma }\cdot (\mathbf{\nabla }\times \mathbf{E%
})+2\mathbf{E}\cdot \mathbf{\nabla }\right) \ .
\end{equation}%
If we could take $\eta $ imaginary, this operator would interfere with the
Darwin and spin-orbit operator $i[\boldsymbol{\gamma }\cdot \mathbf{P},%
\boldsymbol{\gamma }\cdot \mathbf{E}]$, but this would make $\exp (iS_{4})$
non-unitary. Actually, this operator has no impact because $\mathbf{\nabla }%
\cdot \mathbf{E}$ and $2\mathbf{E}\cdot \mathbf{\nabla }$ compensate each
other when acting on wavefunctions (both are standard forms for the Darwin
operator), while the $\boldsymbol{\sigma }\cdot (\mathbf{\nabla }\times 
\mathbf{E})=-\boldsymbol{\sigma }\cdot \mathbf{\dot{B}}$ term drops out for
static fields (and could be rotated away by a dedicated unitary
transformation with $S_{5}\sim \boldsymbol{\sigma }\cdot \mathbf{B}$ anyway).

\subsubsection{Schiff theorem and charged fermion EDMs}
\label{SecNeutSchiff}

Schiff's theorem shows that the energy of a charged particle cannot be
influenced by its EDM at leading order. The naive interpretation of this
result is that a charged particle plunged in an electric field would feel
the Lorentz force and fly away. The Schiff's transformation is then viewed
as a translation that moves us in some sort of rest frame for the charged
fermion in which there is no electric field anymore, hence where the EDM
operator vanishes and cannot\emph{\ }contribute to Stark energy shifts.
Thus, for charged fermions, $[\boldsymbol{\gamma }\cdot \mathbf{P},%
\boldsymbol{\gamma }\cdot \mathbf{B}]$ encodes the leading impact the EDM
has on the particle energies in the non-relativistic limit. Using Eq.~(\ref%
{IdComm}), one can recognize in this term the spin-dependent $\boldsymbol{%
\sigma }\cdot (\mathbf{p}\times \mathbf{B})$ coupling discussed originally
by Schiff~\cite{Schiff:1963zz}. To feel the EDM with electric fields, one
has to go fetch the $\mathcal{O}(1/m^{3})$ operators $\gamma ^{5}\{%
\boldsymbol{\gamma }\cdot \mathbf{P},\{\boldsymbol{\gamma }\cdot \mathbf{P},%
\boldsymbol{\gamma }\cdot \mathbf{E}\}\}$ or $\gamma ^{5}[\boldsymbol{\gamma 
}\cdot \mathbf{P},[\boldsymbol{\gamma }\cdot \mathbf{P},\boldsymbol{\gamma }%
\cdot \mathbf{E}]]$ which, thanks to Eq.~(\ref{RedunEM}), can both be
reduced to $\gamma ^{0}\{\mathbf{P}^{2},\boldsymbol{\sigma }\cdot \mathbf{E}%
\}$. This operator thus encodes the leading relativistic corrections. For
the case of an electron in a heavy atom, significant enhancements of this
operator have been found that guarantee an experimental sensitivity to the
electron EDM~\cite{Khriplovich:1997ga}. Finally, it should be mentioned that
another way to evade the shielding of the EDM is to account for finite-size
effects, that clearly go beyond the current formalism (for a review, see
Ref.~\cite{Sandars:2001nq,Liu:2007zf}).

Schiff's theorem is a statement about the Hamiltonian, and thus applies to
the energy levels of bound charged fermions. It does not mean that the EDM
operator cannot be felt using other observables. In particular, the electric
field does exert a torque on the spin of charged fermions, leading to its
precession in adequate experimental settings. Specifically, to leading
order, the spin operator $\boldsymbol{S}=\boldsymbol{\sigma }/2$ evolves
according to $\mathcal{H}=\gamma ^{0}\left( m+d\boldsymbol{\sigma }\cdot 
\mathbf{E}\right) +e\phi $ as%
\begin{equation}
\boldsymbol{\dot{S}}^{i}=-i[\boldsymbol{S}^{i},\mathcal{H}]=\frac{i}{2}%
d\gamma ^{0}[\boldsymbol{\gamma }^{i},\boldsymbol{\gamma }\cdot \mathbf{E}%
]=d\gamma ^{0}\varepsilon ^{ijk}\boldsymbol{\sigma }^{k}\mathbf{E}%
^{j}=-2d\gamma ^{0}\boldsymbol{S}\times \mathbf{E}\ .
\end{equation}%
which is nothing but one term of the generalized Bargmann-Michel-Telegdi
equation~\cite{Bargmann:1959gz,Nowakowski:2004cv}. After the Schiff
rotation, the $d\boldsymbol{\sigma }\cdot \mathbf{E}$ operator is removed
from $\mathcal{H}$, and the spin operator in that basis satisfies%
\begin{equation}
\boldsymbol{\dot{S}}^{\prime }=-i[\boldsymbol{S}^{\prime },\mathcal{H}%
^{\prime }]=-i[\boldsymbol{S}^{\prime },e\phi ]\ .
\end{equation}%
The crucial point that makes this equation compatible with that of $%
\boldsymbol{S}$ is that the spin operator does not commute with the Schiff
transformation, so that $\boldsymbol{S}^{\prime }=e^{iS_{1}}\boldsymbol{S}%
e^{-iS_{1}}\neq \boldsymbol{S}$. Plugging this in the above equation implies%
\begin{equation}
\frac{d}{dt}(e^{iS_{1}}\boldsymbol{S}e^{-iS_{1}})=-i[e^{iS_{1}}\boldsymbol{S}%
e^{-iS_{1}},e\phi ]\Longrightarrow \boldsymbol{\dot{S}}=i\left[ \boldsymbol{S%
},[iS_{1},e\phi ]-\dot{S}_{1}\right] \ ,
\end{equation}%
which obviously holds by construction, since $[iS_{1},e\phi ]-\dot{S}%
_{1}=-\gamma ^{0}d\boldsymbol{\sigma }\cdot \mathbf{E}$, see Eq.~(\ref%
{SchiffTf3}). This exercise provides another interpretation of Schiff's
theorem. In a gauge in which $\phi =0$, $\boldsymbol{S}^{\prime }$ appears
constant in time, so the Schiff rotation of Eq.~(\ref{SchiffTf1}) is
actually that to the rotating frame in which the spin appears static. This
could have been guessed from the start if one notes that $S_{1}$ actually
involves the helicity operator, $\gamma ^{5}\boldsymbol{\gamma }\cdot 
\mathbf{P}=-\gamma ^{0}\boldsymbol{\sigma }\cdot \mathbf{P}$.

\subsubsection{Schiff theorem and neutral fermion EDMs}

Schiff's theorem cannot apply to neutral fermions. Indeed, one can simply
send $e\rightarrow 0$ to decouple EM fields in Eq.~(\ref{HEMad}) while
keeping an explicit EDM term, but the parameter of the Schiff's
transformation in Eq.$~$(\ref{SchiffTf1}) has to be set to $\alpha =md/e$,
which is undefined in that limit. Said differently, it is only through a
delicate interplay with the couplings to the external EM fields that the
Schiff's transformation can interfere destructively with the EDM term.\
Thus, for the neutron, all one can do is to eliminate the $\mathbf{\dot{E}}$
and $\mathbf{\dot{B}}$ couplings, and starting from Eq.~(\ref{HEMad}) in the 
$e\rightarrow 0$ limit, one ends up with%
\begin{align}
\left. \mathcal{H}_{EM}^{\mathrm{NR}}\right\vert _{e\rightarrow 0}& =\gamma
^{0}\left( m+\frac{\mathbf{p}^{2}}{2m}-\frac{\mathbf{p}^{4}}{8m^{3}}-\delta
_{\mu }\boldsymbol{\sigma }\cdot \mathbf{B}+d\boldsymbol{\sigma }\cdot 
\mathbf{E}+\frac{(\delta _{\mu }\mathbf{E}+d\mathbf{B})^{2}}{2m}\right)  
\notag \\
& \ \ \ \ +\frac{i[\boldsymbol{\gamma }\cdot \mathbf{p},\boldsymbol{\gamma }%
\cdot (\delta _{\mu }\mathbf{E}+d\mathbf{B})]}{2m}+\frac{\gamma ^{5}\{%
\boldsymbol{\gamma }\cdot \mathbf{p},\{\boldsymbol{\gamma }\cdot \mathbf{p},%
\boldsymbol{\gamma }\cdot (\delta _{\mu }\mathbf{B}-d\mathbf{E})\}\}}{8m^{2}}%
+\mathcal{O}(1/m^{4})\ .  \label{HEDMMDM}
\end{align}%
It is not possible to rotate away the EDM. At the fundamental level, $d$ is
induced by all the CP-violating operators involving gluons and/or quarks
(for a review, see e.g. Ref.~\cite{Pospelov:2005pr}). The most important
contribution is that of the $\theta $ term, at the root of the strong CP
puzzle, and which is estimated as~\cite{Pospelov:1999ha,Yamanaka:2017mef}%
\begin{equation}
d_{n}=-(2.7\pm 1.2)\times 10^{-16}\theta ~e\text{ cm\ .}  \label{thetaEDM}
\end{equation}%
In the SM, the CKM contribution is negligible, but in principle, some New
Physics may also induce fundamental EDMs for the quarks (see e.g. Ref.~\cite%
{Smith:2017dtz} and references cited there). As those are certainly far from
non-relativistic inside a neutron, Schiff's theorem should be largely
evaded. In a $SU(6)$ model, the neutron EDM receives then the additional
contribution~\cite{Pospelov:2005pr}%
\begin{equation}
d_{n}=\frac{4}{3}d_{d}-\frac{1}{3}d_{u}\ ,  \label{EDMn}
\end{equation}%
Note that the same rather naive model gives $\delta _{\mu }=(4\mu _{d}-\mu
_{u})/3$ with $\mu _{u,d}=e/m_{u,d}$. With constituent quark masses $%
m_{u}=m_{d}=m_{N}/3$, this gives $\delta _{\mu }=-2e/(2m_{N})$, in fairly
good agreement with the measured $\delta _{\mu }=-1.913e/(2m_{N})$. Though
this hardly suffices to justify Eq.~(\ref{EDMn}) as there is no analog of
Schiff's screening for the magnetic moment, it is in fairly good agreement
with recent lattice calculations~\cite{Dekens:2018bci} $d_{n}\approx
(0.82\pm 0.03)d_{d}-(0.21\pm 0.01)d_{u}$. This shows that Schiff's screening
theorem does not apply to quarks, as could have been expected since those
are bound not by the electromagnetic interactions but by the strong
interactions.

\section{Axion interactions in the non-relativistic limit}
\label{SecAxion}

Nowadays, the equivalence between the pseudoscalar and derivative axial
interactions is understood as particular application of the general
reparametrization theorem to Goldstone bosons~\cite{Kamefuchi:1961sb}. Let
us recall the essence of the argument (see Ref.~\cite{Quevillon:2019zrd} for
more details). For a typical axion model, one starts with a spontaneously
broken chiral symmetry, $U(1)_{PQ}$. Then, the statement that Goldstone
boson $a$ must interact derivatively leads to the unique interaction term
(remember that $g=m/\Lambda$ is absorbed into $a$):%
\begin{equation}
\mathcal{L}_{D}=\bar{\psi}\left( \slashed \partial-m+\frac{%
\gamma^{\mu}\gamma_{5}\partial_{\mu}a}{m}\right) \psi\ .  \label{Lder}
\end{equation}
Obviously, this interaction is invariant under constant shifts of the
Goldstone field. This is the standard form for most axion analyses, but one
should emphasize that the Goldstone field is actually parametrized
non-linearly in this representation, since its simple shifts $a\rightarrow
a+m\theta$ must span the vacuum manifold as $\theta$ varies between $0$ and $%
2k\pi$. Another peculiarity of this representation is that the fermion field $%
\psi$ ends up neutral under the original $U(1)_{PQ}$ symmetry, even though
it must initially be charged otherwise it would not occur in the Noether
current of the $U(1)_{PQ}$ symmetry.

This interpretation is made clearer by adopting a different parametrization
for the fields, that in which the fermion field keeps its original charge.
This is achieved via a chiral reparametrization of the fermion field 
\begin{equation}
\psi \rightarrow \exp (ia\gamma _{5}/m)\psi \ .  \label{FermionRepar}
\end{equation}%
When plugged in $\mathcal{L}_{D}$, the derivative coupling is replaced by a
tower of pseudoscalar interactions%
\begin{equation}
\mathcal{L}_{E}=\bar{\psi}\left( \slashed\partial -m\exp \left( 2i\gamma ^{5}%
\frac{a}{m}\right) \right) \psi \ .  \label{Lpol}
\end{equation}%
A trivial mass term $m\bar{\psi}\psi $ necessarily breaks the axial symmetry 
$U(1)_{PQ}$, so the fermion mass must arise through the symmetry breaking
itself, like in the SM. Such a mass term is still invariant under the
original chiral symmetry because the phase the fermion field acquires under $%
U(1)_{PQ}$, $\psi \rightarrow \exp (i\theta \gamma _{5})\psi $, is
compensated by the shift of the Goldstone field $a\rightarrow a+m\theta $.

Now, to leading order in $a$, this interaction produces a pseudoscalar
coupling of the fermion to the axion:%
\begin{equation}
\mathcal{L}_{E}=\bar{\psi}(i\slashed\partial -m-2ia\gamma _{5})\psi +...\ .
\end{equation}%
Truncating the theory in this way, one should remember that $\mathcal{O}%
(a^{2})$ terms and above are neglected. This approximation is only valid for
on-shell fermions, since by integration by part, $\bar{\psi}(\gamma ^{\mu
}\gamma _{5}\partial _{\mu }a)\psi =-\bar{\psi}(2im\gamma _{5}a)\psi $ upon
enforcing the free equation of motion $(i\overrightarrow{\slashed\partial }%
-m)\psi =\bar{\psi}(i\overleftarrow{\slashed\partial }+m)=0$. As mentioned
in the Introduction, part of the historic controversy on the equivalence
between the axial and pseudoscalar descriptions of nucleon-pion interactions
has to do with this truncation. Nowadays, the equivalence between both
representations is built in chiral effective theories. For axion models, it
is not always fully embedded yet, as we will see. Further, additional care
is needed because the $U(1)_{PQ}$ symmetry being anomalous, so is the chiral
reparametrization Eq.~(\ref{FermionRepar}). As analyzed in details in Ref.~%
\cite{Quevillon:2019zrd} (see also Refs.~\cite%
{Quevillon:2021sfz,Bonnefoy:2020gyh}), the two representations are then
equivalent only up to the presence of specific anomalous contact
interactions of the axion to gauge bosons. In the present section, these
effects are not relevant and will not be discussed further, but we will come back to
them when analyzing the passage from the quark to the nucleon level in Sec.~\ref{AxEDMobs}.

Our goal is to construct and analyze the axion-fermion interactions in the
non-relativistic limit. To treat both representations simultaneously, we
adopt the trick proposed by Friar a long time ago and described in Ref.~\cite%
{Friar:1977xh}. Specifically, let us start from $\mathcal{L}_{D}$. The
Euler-Lagrange equation gives $i\partial _{t}\left\vert \psi \right\rangle =%
\mathcal{H}_{D}\left\vert \psi \right\rangle $ with%
\begin{equation}
\mathcal{H}_{D}=\gamma ^{0}\left( \boldsymbol{\gamma }\cdot \mathbf{p}+m-%
\frac{\gamma ^{0}\gamma _{5}\dot{a}}{m}+\frac{\gamma _{5}\boldsymbol{\gamma }%
\cdot \mathbf{\nabla }a}{m}\right) \ .
\end{equation}%
Then, we partially perform the fermion reparametrization Eq.~(\ref%
{FermionRepar}), which is nothing but a unitary transformation $\psi
\rightarrow \psi =e^{iS(\mu )}\psi $ with%
\begin{equation}
iS(\mu )=-\frac{i\mu }{m}a\gamma _{5}\ .  \label{SFriar}
\end{equation}%
Calculating $\mathcal{H}(\mu )=e^{iS(\mu )}\left( \mathcal{H}_{D}-i\partial
_{t}\right) e^{-iS(\mu )}$ with the help of $\exp (i\alpha \gamma _{5})=\cos
\alpha +i\gamma _{5}\sin \alpha $, we find%
\begin{equation}
\mathcal{H}(\mu )=\gamma ^{0}\left( \boldsymbol{\gamma }\cdot \mathbf{p}+%
\frac{\mu -1}{m}\gamma ^{0}\gamma _{5}\dot{a}+\frac{1-\mu }{m}\gamma _{5}%
\boldsymbol{\gamma }\cdot \mathbf{\nabla }a+m\exp \left( \frac{2i\mu }{m}%
a\gamma ^{5}\right) \right) \ .  \label{muH}
\end{equation}%
So, this form permits to interpolate between the exponential and derivative
representations, with $\mathcal{H}(0)=\mathcal{H}_{D}$ and $\mathcal{H}(1)=%
\mathcal{H}_{E}$. Let us now perform the non-relativistic expansion of this
expression, first as it stands, and then adding electromagnetic interactions.

\subsection{In the absence of EM fields}

\label{SecNeutral}

In the Hamiltonian Eq.~(\ref{muH}), the terms $\boldsymbol{\gamma}\cdot%
\mathbf{p}$ and $\gamma_{5}\boldsymbol{\gamma}\cdot\mathbf{\nabla }%
a=\gamma^{5}[\boldsymbol{\gamma}\cdot\mathbf{p},a]$ are diagonal since $%
\gamma^{5}\boldsymbol{\gamma}=-\gamma^{0}\boldsymbol{\sigma}$, but the $%
\gamma_{5}$ piece coming from the exponential and $\gamma_{5}\dot{a}$ are
not. Splitting the exponential using $\exp(i\alpha\gamma_{5})=\cos
\alpha+i\gamma_{5}\sin\alpha$, the elements to be used for the FW
transformation are%
\begin{equation}
\mathcal{O}=\boldsymbol{\gamma}\cdot\mathbf{p}-\frac{1-\mu}{m}%
\gamma^{0}\gamma^{5}\dot{a}+i\gamma^{5}S_{a}\ ,\ \ \ \mathcal{E}=\frac{1}{m}%
\gamma ^{0}C_{a}+i\frac{1-\mu}{m}\gamma^{0}\gamma^{5}[\boldsymbol{\gamma}%
\cdot\mathbf{p},a]\ ,
\end{equation}
with $S_{a}\equiv m\sin(2\mu a/m)=2\mu a+...$ and $C_{a}\equiv
m^{2}(\cos(2\mu a/m)-1)=-2\mu^{2}a^{2}+...$. The calculation, though
cumbersome, does not present any particular difficulty and we find%
\begin{align}
\mathcal{H}^{\mathrm{NR}}(\mu) & =\gamma^{0}\left( m+\frac{\mathbf{p}^{2}}{2m%
}-\frac{\mathbf{p}^{4}}{8m^{3}}+\frac{i}{2m}\gamma^{5}[\boldsymbol{\gamma }%
\cdot\mathbf{p},S_{a}+2(1-\mu)a]\right)  \notag \\
& \ \ \ \ +\frac{\gamma^{5}\{\boldsymbol{\gamma}\cdot\mathbf{p},\dot{S}%
_{a}+4(1-\mu)\dot{a}\}}{8m^{2}}+\frac{\gamma^{0}\mathcal{H}_{3}}{8m^{3}} 
\notag \\
& \ \ \ \ +\gamma^{0}\left( \frac{S_{a}^{2}+2C_{a}}{2m}-\frac{%
4S_{a}^{2}C_{a}+S_{a}^{4}}{8m^{3}}\right) +\mathcal{O}(1/m^{4})\ ,
\label{AxFW1}
\end{align}
with%
\begin{align}
\mathcal{H}_{3} & =4(1-\mu)^{2}\dot{a}^{2}+2(1-\mu)\dot{a}\dot{S}_{a}+\dot{S}%
_{a}^{2}-i(1-\mu)\gamma^{5}[\boldsymbol{\gamma}\cdot\mathbf{p},\ddot{a}%
]-2(1-\mu)\ddot{a}S_{a}  \notag \\
& \ \ \ \ +i(1-\mu)\gamma^{5}[\boldsymbol{\gamma}\cdot\mathbf{p},[%
\boldsymbol{\gamma}\cdot\mathbf{p},[\boldsymbol{\gamma}\cdot\mathbf{p}%
,a]]]-i\gamma^{5}\{\mathbf{p}^{2},[\boldsymbol{\gamma}\cdot\mathbf{p}%
,S_{a}]\}  \notag \\
& \ \ \ \ -[\boldsymbol{\gamma}\cdot\mathbf{p},S_{a}]^{2}+\{\boldsymbol{%
\gamma}\cdot\mathbf{p},\{\boldsymbol{\gamma}\cdot \mathbf{p},C_{a}\}\}-\{%
\mathbf{p}^{2},S_{a}^{2}\}+(1-\mu)\{S_{a},[\boldsymbol{\gamma}\cdot\mathbf{p}%
,[\boldsymbol{\gamma}\cdot\mathbf{p},a]]\}  \notag \\
& \ \ \ \ -2i\gamma^{5}[\boldsymbol{\gamma}\cdot\mathbf{p}%
,S_{a}C_{a}]+i\gamma^{5}[S_{a},\{\boldsymbol{\gamma}\cdot\mathbf{p}%
,C_{a}\}]-2i\gamma^{5}S_{a}^{2}[\boldsymbol{\gamma}\cdot\mathbf{p},S_{a}]\ .
\label{H3a}
\end{align}
The non-derivative sine and cosine terms are singled out in the last line of
Eq.~(\ref{AxFW1}) because they can be dropped. The specific combination $%
S_{a}^{2}+2C_{a}$ already gives a term of $\mathcal{O}(a^{4})$ and, when
combined with the $\mathcal{O}(1/m^{3})$ terms, gives the totally negligible
contribution 
\begin{equation}
\frac{S_{a}^{2}+2C_{a}}{2m}-\frac{4S_{a}^{2}C_{a}+S_{a}^{4}}{8m^{3}}%
=-2m\sin^{8}\left( \mu a/m\right) =-\frac{2\mu^{8}}{m^{7}}a^{8}+...\ .
\end{equation}
So, even though the polar representation initially involves non-derivative
operators in $a^{n}$, $n>1$, none of them survive in the non-relativistic
limit. This fact was not realized in Ref.~\cite{Friar:1977xh}, where only
terms linear in the pseudoscalar field were kept.

At this stage, we recover the expression in Eq.~(\ref{Intro3}) and~(\ref%
{Intro5}) by setting $\mu=0$ and $\mu=1$, respectively. As stated there, the
axion wind term is independent of the parametrization, and actually%
\begin{equation}
\frac{i}{2m}\gamma^{5}[\boldsymbol{\gamma}\cdot\mathbf{p},S_{a}+2(1-\mu )a]=%
\frac{1}{m}\gamma^{5}\boldsymbol{\gamma}\cdot\mathbf{\nabla}a-2\mu ^{3}\frac{%
a^{2}\mathbf{\nabla}a}{m^{3}}+\mathcal{O}(1/m^{5})\ .  \label{ExpWind}
\end{equation}
On the other hand, the time-derivative term is not~\cite{Friar:1977xh}%
\begin{equation}
\frac{1}{8m^{2}}\gamma^{5}\{\boldsymbol{\gamma}\cdot\mathbf{p},\dot{S}%
_{a}+4(1-\mu)\dot{a}\}=\frac{2-\mu}{4m^{2}}\gamma^{5}\{\boldsymbol{\gamma }%
\cdot\mathbf{p},\dot{a}\}+\mathcal{O}(1/m^{4})\ .
\end{equation}
This coupling even disappear for the specific choice $\mu=2$. Since there
are no other $\mathcal{O}(1/m^{2})$ terms, for this to make sense, this
operator must not embody any real physical effects.

\subsubsection{Schiff's transformations}

Let us first concentrate on the $\mathcal{O}(1/m^{2})$ terms. In analogy
with the transformation done in Sec.~\ref{SecSchiff} to eliminate the EDM
operator, we can perform the unitary transform $\psi \rightarrow
e^{iS_{1}}\psi $ with~\cite{Barnhill:1969ygg,Friar:1974dk}%
\begin{equation}
iS_{1}=\frac{i}{8m^{2}}\gamma ^{5}\{\boldsymbol{\gamma }\cdot \mathbf{p}%
,S_{a}+4(1-\mu )a\}\ .  \label{AxSchiff1}
\end{equation}%
This transformation is unitary and commutes with the mass term, $[\exp (\pm
iS_{1}),\gamma _{0}m]=0$. This means that, with $iS_{1}\sim \mathcal{O}%
(m^{-2})$ and $\mathcal{H}^{\mathrm{NR}}(\mu )-\gamma _{0}m\sim \mathcal{O}%
(m^{-1})$, only the first term of the expansion needs to be kept 
\begin{equation}
\mathcal{H}^{\mathrm{NR}\prime }(\mu )=e^{iS_{1}}\left( \mathcal{H}^{\mathrm{%
NR}}(\mu )-i\partial _{t}\right) e^{-iS_{1}}=\mathcal{H}^{\mathrm{NR}}(\mu
)+[iS_{1},\mathcal{H}^{\mathrm{NR}}(\mu )]-\dot{S}_{1}+\mathcal{O}(m^{-4})\ .
\end{equation}%
Explicitly, plugging in the expression of $S_{1}$,%
\begin{align}
\lbrack iS_{1},\mathcal{H}^{\mathrm{NR}}(\mu )]-\dot{S}_{1}& =-\frac{1}{%
8m^{2}}\gamma ^{5}\{\boldsymbol{\gamma }\cdot \mathbf{p},\dot{S}_{a}+4(1-\mu
)\dot{a}\}  \notag \\
& \ \ \ \ +\frac{1}{16m^{3}}\gamma ^{0}[\{\boldsymbol{\gamma }\cdot \mathbf{p%
},S_{a}+4(1-\mu )a\},[\boldsymbol{\gamma }\cdot \mathbf{p},S_{a}+2(1-\mu )a]]
\notag \\
& \ \ \ \ +\frac{i}{16m^{3}}\gamma ^{0}\gamma ^{5}[\{\boldsymbol{\gamma }%
\cdot \mathbf{p},S_{a}+4(1-\mu )a\},\mathbf{p}^{2}]+\mathcal{O}(m^{-4})\ .
\label{Schiff1Neu}
\end{align}%
The $\dot{S}_{1}$ term cancels precisely the $\mathcal{O}(1/m^{2})$ terms,
by construction. That is Schiff's theorem trick in action. What it means is
that this operator is actually a higher order effect, now embodied in the $%
\mathcal{O}(m^{-3})$ operators. In other words, we have succeeded at
replacing the $\mathcal{O}(1/m^{2})$ terms involving time-derivatives by $%
\mathcal{O}(1/m^{3})$ terms involving only space derivatives, that is, axion
wind operators.

At this stage, it is clear that Schiff's trick can be used to remove or
simplify the terms in $\mathcal{H}_{3}$ involving time derivatives\footnote{Throughout this paper, somewhat abusively, all the unitary transformations done on the non-relativistic Hamiltonians are called "Schiff transformations", by analogy with the original one described in Sec.~\ref{SecSchiff}.}.
Specifically, we can perform%
\begin{equation}
iS_{2}=\frac{1}{8m^{3}}(1-\mu)\gamma^{0}\gamma^{5}[\boldsymbol{\gamma}\cdot%
\mathbf{p},\dot{a}]\ ,  \label{AxSchiff2}
\end{equation}
to remove the term $\gamma^{0}\gamma^{5}[\boldsymbol{\gamma}\cdot \mathbf{p},%
\ddot{a}]$, up to some $\mathcal{O}(1/m^{4})$ contributions. The final
transformation we consider presents us with an alternative. Let us now
rotate $\mathcal{H}^{\mathrm{NR}}(\mu)$ with 
\begin{equation}
iS_{3}=-\frac{i}{4m^{3}}(1-\mu)\gamma^{0}\dot{a}S_{a}\ .  \label{AxSchiff3}
\end{equation}
Since $[iS_{3},\mathcal{H}^{\mathrm{NR}}(\mu)]$ is of $\mathcal{O}(1/m^{4})$%
, the transformed Hamiltonian is just $\mathcal{H}^{\mathrm{NR}\prime}(\mu)=%
\mathcal{H}^{\mathrm{NR}}(\mu)-\dot{S}_{3}$. This kills the $\ddot
{a}S_{a}$
coupling and corrects the $\dot{a}\dot{S}(a)$ in precisely the right way to
make it $\mu$ independent:%
\begin{equation}
4(1-\mu)^{2}\dot{a}^{2}+2(1-\mu)\dot{a}\dot{S}_{a}+\dot{S}_{a}^{2}-2(1-\mu)%
\ddot{a}S_{a}\overset{-\dot{S}_{3}}{\rightarrow}(2\dot{a}(1-\mu )+\dot{S}%
_{a})^{2}=4\dot{a}^{2}+\mathcal{O}(1/m^{2})\ .  \label{ReorgAdot}
\end{equation}
Now, we could have done the opposite, that is, make the $\ddot{a}S_{a}$
coupling $\mu$ independent by removing entirely the $\dot{a}^{2}$ coupling.
This time, the Schiff's transformation is not removing an operator, but
telling us that two of them are redundant, up to higher order corrections.

\subsubsection{Final Hamiltonian in the non-relativistic limit}

All in all, after the unitary transformations $S_{1}$ in Eq.~(\ref{AxSchiff1}%
), $S_{2}$ in Eq.~(\ref{AxSchiff2}), and $S_{3}$ in Eq.~(\ref{AxSchiff3}),
and after expanding $S_{a}$ and $C_{a}$ and keeping only terms up to $%
\mathcal{O}(1/m^{3})$, the Hamiltonian becomes%
\begin{equation}
\mathcal{H}^{\mathrm{NR}}(\mu)=\gamma^{0}\left( m+\frac{\mathbf{p}^{2}}{2m}-%
\frac{\mathbf{p}^{4}}{8m^{3}}+\frac{i}{m}\gamma^{5}[\boldsymbol{\gamma }\cdot%
\mathbf{p},a]\right) +\frac{1}{8m^{3}}\gamma^{0}\mathcal{H}_{3}+\mathcal{O}%
(m^{-4})\ ,
\end{equation}
with%
\begin{align}
\mathcal{H}_{3} & =4\dot{a}^{2}+i(1-\mu)\gamma^{5}[\boldsymbol{\gamma}\cdot%
\mathbf{p},[\boldsymbol{\gamma}\cdot\mathbf{p},[\boldsymbol{\gamma}\cdot%
\mathbf{p},a]]]-2\mu i\gamma^{5}\{\mathbf{p}^{2},[\boldsymbol{\gamma }\cdot%
\mathbf{p},a]\}  \notag \\
& \ \ \ +(2-\mu)i\gamma^{5}[\{\boldsymbol{\gamma}\cdot\mathbf{p},a\},\mathbf{%
p}^{2}]-4\mu^{2}[\boldsymbol{\gamma}\cdot\mathbf{p},a]^{2}-2\mu^{2}\{%
\boldsymbol{\gamma}\cdot\mathbf{p},\{\boldsymbol{\gamma}\cdot\mathbf{p}%
,a^{2}\}\}  \notag \\
& \ \ \ -4\mu^{2}\{\mathbf{p}^{2},a^{2}\}+2\mu(1-\mu)\{a,[\boldsymbol{\gamma 
}\cdot\mathbf{p},[\boldsymbol{\gamma}\cdot\mathbf{p},a]]\}  \notag \\
& \ \ \ +2(2-\mu)[\{\boldsymbol{\gamma}\cdot\mathbf{p},a\},[\boldsymbol{%
\gamma}\cdot\mathbf{p},a]]-4\mu^{3}i\gamma^{5}[a,\{\boldsymbol{\gamma}\cdot%
\mathbf{p},a^{2}\}]  \notag \\
& \ \ \ +8i\mu^{3}\gamma^{5}[\boldsymbol{\gamma}\cdot\mathbf{p}%
,a^{3}]-16i\mu^{3}\gamma^{5}a^{2}[\boldsymbol{\gamma}\cdot\mathbf{p},a]-%
\frac{16}{3}i\mu^{3}\gamma^{5}[\boldsymbol{\gamma}\cdot\mathbf{p},a^{3}]\ ,
\end{align}
where the last term with the 16/3 coefficient comes from the expansion of
the $\mathcal{O}(m^{-1})$ term involving $\mathbf{\nabla}S_{a}$, see Eq.~(%
\ref{ExpWind}). At this stage, algebraic manipulations of $\mathcal{H}_{3}$
using commutator and anticommutator identities, e.g., 
\begin{subequations}
\label{CommId}
\begin{align}
\{a,[\boldsymbol{\gamma}\cdot\mathbf{p},[\boldsymbol{\gamma}\cdot \mathbf{p}%
,a]]\} & =[\{a,\boldsymbol{\gamma}\cdot\mathbf{p}\},[\boldsymbol{\gamma}\cdot%
\mathbf{p},a]]\ , \\
\{\boldsymbol{\gamma}\cdot\mathbf{p},\{\boldsymbol{\gamma}\cdot\mathbf{p}%
,a^{2}\}\} & =-[\boldsymbol{\gamma}\cdot\mathbf{p},[\boldsymbol{\gamma}\cdot%
\mathbf{p},a^{2}]]-2\{a^{2},\mathbf{p}^{2}\}\ , \\
\lbrack\boldsymbol{\gamma}\cdot\mathbf{p},[\boldsymbol{\gamma}\cdot \mathbf{p%
},a^{2}]] & =2[\boldsymbol{\gamma}\cdot\mathbf{p},a]^{2}+\{a,[\boldsymbol{%
\gamma}\cdot\mathbf{p},[\boldsymbol{\gamma}\cdot \mathbf{p},a]]\}\ , \\
\lbrack\boldsymbol{\gamma}\cdot\mathbf{p},[\boldsymbol{\gamma}\cdot \mathbf{p%
},[\boldsymbol{\gamma}\cdot\mathbf{p},a]]] & =-2[\boldsymbol{\gamma }\cdot%
\mathbf{p},\{a,\mathbf{p}^{2}\}]-[\{\boldsymbol{\gamma}\cdot \mathbf{p},a\},%
\mathbf{p}^{2}]\ ,
\end{align}
permit to show that its $\mathcal{O}(a^{3})$ terms cancel out completely,
and its $\mathcal{O}(a^{2})$ and $\mathcal{O}(a)$ terms become independent
of $\mu$. The final Hamiltonian is very simple and contains only five
non-trivial operators: 
\end{subequations}
\begin{equation}
\fbox{$%
\begin{array}{ll}
\mathcal{H}^{\mathrm{NR}}= & \gamma^{0}\left( m+\dfrac{\mathbf{p}^{2}}{2m}-%
\dfrac{\mathbf{p}^{4}}{8m^{3}}+\dfrac{i\gamma^{5}[\boldsymbol{\gamma }\cdot%
\mathbf{p},a]}{m}\right. \medskip \\ 
& \ \ \ -\dfrac{i\gamma^{5}\left( [\mathbf{p}^{2},\{\boldsymbol{\gamma}\cdot%
\mathbf{p},a\}]+2\{\mathbf{p}^{2},[\boldsymbol{\gamma}\cdot \mathbf{p}%
,a]\}\right) }{8m^{3}}\medskip \\ 
& \ \ \ \left. +\dfrac{a[\boldsymbol{\gamma}\cdot\mathbf{p},[\boldsymbol{%
\gamma}\cdot\mathbf{p},a]]}{m^{3}}+\dfrac{\dot{a}^{2}}{2m^{3}}\right) +%
\mathcal{O}(m^{-4})\ ,%
\end{array}
$}  \label{HneutAxion}
\end{equation}
with the further information that $\dot{a}^{2}$ can be freely traded for $%
\ddot{a}a$. Remember that the axion coupling constant has to be put back by $%
a\rightarrow ga$ with $g=m/\Lambda$ and $\Lambda$ the PQ breaking scale.
Three comments are in order.

\begin{itemize}
\item It is remarkable that all $\mu$ dependences have cancelled out, and
this involved highly non-trivial cancellations. In our opinion, it shows
that the essential physical content is correctly identified, and
redundancies kept at a minimum. Interestingly, this Hamiltonian cannot be
obtained by setting $\mu$ to some value in $\mathcal{H}^{\mathrm{NR}}(\mu)$
of Eq.~(\ref{AxFW1}). This is evident since $S_{1}$, $S_{2}$, and $S_{3}$ do
not all vanish for the same value of $\mu$. Said differently, the sequence
of Schiff transformations $S_{1}$, $S_{2}$, and $S_{3}$ does not trivially
undo the original Dyson rotation of Eq.~(\ref{SFriar}). Note though that in
practice, setting $\mu=2$ in Eq.~(\ref{AxFW1}) already goes a long way since 
$S_{1}$ has the most impact but vanishes for that value, at least for
operators up to $\mathcal{O}(m^{-3})$.

\item The $\gamma^{5}\{\boldsymbol{\gamma}\cdot\mathbf{p},\dot{a}\}$ ends up
completely screened, in a way analogous to Schiff's EDM screening. What is
different though is that we do not expect significant violations of this
screening.\ First, finite size effects were relevant for the EDM as the
electric charge density is far from constant in atomic systems.\ By
contrast, the axion background should be relatively homogenous, even on
macroscopic scales. Second, relativistic corrections were found significant
for the EDM. But, as discussed in Sec.~\ref{SecSchiff}, the relativistic
corrections to $\boldsymbol{\gamma}\cdot\mathbf{E}$ were embodied in the
very similar $\{\mathbf{P}^{2},\boldsymbol{\gamma}\cdot\mathbf{E}\}$
operator. Here, the relativistic corrections replacing $\gamma^{5}\{%
\boldsymbol{\gamma}\cdot\mathbf{p},\dot{a}\}$ are totally different in
nature: they all involve the axion wind and even vanish if $\mathbf{\nabla}%
a=0$ (note that $\left[ \mathbf{p}^{2},\{\boldsymbol{\gamma}\cdot\mathbf{p}%
,a\}\right] =\{\boldsymbol{\gamma}\cdot\mathbf{p},[\mathbf{p}^{2},a]\}$). In
that $\mathbf{\nabla}a=0$ scenario, the relativistic corrections replacing $%
\gamma^{5}\{\boldsymbol{\gamma}\cdot\mathbf{p},\dot{a}\}$ would at best
arise at $\mathcal{O}(m^{-4})$. For these reasons, we expect the screening
of $\gamma^{5}\{\boldsymbol{\gamma}\cdot\mathbf{p},\dot{a}\}$ to be
particularly effective.

\item The leading fermionic coupling in a $\mathbf{\nabla }a=0$ scenario is $%
\dot{a}^{2}/(2m^{3})$, which is not a relativistic correction to $\gamma
^{5}\{\boldsymbol{\gamma }\cdot \mathbf{p},\dot{a}\}$ but a genuine
independent coupling. In this case though, being quadratic in the axion
field, it is presumably totally negligible, and better windows could exist.
In particular, in most scenarios, the axion also couples to photons.
Classically, the $aF_{\mu \nu }\tilde{F}^{\mu \nu }$ coupling can generates
a $\dot{a}\mathbf{B}$ term that act as a current density~\cite%
{Sikivie:1983ip}.

\item On a technical note, let us stress that it is crucial to use the full
exponential parametrization to correctly identify the final operators. Had
we truncated the polar representation to its leading term by setting $%
S_{a}=2\mu a$ and $C_{a}=0$, not only would there still be $\mathcal{O}%
(a^{3})$ operators in the final Hamiltonian, but the $\mu $ dependence would
not have cancelled completely~\cite{Friar:1977xh}. This explains why
historically, the $\mu $ dependence was interpreted as an ambiguity. Now, we
see that requiring reparametrization invariance actually points to a
preferred basis of operators for $\mathcal{H}^{\mathrm{NR}}$.
\end{itemize}

The fact that the axioelectric operator is screened can be demonstrated in
an alternative way, shedding a different light on the mechanism at play
behind the Schiff transformation. Let us assume for now that $[\boldsymbol{%
\gamma }\cdot \mathbf{p},a]=\boldsymbol{\gamma }\cdot \mathbf{\nabla }a=0$,
and define%
\begin{equation}
\mathcal{H}_{0}^{\mathrm{NR}}=\gamma ^{0}\left( m+\frac{\mathbf{p}^{2}}{2m}%
\right) \ ,\ \ \ \mathcal{V}(t)=\frac{2-\mu }{4m^{2}}\gamma ^{5}\{%
\boldsymbol{\gamma }\cdot \mathbf{p},\dot{a}\}\ .  \label{TDPT}
\end{equation}%
In the interaction picture, $\left\vert \psi ^{I}(t)\right\rangle =\exp (i%
\mathcal{H}_{0}^{\mathrm{NR}}t)\left\vert \psi (t)\right\rangle $, the
time-evolution of $\left\vert \psi ^{I}(t)\right\rangle $ can be encoded
into the evolution operator%
\begin{equation}
U(t,t_{0})=T\exp \left[ -i\int_{t_{0}}^{t}dt^{\prime }\mathcal{V}%
^{I}(t^{\prime })\right] \ ,  \label{Uop}
\end{equation}%
with $T$ the time-ordered product, and such that $\left\vert \psi
^{I}(t)\right\rangle =U(t,t_{0})\left\vert \psi ^{I}(t_{0})\right\rangle $.
The interaction picture perturbation is the same as the Schr\"{o}dinger one
at leading order 
\begin{equation}
\mathcal{V}^{I}(t)=\exp (i\mathcal{H}_{0}^{\mathrm{NR}}t)\mathcal{V}(t)\exp
(-i\mathcal{H}_{0}^{\mathrm{NR}}t)=\mathcal{V}(t)+\mathcal{O}(m^{-3})\ ,
\end{equation}%
since $[\gamma ^{0}m,\mathcal{V}(t)]=0$. Now, we see that whenever $\mathcal{%
V}(t)=\partial _{t}\mathcal{X}(t)$, the evolution operator collapses to the
universal $U(t,t_{0})=\exp (i(\mathcal{X}(t)-\mathcal{X}(t_{0})))$, and
drops out when the initial and final times are set by the experimental
conditions. Thus, perturbations that are total time derivatives do not
change energy levels\footnote{%
They could affect other observales though. For instance, $U(t,t_{0})$ with $%
\mathcal{V}(t)$ in Eq.~(\ref{TDPT}) is essentially a spin rotation, see the
discussion of Sec.~\ref{SecSchiff}.}. Note also that $U(t,t_{0})$ is precisely the Schiff
transformation done in Eq.~(\ref{AxSchiff1}) to get rid of the perturbation
in the first place. It corresponds to $\left\vert \psi ^{I}(t)\right\rangle
\rightarrow \left\vert \psi ^{\prime I}(t)\right\rangle =\exp (i\mathcal{X}%
(t))\left\vert \psi ^{I}(t)\right\rangle $, with then $i\partial
_{t}\left\vert \psi ^{\prime I}(t)\right\rangle =0$ since $\mathcal{V}%
^{\prime I}(t)=\mathcal{V}(t)-\partial _{t}\mathcal{X}(t)=0$. In this
picture (in the quantum mechanical sense), since $\mathcal{X}(t)$ commutes
with $\mathcal{H}_{0}^{\mathrm{NR}}$ up to terms of $\mathcal{O}(m^{-3})$, $%
\left\vert \psi ^{\prime I}(t)\right\rangle $ stays fixed to some linear
combination of eigenstates of the free Hamiltonian $\mathcal{H}_{0}^{\mathrm{%
NR}}$. In this sense, performing the Schiff transformation to get rid of $%
\mathcal{V}(t)$ produces a non-relativistic Hamiltonian that better reflects
the physics of the system. That is the same idea as the original Schiff
transformation for EDMs: $\mathcal{H}_{EM}^{\mathrm{NR}}$ in Eq.~(\ref%
{HEMSch}) better reflects the energy level of the system than that in Eq.~(%
\ref{HEM}).

\subsection{For charged fermions in an external EM field}

\label{SecCharged}

The situation described in the previous section changes in a crucial way in
the presence of minimally coupled electromagnetic fields. To show this, let
us repeat all the steps of the previous section, but starting from%
\begin{equation}
\mathcal{H}(\mu )=\gamma ^{0}\left( \boldsymbol{\gamma }\cdot \mathbf{P}+m-%
\frac{1-\mu }{m}\gamma ^{0}\gamma _{5}\dot{a}+i\frac{1-\mu }{m}\gamma _{5}[%
\boldsymbol{\gamma }\cdot \mathbf{P},a]+\left( \exp \left( 2i\frac{\mu }{m}%
a\gamma ^{5}\right) -1\right) m\right) +e\phi \ .
\end{equation}%
Note that $[\boldsymbol{\gamma }\cdot \mathbf{P},a]=[\boldsymbol{\gamma }%
\cdot \mathbf{p},a]=-i\boldsymbol{\gamma }\cdot \mathbf{\nabla }a$ since $a$
is electrically neutral. This Hamiltonian can be block-diagonalized by
plugging%
\begin{align}
\mathcal{O}& =\boldsymbol{\gamma }\cdot \mathbf{P}-\frac{1-\mu }{m}\gamma
^{0}\gamma ^{5}\dot{a}+i\gamma ^{5}S_{a}\ ,\  \\
\mathcal{E}& =\gamma ^{0}\frac{1}{m}C_{a}+i\frac{1-\mu }{m}\gamma ^{0}\gamma
^{5}[\boldsymbol{\gamma }\cdot \mathbf{P},a]+e\phi \ ,
\end{align}%
in Eq.~(\ref{FW}). This produces%
\begin{align}
\mathcal{H}^{\mathrm{NR}}(\mu )& =\mathcal{H}_{EM}^{\mathrm{NR}}+\frac{%
i\gamma ^{0}\gamma ^{5}[\boldsymbol{\gamma }\cdot \mathbf{P},S_{a}+2(1-\mu
)a]}{2m}  \notag \\
& \ \ \ \ +\frac{\gamma ^{5}\{\boldsymbol{\gamma }\cdot \mathbf{P},\dot{S}%
_{a}+4(1-\mu )\dot{a}\}}{8m^{2}}-\frac{eS_{a}\gamma ^{5}\boldsymbol{\gamma }%
\cdot \mathbf{E}}{4m^{2}}+\frac{1}{8m^{3}}\gamma ^{0}\mathcal{H}_{3}\ ,
\label{HCharBS}
\end{align}%
where $\mathcal{H}_{EM}^{\mathrm{NR}}$ is the electromagnetic Hamiltonian,
Eq.~(\ref{PauliEM}), and $\mathcal{H}_{3}$ is obtained from the neutral one
in Eq.~(\ref{H3a}) by replacing $\boldsymbol{\gamma }\cdot \mathbf{p}%
\rightarrow \boldsymbol{\gamma }\cdot \mathbf{P}$ and $\mathbf{p}%
^{2}\rightarrow \mathbf{P}^{2}+e\gamma ^{0}\gamma ^{5}\boldsymbol{\gamma }%
\cdot \mathbf{B}$ (which is nothing but $(\boldsymbol{\gamma }\cdot \mathbf{p%
})^{2}\rightarrow (\boldsymbol{\gamma }\cdot \mathbf{P})^{2}$). Compared to
the neutral case, the only unexpected new addition is the EDM coupling $%
S_{a}\gamma ^{5}\boldsymbol{\gamma }\cdot \mathbf{E}=2\mu a\gamma ^{5}%
\boldsymbol{\gamma }\cdot \mathbf{E}+...$. Because it does not arise
starting from the axion derivative interaction, it does not appear in the
literature (though it is present in Ref.~\cite{Friar:1977xh}).

As in the free case, to get a better handle on the physical couplings, let
us perform the sequence of Schiff transformations: 
\begin{subequations}
\begin{align}
iS_{1}& =\frac{i}{8m^{2}}\gamma ^{5}\{\boldsymbol{\gamma }\cdot \mathbf{P}%
,S_{a}+4(1-\mu )a\}\ ,  \label{SchiffEM1} \\
iS_{2}& =\frac{1}{8m^{3}}(1-\mu )\gamma ^{0}\gamma ^{5}[\boldsymbol{\gamma }%
\cdot \mathbf{P},\dot{a}]\ ,\   \label{SchiffEM2} \\
iS_{3}& =-\frac{i}{4m^{3}}(1-\mu )\gamma ^{0}\dot{a}S_{a}\ .
\label{SchiffEM3}
\end{align}%
After this, the Hamiltonian becomes 
\end{subequations}
\begin{equation}
\mathcal{H}^{\mathrm{NR}}(\mu )=\mathcal{H}_{EM}^{\mathrm{NR}}+\frac{i\gamma
^{0}\gamma ^{5}[\boldsymbol{\gamma }\cdot \mathbf{P},S_{a}+2(1-\mu )a]}{2m}-%
\frac{e\gamma ^{5}\{\boldsymbol{\gamma }\cdot \mathbf{E},S_{a}+2(1-\mu )a\}}{%
2m^{2}}+\frac{1}{8m^{3}}\gamma ^{0}\mathcal{H}_{3}\ ,
\end{equation}%
with 
\begin{align}
\mathcal{H}_{3}& =(2\dot{a}(1-\mu )+\dot{S}_{a})^{2}+i(1-\mu )\gamma ^{5}[%
\boldsymbol{\gamma }\cdot \mathbf{P},[\boldsymbol{\gamma }\cdot \mathbf{P},[%
\boldsymbol{\gamma }\cdot \mathbf{P},a]]]+\{\boldsymbol{\gamma }\cdot 
\mathbf{P},\{\boldsymbol{\gamma }\cdot \mathbf{P},C_{a}\}\}  \notag \\
& \ \ \ \ -[\boldsymbol{\gamma }\cdot \mathbf{P},S_{a}]^{2}-i\gamma ^{5}\{%
\mathbf{P}^{2}+e\gamma ^{0}\gamma ^{5}\boldsymbol{\gamma }\cdot \mathbf{B},[%
\boldsymbol{\gamma }\cdot \mathbf{P},S_{a}]\}-\{\mathbf{P}^{2}+e\gamma
^{0}\gamma ^{5}\boldsymbol{\gamma }\cdot \mathbf{B},S_{a}^{2}\}  \notag \\
& \ \ \ \ +(1-\mu )\{S_{a},[\boldsymbol{\gamma }\cdot \mathbf{P},[%
\boldsymbol{\gamma }\cdot \mathbf{P},a]]\}-2i\gamma ^{5}[\boldsymbol{\gamma }%
\cdot \mathbf{P},S_{a}C_{a}]+i\gamma ^{5}[S_{a},\{\boldsymbol{\gamma }\cdot 
\mathbf{P},C_{a}\}]  \notag \\
& \ \ \ \ -2iS_{a}^{2}\gamma ^{5}[\boldsymbol{\gamma }\cdot \mathbf{P}%
,S_{a}]+\frac{i}{2}\gamma ^{5}[\{\boldsymbol{\gamma }\cdot \mathbf{P}%
,S_{a}+4(1-\mu )a\},\mathbf{P}^{2}+e\gamma ^{0}\gamma ^{5}\boldsymbol{\gamma 
}\cdot \mathbf{B}]  \notag \\
& \ \ \ \ +\frac{1}{2}[\{\boldsymbol{\gamma }\cdot \mathbf{P},S_{a}+4(1-\mu
)a\},[\boldsymbol{\gamma }\cdot \mathbf{P},S_{a}+2(1-\mu )a]]\ .
\end{align}%
Let us now expand $S_{a}$ and $C_{a}$ and keep only terms up to $\mathcal{O}%
(m^{-3})$. This calculation is simpler than it seems because most of the
algebra done in the neutral case relied on the use of commutator and
anticommutator identities, see Eq.~(\ref{CommId}), which remain essentially
valid. One only has to pay attention to the extra $\boldsymbol{\gamma }\cdot 
\mathbf{B}$ terms coming from $(\boldsymbol{\gamma }\cdot \mathbf{p}%
)^{2}\rightarrow (\boldsymbol{\gamma }\cdot \mathbf{P})^{2}=-2\mathbf{P}%
^{2}-2e\gamma ^{0}\gamma ^{5}\boldsymbol{\gamma }\cdot \mathbf{B}$, which
implies for example $[\mathbf{P}^{2},\boldsymbol{\gamma }\cdot \mathbf{P}%
]=e\gamma ^{0}\gamma ^{5}[\boldsymbol{\gamma }\cdot \mathbf{P},\boldsymbol{%
\gamma }\cdot \mathbf{B}]$ from $[\boldsymbol{\gamma }\cdot \mathbf{P},\{%
\boldsymbol{\gamma }\cdot \mathbf{P},\boldsymbol{\gamma }\cdot \mathbf{P}%
\}]=-2[\boldsymbol{\gamma }\cdot \mathbf{P},\mathbf{P}^{2}]-2e\gamma
^{0}\gamma ^{5}[\boldsymbol{\gamma }\cdot \mathbf{P},\boldsymbol{\gamma }%
\cdot \mathbf{B}]$ and $[\boldsymbol{\gamma }\cdot \mathbf{P},\{\boldsymbol{%
\gamma }\cdot \mathbf{P},\boldsymbol{\gamma }\cdot \mathbf{P}\}]=0$ since $%
[A,\{B,C\}]=\{C,[A,B]\}-\{B,[C,A]\}$. Putting all together, the $\mu $
dependence again cancels out precisely, $\mathcal{H}^{\mathrm{NR}}(\mu )=%
\mathcal{H}^{\mathrm{NR}}$, and $\mathcal{H}_{3}$ greatly simplifies to only
a few operators:%
\begin{align}
\mathcal{H}^{\mathrm{NR}}& =\mathcal{H}_{EM}^{\mathrm{NR}}+\frac{i\gamma
^{0}\gamma ^{5}[\boldsymbol{\gamma }\cdot \mathbf{P},a]}{m}-\frac{ea\gamma
^{5}\boldsymbol{\gamma }\cdot \mathbf{E}}{m^{2}}  \notag \\
& \ \ \ \ +i\gamma ^{0}\gamma ^{5}\frac{2\{\{\boldsymbol{\gamma }\cdot 
\mathbf{P},\boldsymbol{\gamma }\cdot \mathbf{P}\},[\boldsymbol{\gamma }\cdot 
\mathbf{P},a]\}+[\{\boldsymbol{\gamma }\cdot \mathbf{P},\boldsymbol{\gamma }%
\cdot \mathbf{P}\},\{\boldsymbol{\gamma }\cdot \mathbf{P},a\}]}{16m^{3}} 
\notag \\
& \ \ \ \ +\frac{\gamma ^{0}a[\boldsymbol{\gamma }\cdot \mathbf{P},[%
\boldsymbol{\gamma }\cdot \mathbf{P},a]]}{m^{3}}+\frac{\gamma ^{0}\dot{a}^{2}%
}{2m^{3}}+\mathcal{O}(m^{-4})\ .
\end{align}%
where $\{\boldsymbol{\gamma }\cdot \mathbf{P},\boldsymbol{\gamma }\cdot 
\mathbf{P}\}=-2\mathbf{P}^{2}-2e\gamma ^{0}\gamma ^{5}\boldsymbol{\gamma }%
\cdot \mathbf{B}$. Apart from the new EDM coupling, this expression is
identical to the neutral case, but for $\boldsymbol{\gamma }\cdot \mathbf{p}%
\rightarrow \boldsymbol{\gamma }\cdot \mathbf{P}$.

This is not our final form for the Hamiltonian. Because of their importance,
we think it is crucial to keep track of the redundancies when they involve
operators of the same order. So, let us reintroduce two free parameters
explicitly and perform a final unitary transformation%
\begin{equation}
iS_{4}=-\frac{i\alpha }{2m^{2}}\gamma ^{5}\{\boldsymbol{\gamma }\cdot 
\mathbf{P},a\}+\frac{i\beta }{2m^{3}}\gamma ^{0}a\dot{a}\ .
\end{equation}%
Then, we obtain:%
\begin{equation}
\fbox{$%
\begin{array}{ll}
\mathcal{H}^{\mathrm{NR}}(\alpha ,\beta )= & \mathcal{H}_{EM}^{\mathrm{NR}}+%
\dfrac{i\gamma ^{0}\gamma ^{5}[\boldsymbol{\gamma }\cdot \mathbf{P},a]}{m}%
+\alpha \dfrac{\gamma ^{5}\{\boldsymbol{\gamma }\cdot \mathbf{P},\dot{a}\}}{%
2m^{2}}-(1-\alpha )\dfrac{ea\gamma ^{5}\boldsymbol{\gamma }\cdot \mathbf{E}}{%
m^{2}}\medskip  \\ 
& +i\gamma ^{0}\gamma ^{5}\dfrac{\{\{\boldsymbol{\gamma }\cdot \mathbf{P},%
\boldsymbol{\gamma }\cdot \mathbf{P}\},[\boldsymbol{\gamma }\cdot \mathbf{P}%
,a]\}}{8m^{3}}+i(1-2\alpha )\gamma ^{0}\gamma ^{5}\dfrac{[\{\boldsymbol{%
\gamma }\cdot \mathbf{P},\boldsymbol{\gamma }\cdot \mathbf{P}\},\{%
\boldsymbol{\gamma }\cdot \mathbf{P},a\}]}{16m^{3}}\medskip  \\ 
& -\beta \dfrac{\gamma ^{0}a\ddot{a}}{2m^{3}}+(1-\beta )\dfrac{\gamma ^{0}%
\dot{a}^{2}}{2m^{3}}+(1-\alpha )\dfrac{\gamma ^{0}a[\boldsymbol{\gamma }%
\cdot \mathbf{P},[\boldsymbol{\gamma }\cdot \mathbf{P},a]]}{m^{3}}+\mathcal{O%
}(m^{-4})\ ,%
\end{array}%
$}  \label{HfinalEM}
\end{equation}%
with the understanding that the choice of $\alpha $ and $\beta $ is totally
free. Remember that the axion scale has to be put back in these operators by
writing $a\rightarrow ga$ with $g=m/\Lambda $ and $\Lambda $ the PQ breaking
scale. To be more explicit, the leading operators read%
\begin{equation}
\mathcal{H}^{\mathrm{NR}}(\alpha )=\mathcal{H}_{EM}^{\mathrm{NR}}-\dfrac{%
\boldsymbol{\sigma }\cdot \mathbf{\nabla }a}{\Lambda }+\alpha \gamma ^{0}%
\dfrac{i\boldsymbol{\sigma }\cdot \mathbf{\nabla }\dot{a}-2\dot{a}%
\boldsymbol{\sigma }\cdot \mathbf{P}}{2m\Lambda }+(1-\alpha )\dfrac{ea}{%
m\Lambda }\gamma ^{0}\boldsymbol{\sigma }\cdot \mathbf{E}+\mathcal{O}%
(m^{-3})\ ,
\end{equation}%
where we used $\gamma ^{5}\boldsymbol{\gamma }=-\gamma ^{0}\otimes 
\boldsymbol{\sigma }$ to put the operator in the standard form.

\subsubsection{On the axioelectric -- axionic EDM equivalence}

\label{SecEquiv}

At this stage, we have two operators at $\mathcal{O}(1/m^{2})$ whose
relative weight can be freely tuned, but whose overall impact must be
identitical. In other words, starting from $\mathcal{H}^{\mathrm{NR}}(\alpha
,\beta )$, $\alpha $ and $\beta $ must drop out of physical observables.
Clearly, this means that the covariant axioelectric operator and the axionic
EDM operators must be equivalent:%
\begin{equation}
\dfrac{\gamma ^{5}\{\boldsymbol{\gamma }\cdot \mathbf{P},\dot{a}\}}{2m^{2}}%
\Leftrightarrow -\dfrac{ea\gamma ^{5}\boldsymbol{\gamma }\cdot \mathbf{E}}{%
m^{2}}\ .  \label{AxioEquiv}
\end{equation}%
At first sight, these operators appear to encode different physics. One
depends on $\dot{a}$ but not on $\mathbf{E}$, while the other depends on $%
\mathbf{E}$ but not on $\dot{a}$. Yet, as we now discuss, there are several
ways to interpret this equivalence, and to understand that at the level of
observables, both operators always end up being strictly equivalent.

\begin{table}[t] \centering
$
\begin{tabular}{lccccr}
\hline
\multicolumn{5}{r}{$-\dfrac{e}{m^{2}}\gamma ^{5}\boldsymbol{\gamma }\cdot
\int \dot{a}\mathbf{A}dt=\dfrac{e}{m^{2}}\gamma ^{5}\boldsymbol{\gamma }%
\cdot \int a\mathbf{\dot{A}}dt\rule[-0.14in]{0in}{0.35in}$} & \  \\ 
&  & $\ \ \ \ \ \ \ \ \ \ \ \ \ \ \ \ \ \ \updownarrow 
\rule[-0.14in]{0in}{0.37in}$ &  & $\updownarrow $ & \  \\ 
\multicolumn{3}{l}{$\dfrac{\gamma ^{5}\{\boldsymbol{\gamma }\cdot \mathbf{P}%
,\dot{a}\}}{2m^{2}}=\dfrac{\gamma ^{5}\{\boldsymbol{\gamma }\cdot \mathbf{p},%
\dot{a}\}}{2m^{2}}-\dfrac{e\dot{a}\gamma ^{5}\boldsymbol{\gamma }\cdot 
\mathbf{A}}{m^{2}}$} & \multicolumn{1}{l}{$\ \Leftrightarrow \ $} & 
\multicolumn{2}{r}{$\dfrac{ea\gamma ^{5}\boldsymbol{\gamma }\cdot (\mathbf{%
\nabla }\phi +\mathbf{\dot{A}})}{m^{2}}=-\dfrac{ea\gamma ^{5}\boldsymbol{%
\gamma }\cdot \mathbf{E}}{m^{2}}$} \\ 
(axioelectric) &  & $\updownarrow \rule[-0.14in]{0in}{0.37in}\ \ \ \ \ \ \ \
\ \ \ \ \ \ \ \ \ \ \ \ \ $ &  & $\updownarrow $ & (axionic EDM) \\ 
\multicolumn{5}{r}{$\ \ \ \ \ \ \ \ \ \ \ \ \ \ \ \ \ -\dfrac{\gamma ^{5}a\{%
\mathbf{\gamma }\cdot \mathbf{p},\partial _{t}\}}{2m^{2}}\overset{EOM}{=}%
\dfrac{i\gamma ^{5}a[\boldsymbol{\gamma }\cdot \mathbf{p},\mathcal{H}^{NR}]}{%
2m^{2}}=\dfrac{ea\gamma ^{5}\boldsymbol{\gamma }\cdot \mathbf{\nabla }\phi }{%
m^{2}}\rule[-0.14in]{0in}{0.39in}$} & \  \\ \hline
\end{tabular}
$
\caption
{Schematic representation of the equivalence of Eq. (\ref{AxioEquiv}) at the level of observables.
In the top line, time-dependent perturbation theory is understood, while for the bottom line, the operators are understood to be sandwiched between bound fermionic states.}
\label{TableEquiv}
\end{table}

\begin{itemize}
\item \textbf{Double screening}: The interplay between the $\gamma ^{5}\{%
\boldsymbol{\gamma }\cdot \mathbf{P},\dot{a}\}$ and $a\gamma ^{5}\boldsymbol{%
\gamma }\cdot \mathbf{E}$ operators should have been expected. We know from
the previous section that in the absence of electromagnetic fields, $\gamma
^{5}\{\boldsymbol{\gamma }\cdot \mathbf{P},\dot{a}\}\rightarrow \gamma ^{5}\{%
\boldsymbol{\gamma }\cdot \mathbf{p},\dot{a}\}$ can be eliminated. And,
Schiff's theorem is telling us that if the axion field is constant, $a\gamma
^{5}\boldsymbol{\gamma }\cdot \mathbf{E}$ becomes a fixed EDM coupling that
can be rotated away. So, we see that both a time-varying axion field and
minimal couplings to the external electromagnetic fields are required to get
a physical effect. Each form of the operator makes one of these screening
manifest, but only their equivalence embodies their true impact on the
physics.

\item \textbf{Duality in the Dirac equation}: Before the Schiff
transformation, the derivative and polar representations do not match and
rather produce, from Eq.~(\ref{HCharBS}), 
\begin{eqnarray}
\text{Derivative} &:&\mathcal{H}^{\mathrm{NR}}(0)=\mathcal{H}_{EM}^{\mathrm{%
NR}}+\frac{i\gamma ^{0}\gamma ^{5}[\boldsymbol{\gamma }\cdot \mathbf{P},a]}{m%
}+\frac{\gamma ^{5}\{\boldsymbol{\gamma }\cdot \mathbf{P},\dot{a}\}}{2m^{2}}%
+...\ , \\
\text{Polar} &:&\mathcal{H}^{\mathrm{NR}}(1)=\mathcal{H}_{EM}^{\mathrm{NR}}+%
\frac{i\gamma ^{0}\gamma ^{5}[\boldsymbol{\gamma }\cdot \mathbf{P},a]}{m}+%
\frac{\gamma ^{5}\{\boldsymbol{\gamma }\cdot \mathbf{P},\dot{a}\}}{4m^{2}}-%
\frac{ea\gamma ^{5}\boldsymbol{\gamma }\cdot \mathbf{E}}{2m^{2}}+...\ .
\end{eqnarray}%
Yet, upon the equivalence of Eq.~(\ref{AxioEquiv}), one can choose to put
the whole $\mathcal{O}(1/m^{2})$ part of the Hamiltonian into the form of an
EDM coupling $-ea\gamma ^{5}\boldsymbol{\gamma }\cdot \mathbf{E/}m^{2}$,
which then has a very straightforward interpretation. For a charged
particle, it is well-known that the Dirac equation predicts a magnetic
moment $g=2$ via the Zeeman term,%
\begin{equation}
m\bar{\psi}\psi \rightarrow \frac{e}{2m}\gamma ^{0}\boldsymbol{\sigma }\cdot 
\mathbf{B}\ .  \label{dual1}
\end{equation}%
The axion coupling to fermion can be viewed as a pseudoscalar mass term,
dual to the scalar mass term. As a result, the Dirac equation then predicts
an electric moment,%
\begin{equation}
m(2a/m)\bar{\psi}i\gamma ^{5}\psi \rightarrow \frac{e}{2m}(2a/m)\gamma ^{0}%
\boldsymbol{\sigma }\cdot \mathbf{E}\ ,  \label{dual2}
\end{equation}%
since duality interchanges $\mathbf{B}$ and $\mathbf{E}$. In this sense, the
prediction $d=ea/m^{2}$ for the axionic EDM is the exact analogue of $g=2$
for the magnetic moment. It represents an inescapable consequence of the
Dirac equation whenever the charged fermion has a pseudoscalar coupling to
the axion\footnote{%
This provides another perspective on the original Schiff screening of
constant EDM: it can always be eliminated by a suitable chiral rotation of
the fermion mass term.}. The observability of this oscillating EDM is
another question though, because one must fight the various screening
effects, and will be discussed in Sec.~\ref{AxEDMobs}.

\item \textbf{Time-dependent perturbation theory}: If we set $\phi =0$ and
write $\gamma ^{5}\{\boldsymbol{\gamma }\cdot \mathbf{P},\dot{a}\}=\gamma
^{5}\{\boldsymbol{\gamma }\cdot \mathbf{p},\dot{a}\}-2e\dot{a}\gamma ^{5}%
\boldsymbol{\gamma }\cdot \mathbf{A}$, the $\gamma ^{5}\{\boldsymbol{\gamma }%
\cdot \mathbf{p},\dot{a}\}$ piece can be rotated away as in the neutral
case. Basically, this contribution is kinematically suppressed, and encoded
into axion wind operators of $\mathcal{O}(1/m^{3})$. Then, at leading order,
what the Schiff transformation Eq.~(\ref{SchiffEM1}) is telling us is that a
coupling $\dot{a}\boldsymbol{\sigma }\cdot \mathbf{A}$ is equivalent%
\footnote{%
In Refs.~\cite{Domcke:2018gfr,Domcke:2021fee}, the impact of a
time-dependent axionic background on the Schwinger effect, i.e.,
fermion-antifermion pair creation by a strong electric field, was analyzed.
Though the non-relativistic approximation is obviously inadequate to
describe that phenomenon, it must be noted that there also, an interplay
between time-dependence and electric field was identified and exploited.} to
a coupling $a\boldsymbol{\sigma }\cdot \mathbf{\dot{A}}$, that is, $a%
\boldsymbol{\sigma }\cdot \mathbf{E}$, exactly like the transformation Eq.~(%
\ref{SchiffEM3}) is telling us that $\dot{a}^{2}$ encodes the same physics
as $a\ddot{a}$. These pairs of operators must give the same result when
acting on fermion wavefunctions within physical observables (see Table~\ref%
{TableEquiv}). This is clearly in accordance with the interaction picture
evolution of Eq.~(\ref{Uop}), where a perturbation like $a\boldsymbol{\sigma 
}\cdot \mathbf{E}$ or $a\ddot{a}$ is to be integrated over time. This also
shows how the original Schiff screening comes back if $a(t)$ becomes
constant, as the time-integral of $a\boldsymbol{\sigma }\cdot \mathbf{E}%
=\partial _{t}(a\boldsymbol{\sigma }\cdot \mathbf{A})$ then sums up to an
unobservable constant rephasing of the fermion wavefunction.

\item \textbf{The axioelectric effects as EDM-induced}: If one encodes
entirely the $\mathcal{O}(1/m^{2})$ terms into an EDM operator, the usual
axioelectric effect is nevertheless still there. Because the two forms in
Eq.~(\ref{AxioEquiv}) are equivalent, the same matrix elements as in Ref.~%
\cite{Pospelov:2008jk} must be recovered. The equivalence in the $\phi =0$
gauge was discussed above, so let us now concentrate instead on that with $%
\mathbf{A}=0$, so that $\mathbf{E}=-\mathbf{\nabla }\phi $. In this case,
the covariant axioelectric operator reduces as $\gamma ^{5}\{\boldsymbol{%
\gamma }\cdot \mathbf{P},\dot{a}\}=\gamma ^{5}\{\boldsymbol{\gamma }\cdot 
\mathbf{p},\dot{a}\}$. Though identical in form with the neutral fermion
axioelectric operator, it cannot be rotated away when $\phi \neq 0$ (there
is an extra term in Eq.~(\ref{Schiff1Neu}) from $[iS_{1},\phi ]\neq 0$ with $%
S_{1}$ in Eq.~(\ref{AxSchiff1})). As explicitly calculated in Ref.~\cite%
{Pospelov:2008jk}, the $\gamma ^{5}\{\boldsymbol{\gamma }\cdot \mathbf{p},%
\dot{a}\}$ can then induce transitions between energy levels for an electron
bound in the potential $\phi $. Now, starting instead from the EDM operator,
we can write for an electrostatically bound electron\footnote{%
The author is indebted to Maxim Pospelov for this important clarifying
observation.}, $ea\boldsymbol{\sigma }\cdot \mathbf{E}=-ea\boldsymbol{\sigma 
}\cdot \mathbf{\nabla }\phi =-ia\boldsymbol{\sigma }\cdot \mathbf{[p},%
\mathcal{H}^{\mathrm{NR}}]+...$. With this, the transition matrix element of
Ref.~\cite{Pospelov:2008jk} is trivially recovered as $\left\langle \psi
_{f}\right\vert ea\boldsymbol{\sigma }\cdot \mathbf{E}\left\vert \psi
_{i}\right\rangle =\partial _{t}a\left\langle \psi _{f}\right\vert 
\boldsymbol{\sigma }\cdot \mathbf{p}\left\vert \psi _{i}\right\rangle $ upon
using $i\partial _{t}\left\vert \psi _{i,f}\right\rangle =\mathcal{H}^{%
\mathrm{NR}}\left\vert \psi _{i,f}\right\rangle $ and integrating by part
over time (see Table~\ref{TableEquiv}). Note that these mathematical steps
essentially undo the Schiff transformation of Eq.~(\ref{SchiffEM1}). This
shows that whatever the operator, the \textit{same} matrix element for the
axioelectric effect is obtained, as it should since physics must not depend
on the representation. Yet, we think it sheds new light to interpret the
axioelectric effects rather as a manifestation of an axion-induced EDM,
especially in view of the other points discussed previously.
\end{itemize}

\subsection{For neutral fermions having an EDM interaction}

\label{SecEDM}

The final application is the non-relativistic limit of the Hamiltonian for a
neutral state coupled to the axion, but in the presence of both the magnetic
and electric dipole operators. Those are not invariant under the PQ
symmetry, so one has to decide how they should be introduced. We consider
that they arise in the same way as the mass term, through the PQ symmetry
breaking. The fermion field is neutral under the PQ symmetry only in the
derivative representation, so those effective operators can be added to Eq.~(%
\ref{Lder}) as%
\begin{equation}
\mathcal{L}_{D}=\bar{\psi}\left( i\slashed\partial -m+\frac{\gamma ^{\mu
}\gamma _{5}\partial _{\mu }a}{m}-\frac{\delta _{\mu }}{2}\sigma ^{\mu \nu
}F_{\mu \nu }+i\frac{d}{2}\sigma ^{\mu \nu }\gamma ^{5}F_{\mu \nu }\right)
\psi \ .
\end{equation}%
Then, if we use again Friar's trick, Eq.~(\ref{SFriar}), the Lagrangian
interpolating between derivative and polar representations is 
\begin{equation}
\mathcal{L}_{D}=\bar{\psi}\left( i\gamma ^{\mu }\partial _{\mu }-m\left( 1+%
\frac{C_{a}}{m^{2}}\right) -i\gamma _{5}S_{a}+\frac{1-\mu }{m}\gamma ^{\mu
}\gamma _{5}\partial _{\mu }a-\frac{\tilde{\delta}_{\mu }}{2}\sigma ^{\mu
\nu }F_{\mu \nu }+i\frac{\tilde{d}}{2}\sigma ^{\mu \nu }\gamma ^{5}F_{\mu
\nu }\right) \psi \ ,  \label{LagrFriarDM}
\end{equation}%
where%
\begin{equation}
\tilde{\delta}_{\mu }=\delta _{\mu }+\frac{S_{a}}{m}d+\frac{C_{a}}{m^{2}}%
\delta _{\mu }\ ,\ \ \tilde{d}=d-\frac{S_{a}}{m}\delta _{\mu }+d\frac{C_{a}}{%
m^{2}}\ ,   \label{ShiftDipole}
\end{equation}%
and $S_{a}=m\sin (2\mu a/m)$, $C_{a}=m^{2}(\cos (2\mu a/m)-1)$ as before.
The fact that the dipole operators end up proportional to the axion field in
the exponential representation is similar as in the SM, where they
necessarily involve the Higgs boson field~\cite{Buchmuller:1985jz}.

The Hamiltonian in the non-relativistic limit can be obtained by plugging
the odd and even elements%
\begin{align}
\mathcal{O} & =\boldsymbol{\gamma}\cdot\mathbf{p}-\frac{1-\mu}{m}\gamma
^{0}\gamma^{5}\dot{a}+i\gamma^{5}S_{a}+i\gamma^{0}\boldsymbol{\gamma}\cdot(%
\tilde{\delta}_{\mu}\mathbf{E}+\tilde{d}\mathbf{B})\ ,\  \\
\mathcal{E} & =\gamma^{0}\frac{1}{m}C_{a}+\frac{1-\mu}{m}i\gamma^{0}%
\gamma^{5}[\boldsymbol{\gamma}\cdot\mathbf{p},a]+\gamma^{5}\boldsymbol{%
\gamma }\cdot(\tilde{\delta}_{\mu}\mathbf{B}-\tilde{d}\mathbf{E})\ ,
\end{align}
in Eq.~(\ref{FW}). After some algebra, and noting that $\tilde{\delta}_{\mu
}=\delta_{\mu}+\mathcal{O}(1/m^{2})$ and $\tilde{d}=d+\mathcal{O}(1/m^{2})$,
we arrive at%
\begin{align}
\mathcal{H}^{\mathrm{NR}}(\mu) & =\gamma^{0}\left( m+\frac{\mathbf{p}^{2}}{2m%
}-\frac{\mathbf{p}^{4}}{8m^{3}}+\frac{i\gamma^{5}[\boldsymbol{\gamma}\cdot%
\mathbf{p},S_{a}+2(1-\mu)a]}{2m}\right.  \notag \\
& \ \ \ \ \ \ \ \ \ \ \left. +\frac{(\delta_{\mu}\mathbf{E}+d\mathbf{B})^{2}%
}{2m}+\frac{\{\boldsymbol{\gamma}\cdot\mathbf{p},\boldsymbol{\gamma}%
\cdot(\delta_{\mu}\mathbf{\dot{E}}+d\mathbf{\dot{B}})\}}{8m^{2}}\right) 
\notag \\
& \ \ \ +\gamma^{5}\boldsymbol{\gamma}\cdot(\delta_{\mu}\mathbf{B}-d\mathbf{E%
})\left( 1+\frac{S_{a}^{2}+2C_{a}}{2m^{2}}\right) +\frac {i[\boldsymbol{%
\gamma}\cdot\mathbf{p},\boldsymbol{\gamma}\cdot(\delta_{\mu }\mathbf{E}+d%
\mathbf{B})]}{2m}  \notag \\
& \ \ \ -i\frac{\{S_{a},[\boldsymbol{\gamma}\cdot\mathbf{p},\boldsymbol{%
\gamma}\cdot(\delta_{\mu}\mathbf{B}-d\mathbf{E})]\}}{8m^{2}}+\frac{%
\gamma^{5}\{\boldsymbol{\gamma}\cdot\mathbf{p},\dot{S}_{a}+4(1-\mu)\dot{a}\}%
}{8m^{2}}  \notag \\
& \ \ \ +\frac{\gamma^{5}\{\boldsymbol{\gamma}\cdot\mathbf{p},\{\boldsymbol{%
\gamma}\cdot\mathbf{p},\boldsymbol{\gamma}\cdot(\delta_{\mu }\mathbf{B}-d%
\mathbf{E})\}\}}{8m^{2}}+\frac{1}{8m^{3}}\gamma^{0}\mathcal{H}_{3}\ ,
\label{HaxEDM}
\end{align}
with the same $\mathcal{H}_{3}$ as before, Eq.~(\ref{H3a}). We have used $%
[A,\{B,C\}]=\{C,[A,B]\}-\{B,[C,A]\}$ to rewrite some operators,%
\begin{align}
\lbrack S_{a},\{\boldsymbol{\gamma}\cdot\mathbf{p},\boldsymbol{\gamma}%
\cdot(\delta_{\mu}\mathbf{B}-d\mathbf{E})\}] & =\{\boldsymbol{\gamma}%
\cdot(\delta_{\mu}\mathbf{B}-d\mathbf{E}),[S_{a},\boldsymbol{\gamma}\cdot%
\mathbf{p}]\}\ , \\
\lbrack\boldsymbol{\gamma}\cdot\mathbf{p},\{\boldsymbol{\gamma}\cdot
(\delta_{\mu}\mathbf{B}-d\mathbf{E}),S_{a}\}] & =\{S_{a},[\boldsymbol{\gamma 
}\cdot\mathbf{p},\boldsymbol{\gamma}\cdot(\delta_{\mu}\mathbf{B}-d\mathbf{E}%
)]\}-\{\boldsymbol{\gamma}\cdot(\delta_{\mu}\mathbf{B}-d\mathbf{E}),[S_{a},%
\boldsymbol{\gamma}\cdot\mathbf{p}]\}\ .
\end{align}
Notice that the same combination of $S_{a}$ and $C_{a}$ as in the last line
of Eq.~(\ref{AxFW1}) has already been dropped. Similarly, $%
S_{a}^{2}+2C_{a}=-4m^{2}\sin^{4}(\mu a/m)\sim\mathcal{O}(1/m^{2})$ can be
discarded. Again, we observe that the infinite towers of interactions in the
polar representation, the $\exp(2ia/m)\sigma^{\mu\nu}F_{\mu\nu}$ and $\exp
(2ia/m)\sigma^{\mu\nu}\tilde{F}_{\mu\nu}$ terms in Eq.~(\ref{LagrFriarDM})
for $\mu=1$, are automatically truncated when expanded in the
non-relativistic limit. This fact would have been totally missed if we had
truncated the series already in Eq.~(\ref{LagrFriarDM}). There is another
interesting aspect of this truncation. Setting $\mu=1$ in Eq.~(\ref%
{LagrFriarDM}), a direct coupling of the axion to the electric dipole
operator $a\bar{\psi}\sigma^{\mu\nu}\gamma^{5}\psi F_{\mu\nu}$ is present in
the polar representation, but not in the derivative one, and with a
coefficient proportional to the magnetic moment of $\psi$. We now see that
this coupling disappears in the non-relativistic limit, making both
representations compatible. This information will play an important role in
analyzing nucleon EDMs in Sec.~\ref{AxEDMobs}.

At this point, we start to perform some Schiff transformations,
specifically, $S_{1}$ as given in Eq.~(\ref{AxSchiff1}), $S_{2}$ in Eq.~(\ref%
{AxSchiff2}), $S_{3}$ in Eq.~(\ref{AxSchiff3}), and finally, 
\begin{equation}
iS_{4}=\frac{i}{8m^{2}}\gamma ^{0}\{\boldsymbol{\gamma }\cdot \mathbf{p},%
\boldsymbol{\gamma }\cdot (\delta _{\mu }\mathbf{E}+d\mathbf{B})\}\ ,
\end{equation}%
to remove the operator involving $\delta _{\mu }\mathbf{\dot{E}}+d\mathbf{%
\dot{B}}$. The $S_{2}$ and $S_{3}$ transformations reorganize the terms in $%
\ddot{a}$ occurring in $\mathcal{H}_{3}$, exactly as in Eq.~(\ref{ReorgAdot}%
). For $S_{1}$, an additional term appears (compare with Eq.~(\ref%
{Schiff1Neu}))%
\begin{align}
\lbrack iS_{1},\mathcal{H}^{\mathrm{NR}}(\mu )]-\dot{S}_{1}& =-\frac{1}{%
8m^{2}}\gamma ^{5}\{\boldsymbol{\gamma }\cdot \mathbf{p},\dot{S}_{a}+4(1-\mu
)\dot{a}\}  \notag \\
& \ \ \ +\frac{i}{16m^{3}}\gamma ^{0}\gamma ^{5}[\{\boldsymbol{\gamma }\cdot 
\mathbf{p},S_{a}+4(1-\mu )a\},\mathbf{p}^{2}]  \notag \\
& \ \ \ +\frac{1}{16m^{3}}\gamma ^{0}[\{\boldsymbol{\gamma }\cdot \mathbf{p}%
,S_{a}+4(1-\mu )a\},[\boldsymbol{\gamma }\cdot \mathbf{p},S_{a}+2(1-\mu )a]]
\notag \\
& \ \ \ +\frac{i}{8m^{2}}[\boldsymbol{\gamma }\cdot (\delta _{\mu }\mathbf{B}%
-d\mathbf{E}),\{\boldsymbol{\gamma }\cdot \mathbf{p},S_{a}+4(1-\mu )a\}]+%
\mathcal{O}(m^{-4})\ .
\end{align}%
This new term combines with the $\{S_{a},[\boldsymbol{\gamma }\cdot \mathbf{p%
},\boldsymbol{\gamma }\cdot (\delta _{\mu }\mathbf{B}-d\mathbf{E})]\}$
operator of Eq.~(\ref{HaxEDM}) to make it $\mu $ independent. The other
terms combine with those in $\mathcal{H}_{3}$ as in Sec.~\ref{SecNeutral},
and the final Hamiltonian no longer depends on $\mu $ at all in the
non-relativistic limit:%
\begin{equation}
\fbox{$%
\begin{array}{ll}
\mathcal{H}^{\mathrm{NR}}= & \gamma ^{0}\left( m+\dfrac{\mathbf{p}^{2}}{2m}-%
\dfrac{\mathbf{p}^{4}}{8m^{3}}-\delta _{\mu }\boldsymbol{\sigma }\cdot 
\mathbf{B}+d\boldsymbol{\sigma }\cdot \mathbf{E}+\dfrac{(\delta _{\mu }%
\mathbf{E}+d\mathbf{B})^{2}}{2m}\right. \medskip \\ 
& \ \ \ \ \dfrac{i\gamma ^{5}[\boldsymbol{\gamma }\cdot \mathbf{p},a]}{m}-%
\dfrac{i\gamma ^{5}\left( [\mathbf{p}^{2},\{\boldsymbol{\gamma }\cdot 
\mathbf{p},a\}]+2\{\mathbf{p}^{2},[\boldsymbol{\gamma }\cdot \mathbf{p}%
,a]\}\right) }{8m^{3}}\medskip \\ 
& \ \ \ \ \left. +\dfrac{a[\boldsymbol{\gamma }\cdot \mathbf{p},[\boldsymbol{%
\gamma }\cdot \mathbf{p},a]]}{m^{3}}+\dfrac{\dot{a}^{2}}{2m^{3}}\right)
\medskip \\ 
& +\dfrac{i[\boldsymbol{\gamma }\cdot \mathbf{p},\boldsymbol{\gamma }\cdot
(\delta _{\mu }\mathbf{E}+d\mathbf{B})]}{2m}+\dfrac{\gamma ^{5}\{\boldsymbol{%
\gamma }\cdot \mathbf{p},\{\boldsymbol{\gamma }\cdot \mathbf{p},\boldsymbol{%
\gamma }\cdot (\delta _{\mu }\mathbf{B}-d\mathbf{E})\}\}}{8m^{2}}\medskip \\ 
& -\dfrac{i\{a,[\boldsymbol{\gamma }\cdot \mathbf{p},\boldsymbol{\gamma }%
\cdot (\delta _{\mu }\mathbf{B}-d\mathbf{E})]\}}{2m^{2}}+\mathcal{O}%
(m^{-4})\ .%
\end{array}%
$}  \label{Hfinal3}
\end{equation}%
One can recognize in the first and fourth lines the terms of $\mathcal{H}%
_{EM}^{\mathrm{NR}}$ in the $e\rightarrow 0$ limit, Eq.~(\ref{HEDMMDM}), the
terms in the second and third as those of the neutral fermion, Eq.~(\ref%
{HneutAxion}), so the only new feature is the operator in the last line. It
encodes higher order effects induced by the Schiff transformation of the $%
\dot{a}$ term, and can be worked out using 
\begin{equation}
i[\boldsymbol{\gamma }\cdot \mathbf{p},\boldsymbol{\gamma }\cdot (\delta
_{\mu }\mathbf{B}-d\mathbf{E})]=d\mathbf{\nabla \cdot E}-i\delta _{\mu }%
\boldsymbol{\sigma }\cdot \mathbf{\nabla }\times \mathbf{B}-2\boldsymbol{%
\sigma }\cdot ((\delta _{\mu }\mathbf{B}-d\mathbf{E})\times \mathbf{P})\ ,
\end{equation}%
where we have set $\mathbf{\nabla \cdot B}=0$ and $\mathbf{\nabla }\times 
\mathbf{E}=0$. These couplings are rather similar to the Darwin and
spin-orbit couplings, and for a neutral fermion, should not be directly
accessible. Further, whenever $d$ is already induced by the axionic
background, these couplings represent a negligible second order effect. So,
the important conclusion of this calculation is that for a neutral state,
there is no coupling to the time-derivative of the axion background, but the
leading non-axionic EDM term is physical.

\section{Axionic EDM observability and estimates}

\label{AxEDMobs}

Whether in its axioelectric or axionic EDM form, the $\mathcal{O}(m^{-2})$
operators lead to oscillating EDM for charged states, i.e., to charged
leptons and quarks, and thereby presumably to nucleons. Naively, if the analogy with
the magnetic moments is valid, the expected size of these oscillating EDMs
should be larger than those due to the axion-gluon or axion-photon local
anomalous interactions, $aG_{\mu \nu }\tilde{G}^{\mu \nu }$ or $aF_{\mu \nu }%
\tilde{F}^{\mu \nu }$, respectively. In some sense, the $\mathcal{O}(m^{-2})$
operators are the equivalent of the Dirac prediction $g=2$ (see Eqs.~(\ref%
{dual1},~\ref{dual2})), while the loop-level anomalous contributions play
the role of the anomalous magnetic moment. At the same time, there are also
reasons to think that this analogy does not hold because the $\mathcal{O}%
(m^{-2})$ operators and the anomalous contributions to a given particle EDM
appear different in their scaling with the axion mass, in their connection
with Schiff's screening, and in the case of the nucleons, in their
hadronization. So, to confirm the naive expectation, it is necessary to
delve into the details of how these operators translate into observables,
from which oscillating EDMs could in principle be accessed. We will not do a
systematic analysis of all the possible experimental settings, and to avoid
complications arising from nuclear and atomic effects, we will focuss on the
simplest system consisting of a single particle precessing in external EM
fields, first in the leptonic case, and then in the more intricate nucleon
case. This treatment will prove sufficient to establish once more the
equivalence between the axioelectric and axionic EDM operators, and to show
and characterize the situations in which they do lead to much larger EDMs
than expected on the basis of the anomalous axion couplings alone.

\subsection{Leptons}

Imagine a charged lepton, not bound in an atom, travelling in a region where 
$\mathbf{\nabla }a$ is negligible and only some electric fields are present.
Then, its spin precession is dictated by the Hamiltonian in Eq.~(\ref%
{HfinalEM}) as, to leading order in $1/m$,%
\begin{equation}
\boldsymbol{\dot{S}}=-i[\boldsymbol{S},\mathcal{H}^{\mathrm{NR}}(\alpha
,\beta )]=-i\left[ \boldsymbol{S},\alpha \dfrac{\gamma ^{5}\{\boldsymbol{%
\gamma }\cdot \mathbf{P},\dot{a}\}}{2m^{2}}-(1-\alpha )\frac{ea\gamma ^{5}%
\boldsymbol{\gamma }\cdot \mathbf{E}}{m^{2}}\right] \ .  \label{LepPrec1}
\end{equation}%
In writing this equation, we implicitly use the fact that the spin operator is the
same for all $\alpha $ (but for the gauge variance due to $\mathbf{P}$,
which cancels only when acting on the fermion wavefunction). This makes
sense because, compared to the discussion in Sec.~\ref{SecNeutSchiff}, the
Schiff transformation of Eq.~(\ref{SchiffEM1}) relates two equivalent forms
of the axion coupling. It does not affect the mass term, which anchors the
rest-frame in which $\boldsymbol{S}$ is defined. In other words, at the
level of the Lagrangian, the original Schiff transformation of Eq.~(\ref%
{SchiffTf1}) is related to the purely chiral rotation $\exp (i\alpha \gamma
^{5})$, as was made clear in Eqs.~(\ref{dual1},~\ref{dual2}), while that for
the axion coupling, Eq.~(\ref{SchiffEM1}), is rather related to the
Goldstone boson reparametrization $\exp (i\alpha \gamma ^{5}a/m)$.

\paragraph{Reparametrization invariance.}

Let us first prove that observables do not depend on $\alpha $. In the $\phi
=0$ gauge, the $\boldsymbol{\gamma }\cdot \mathbf{p}$ term does not
contribute and $\mathbf{E}=-\mathbf{\dot{A}}$, so the equation for $%
\boldsymbol{S}$ reduces to%
\begin{equation}
\boldsymbol{\dot{S}}=\frac{2e}{m^{2}}\gamma ^{0}\boldsymbol{S}\times
(-\alpha \dot{a}\mathbf{A}-(1-\alpha )a\mathbf{E})\ .
\end{equation}%
Since this equation is of the $\dot{f}=f\times g$ type, for which a solution
involves the time integral of $g$, the two operators give the same
contribution and $\alpha $ drops out (under appropriate boundary and gauge
conditions). We recover the equivalence of the two representations for the $%
\mathcal{O}(m^{-2})$ axion couplings, as depicted on the top line of Table~%
\ref{TableEquiv}. 

If we take instead a gauge where $\mathbf{A}=0$, the $\boldsymbol{\gamma }%
\cdot \mathbf{p}$ produces the required $\mathbf{\nabla }\phi $ term
of $\mathbf{E}$, as depicted in the bottom line of Table~\ref{TableEquiv}.
Let us check this explicitly, assuming $\mathbf{\nabla }a=0$. We
start with the EDM form of the operator in the equation of motion of $%
\boldsymbol{S}$, and first replace $e\mathbf{E}=-e\mathbf{\nabla }\phi =-i[%
\mathbf{p},\mathcal{H}_{0}^{\mathrm{NR}}]+\mathcal{O}(1/m)$ with $\mathcal{H}%
_{0}^{\mathrm{NR}}$ the Hamiltonian with $a=0$: 
\begin{equation}
\left. \boldsymbol{\dot{S}}\right\vert _{a\mathbf{E}}=-i\left[ \boldsymbol{S}%
,-\frac{ea\gamma ^{5}\boldsymbol{\gamma }\cdot \mathbf{E}}{m^{2}}\right] =%
\left[ \boldsymbol{S},\frac{a\gamma ^{5}\boldsymbol{\gamma }\cdot \lbrack 
\mathbf{p},\mathcal{H}_{0}^{\mathrm{NR}}]}{m^{2}}\right] \ .
\end{equation}%
Acting with $\mathcal{H}_{0}^{\mathrm{NR}}$ on the external spinors gives
the energy difference, which comes entirely from the difference in electric
potential felt by the initial and final states. This must match the energy
brought in by the axion, i.e. $m_{a}$, so that $a\gamma ^{5}\boldsymbol{%
\gamma }\cdot \lbrack \mathbf{p},\mathcal{H}_{0}^{\mathrm{NR}}]$ collapses
to $\dot{a}\gamma ^{5}\boldsymbol{\gamma }\cdot \mathbf{p}$, as it should%
\footnote{%
These manipulations do not apply to neutral states. For them, if there is an
EDM term in the equation of motion of $\boldsymbol{S}$, it cannot be
expressed as an axioelectric operator since $\mathbf{E}$ cannot be expressed
in terms of $[\mathbf{p},\mathcal{H}^{\mathrm{NR}}]$. This is in accordance
with the fact that the axioelectric operator can be rotated away for neutral
states.}. In full details, the correspondance is verified by writing
explicitly the external spinors. According to the Ehrenfest theorem, 
\begin{eqnarray}
\frac{d}{dt}\left. \left\langle \psi \right\vert \boldsymbol{S}\left\vert
\psi \right\rangle \right\vert _{a\mathbf{E}} &=&\frac{1}{m^{2}}\left\langle
\psi \right\vert \left[ \boldsymbol{S},a\gamma ^{5}\boldsymbol{\gamma }\cdot
\lbrack \mathbf{p},\mathcal{H}_{0}^{\mathrm{NR}}]\right] \left\vert \psi
\right\rangle   \notag \\
&=&\frac{1}{m^{2}}\left\langle \psi \right\vert -[\mathcal{H}_{0}^{\mathrm{NR%
}},[\boldsymbol{S},a\gamma ^{5}\boldsymbol{\gamma }\cdot \mathbf{p}]]-[[%
\boldsymbol{S},\mathcal{H}_{0}^{\mathrm{NR}}],a\gamma ^{5}\boldsymbol{\gamma 
}\cdot \mathbf{p}]\left\vert \psi \right\rangle   \notag \\
&=&\frac{1}{m^{2}}\left\langle \psi \right\vert i\overleftarrow{\partial }%
_{t}[\boldsymbol{S},a\gamma ^{5}\boldsymbol{\gamma }\cdot \mathbf{p}]]+[%
\boldsymbol{S},a\gamma ^{5}\boldsymbol{\gamma }\cdot \mathbf{p}]]i%
\overrightarrow{\partial }_{t}\left\vert \psi \right\rangle   \notag \\
&=&-i\left\langle \psi \right\vert [\boldsymbol{S},\frac{\gamma ^{5}\{%
\boldsymbol{\gamma }\cdot \mathbf{p},\dot{a}\}}{2m^{2}}]\left\vert \psi
\right\rangle =\frac{d}{dt}\left. \left\langle \psi \right\vert \boldsymbol{S%
}\left\vert \psi \right\rangle \right\vert _{\dot{a}\mathbf{p}}\ ,
\label{PrecessAl}
\end{eqnarray}%
using the equation of motion of the external states $i\partial
_{t}\left\vert \psi \right\rangle =\mathcal{H}_{0}^{\mathrm{NR}}\left\vert
\psi \right\rangle $, the Jacobi identity $[A,[B,C]]+[C,[A,B]]+[B,[C,A]]=0$,
and integrating by part over time, i.e., imposing the conservation of
energy. We do not include $\left\langle \psi \right\vert \boldsymbol{\dot{S}}%
\left\vert \psi \right\rangle $ in the evolution of $\left\langle \psi
\right\vert \boldsymbol{S}\left\vert \psi \right\rangle $ and used $[%
\boldsymbol{S},\mathcal{H}_{0}^{\mathrm{NR}}]=i\boldsymbol{\dot{S}}=0$
because as an operator, $\boldsymbol{S}$ does not vary in time. This is
consistent since in this picture, $\mathcal{H}_{0}^{\mathrm{NR}}=\gamma
^{0}(m+...)$ and $\boldsymbol{S}=\boldsymbol{\sigma }/2=-\gamma ^{0}\gamma
^{5}\boldsymbol{\gamma }/2$, so clearly $[\boldsymbol{S},\mathcal{H}_{0}^{%
\mathrm{NR}}]=0$. We thus recover the equivalence between the axionic EDM
and axioelectric operators of Eq.~(\ref{AxioEquiv}). All in all, the
situation is totally analogous to that for the axioelectric effect discussed
in Sec.~\ref{SecEquiv}: observables are gauge invariant and independent of
the choice of operators in $\mathcal{H}^{\mathrm{NR}}(\alpha ,\beta )$.

\paragraph{Numerical estimates. }

Altogether, the $\alpha $ parameters drops out from the equation of motion
of $\boldsymbol{S}$ which can be written as%
\begin{equation}
\boldsymbol{\dot{S}}=-\frac{2ea}{m^{2}}\gamma ^{0}\boldsymbol{S}\times 
\mathbf{E}\ .  \label{aEDMprec}
\end{equation}%
Naively, this could represent a significant effect. For an electron, taking
the coherent classical axion background $a(t)=a_{0}\cos (m_{a}t)$ with $%
m_{a}a_{0}=\sqrt{2\rho _{DM}}$ and $\rho _{DM}=0.4\ $GeV$/cm^{3}~$\cite%
{Catena:2009mf}%
\begin{equation}
d_{e}(t)=\frac{ea(t)}{m_{e}\Lambda }\approx 10^{-11}\frac{a(t)}{\Lambda }~e%
\text{ cm\ .}  \label{atEDM}
\end{equation}%
With in addition the QCD axion mass and scale related by $m_{a}\Lambda
\approx f_{\pi }m_{\pi }\approx m_{\pi }^{2}$, we find%
\begin{equation}
d_{e}(t)\approx 10^{-11}\frac{\sqrt{2\rho _{DM}}}{m_{\pi }^{2}}\cos
(m_{a}t)\approx 10^{-30}\cos (m_{a}t)~e\text{ cm\ ,}
\end{equation}%
independently of $m_{a}$ and $\Lambda $~\cite{Graham:2011qk,Graham:2013gfa}.
Alternatively, this same estimate can be expressed in terms of the
axion-electron coupling (which corresponds to taking $a\rightarrow g_{ae}a$
in Eq.~(\ref{HfinalEM})),%
\begin{equation}
d_{e}(t)=eg_{ae}\dfrac{\sqrt{2\rho _{DM}}}{m_{e}^{2}m_{a}}\cos
(m_{a}t)\approx (10^{-19}~\text{eV})\text{\ }\frac{g_{ae}}{m_{a}}\cos
(m_{a}t)~e\text{ cm}\ .
\end{equation}%
The electron EDM thus appears very promising to set competitive bounds on
the axion (or ALP) couplings, especially when compared to the current limit $%
d_{e}^{\exp }<1.1\times 10^{-29}$ $e$ cm~\cite{ACME:2018yjb} for a fixed $%
d_{e}\boldsymbol{\sigma }\cdot \mathbf{E}$ coupling. Yet, the two cannot be
immediately related, not least because $d_{e}^{\exp }$ is extracted from
atoms, i.e., bound electrons, but also because even for freely precessing
leptons, there are conditions hidden in Eq.~(\ref{aEDMprec}). Also,
trivially, the axionic EDM integrates to zero if the observation time $T$
runs over several oscillation cycles, and the electric field $\left\vert 
\mathbf{E}\right\vert =E_{0}$ is constant. 

An immediate question looking at Eq.~(\ref{aEDMprec}) is what happens if the axion is sufficiently slowly
varying compared to $T$, i.e., when $T$ is small enough compared to $1/m_{a}$.
In effect, the axion field is constant, and it may seem $\boldsymbol{\dot{S%
}}\sim a\boldsymbol{S}\times \mathbf{E}$ survives as $m_{a}\rightarrow 0$
since it is linear in $a$. This is not true though, because if the axion
field is constant, then $\boldsymbol{S}$ is no longer the right spin
operator. With $a(t)=a_{0}$, the mass term becomes complex (see Eq.~(\ref%
{Lpol})), and a chiral rotation of the wavefunction becomes necessary to
identify the true spin operator. This additional change of basis, required
only if the axion field is constant over the observation time (or other
relevant time scale), is what distinguish the situation with $\alpha =0$
from $\alpha =1$ in Eq.~(\ref{LepPrec1}). More generally, if we write $%
a(t)=a_{0}+\dot{a}t+...$, then the $a_{0}$ term disappears when the fermion
mass is made real. When $T\lesssim 1/m_{a}$, the observable change in the
spin orientation ends up linear in $m_{a}$, with $a\left\vert \mathbf{E}%
\right\vert \approx E_{0}a_{0}m_{a}T$. This makes the precession
inobservable for very small axion masses, reproduced the expected decoupling
of the axion in the $m_{a}\rightarrow 0$ limit, but makes the exploitation
of $d_{e}^{\exp }$ for constraining axion interactions impossible.

More promising are situations in which the electric field is oscillating in
time at some frequency $\omega \approx m_{a}$. Then, no change of basis is
needed, $a\left\vert \mathbf{E}\right\vert =E_{0}a_{0}\cos (m_{a}t)\cos
(\omega t)$ is not linear in $m_{a}$, and it integrates to a non-zero value
over some long enough observation time. Clearly, one can no longer take the $%
m_{a}\rightarrow 0$ limit here, it is ill-defined for this situation. Also,
the matching with the axioelectric form is trivial since to $\left\vert 
\mathbf{E}\right\vert =E_{0}\cos (\omega t)$ corresponds $\left\vert \mathbf{%
A}\right\vert =(E_{0}/\omega )\sin \omega t$, so $\dot{a}\left\vert \mathbf{A%
}\right\vert $ matches $a\left\vert \mathbf{E}\right\vert $, up to fixed
boundary terms. This shows how oscillating electric fields really takes the
full advantage of the fact that the derivative of the axion field, $\dot{a}$%
, is coupled to the vector potential, and not to the electric field\footnote{%
Actually, there can be no coupling $\dot{a}\boldsymbol{\gamma }\cdot \mathbf{%
E}$ at $\mathcal{O}(m^{-3})$, only axion wind operators. At $\mathcal{O}%
(m^{-4})$, one finds the pair of equivalent couplings $\ddot{a}\boldsymbol{%
\gamma }\cdot \mathbf{E}$ and $\dot{a}\boldsymbol{\gamma }\cdot \mathbf{\dot{%
E}}$, which are negligible compared to the $\mathcal{O}(m^{-2})$ term
discussed here.}. Thus, provided the axion mass is not too small so that $T$
covers more than a fraction of an oscillation, it is in principle possible
to access directly to the axion-induced EDM, as predicted by Eq.~(\ref{atEDM}%
). This could be particularly interesting with electric microwaves, covering
the $\mu $eV range of axion masses favored by the misalignment mechanism.
Note also that this reasoning remains valid in a CASPER-like situation\cite%
{Budker:2013hfa} in which the spin of the charged lepton is precessing at a
Larmor frequency $\omega _{L}\approx m_{a}$ in a magnetic field. In that
case, a constant electric field is seen in the rotating frame as oscillating
at that $\omega _{L}$ frequency and Eq.~(\ref{atEDM}) applies.

Let us stress though that Eq.~(\ref{atEDM}) does not apply to CASPER itself as currently designed, since we are dealing with free charged leptons here. Further, we do expect very significant suppressions for charged fermions bound into a neutral atomic system, as will be shown in the next section for the neutron. Further work is needed to cover these systems, to see whether some sensitivity can be retained for some range of axion masses. In that respect, the equivalence between the axionic EDM and axioelectric forms of the operator may turn out to be useful. We already know that they both have precisely the same capability to kick bound electrons out when the axion brings enough energy, and we have seen above that they act in the same way on free lepton spins, so there is no reason to think they could act differently on bound electrons. As a tool, this equivalence could thus help in obtaining realistic numerical estimates for atomic systems. This is left for future work.

\subsection{Nucleons}

Let us now turn to the nucleons, for which the situation is much more
complicated because of hadronic effects, and because the quarks are not
expected to be non-relativistic inside a nucleon. We concentrate on the
connection between the quark and nucleon levels here, with an idealized
nucleon precession experiment in mind, and leave the discussions about the
nuclear or atomic levels to future work. To proceed, let us characterize the
various contributions to the nucleon EDMs in terms of effective Lagrangian
couplings.

\begin{description}
\item[\textbf{1- Quark constant EDMs}.] The simplest mechanism to generate a
nucleon EDM occurs when quarks develop constant EDMs, like in the presence
of some new CP violating sources. The quark electric moment operators then
translate naturally into the corresponding nucleon operators%
\begin{equation}
\mathcal{L}_{q,g}\supset \bar{\psi}_{q}\left( i\frac{d_{q}}{2}\sigma ^{\mu
\nu }\gamma ^{5}F_{\mu \nu }\right) \psi _{q}\rightarrow \mathcal{L}%
_{N}\supset \bar{\psi}_{N}\left( i\frac{d_{N}}{2}\sigma ^{\mu \nu }\gamma
^{5}F_{\mu \nu }\right) \psi _{N}\ ,
\end{equation}%
with $q=u,d$. Given current lattice estimates~\cite{Dekens:2018bci},
hadronization appears essentially transparent on these local operators, and
the naive $SU(6)$ estimate in Eq.~(\ref{EDMn}) is actually quite good. In
case the short-distance quark EDMs scale as $1/m_{q}$, analogy with the
magnetic moment would suggest a similarly transparent hadronization, but
with running quark masses replaced by constituent masses. The important
information is that even if the neutron is neutral, it is not
\textquotedblleft EDM neutral\textquotedblright , because the EDM
interaction involves a combination of spins and electric charges. As a
result, even very soft photons, insensitive to the quark structure, do
interact with the neutron via its spin.

\item[\textbf{2- Gluonic contributions}.] The fundamental axion interactions
with quarks or gluons do not involve the photon field at leading order. Yet,
the quarks being electrically charged, non-local processes at the partonic
level can induce local EDM operators at the nucleon level. The most
well-known such non-local EDM contribution comes from the $\theta $ term of
QCD. Indeed, in the presence of an axion field, one expects a coupling%
\begin{equation}
\mathcal{L}_{q,g}\supset \frac{g^{2}}{32\pi ^{2}}\left( \frac{a}{\Lambda }%
+\theta \right) G_{\mu \nu }\tilde{G}^{\mu \nu }\rightarrow \mathcal{L}%
_{N}\supset \bar{\psi}_{N}\left( i\frac{d_{N}}{2}\frac{a}{\Lambda }\sigma
^{\mu \nu }\gamma ^{5}F_{\mu \nu }\right) \psi _{N}\ .  \label{StrongCP}
\end{equation}%
The constant $\theta $ term is cancelled by the axion field falling to its
true minimum, but this leaves a $aG_{\mu \nu }\tilde{G}^{\mu \nu }$
coupling. In the presence of a dark matter axion background, $a(t,\mathbf{x}%
)=a_{0}\cos (m_{a}t-\mathbf{k}\cdot \mathbf{x}+\phi )$, and from Eq.~(\ref%
{thetaEDM}), this term then induces an EDM for the nucleons~\cite%
{Graham:2011qk,Graham:2013gfa}. Using the matrix element estimates quoted in
Ref.~\cite{Yamanaka:2017mef}: 
\begin{subequations}
\label{GlueEDM}
\begin{eqnarray}
d_{n}(t) &=&-(2.7\pm 1.2)\times 10^{-16}\frac{a(t)}{\Lambda }\ ~e\text{ cm\ ,%
} \\
d_{p}(t) &=&+(2.1\pm 1.2)\times 10^{-16}\frac{a(t)}{\Lambda }\ ~e\text{ cm\ ,%
}
\end{eqnarray}%
(we use $d_{N}(t)$ to denote the total nucleon EDM, and $d_{N}$ the
coefficient of the operator in Eq.~(\ref{StrongCP})). Note though that
recent lattice estimates reduce this matrix element by about a factor of two~%
\cite{Dragos:2019oxn,Liang:2023jfj}. Anyway, this prediction has motivated
dedicated experimental searches~\cite{Budker:2013hfa,Abel:2017rtm}, with
specific strategies designed to tackle the oscillatory nature of the EDM. At
this stage, we should point out though that strictly speaking, the matrix
elements was extracted for a constant $\theta $, by extrapolating the
form-factor for $N\rightarrow N\gamma (q)$ to $q^{2}\rightarrow 0$. The idea
is that provided the axion background is not varying too quickly, QCD has
time to account for the presence of the action field as a kind of effective
axionic $\theta $ term. Yet, notice that in the general case, the axion
field is also injecting some energy, albeit a small amount, and $d_{N}$ in
Eq.~(\ref{StrongCP}) is actually a form-factor that depends on both the
photon and axion momenta. When applying this estimate to the axion EDM
coupling, one implicitly makes the assumption that the axion is very soft
and that the limit $m_{a}\rightarrow 0$ is smooth. We will come back to this
point below.
\begin{figure}[t]
\centering\includegraphics[width=0.90\textwidth]{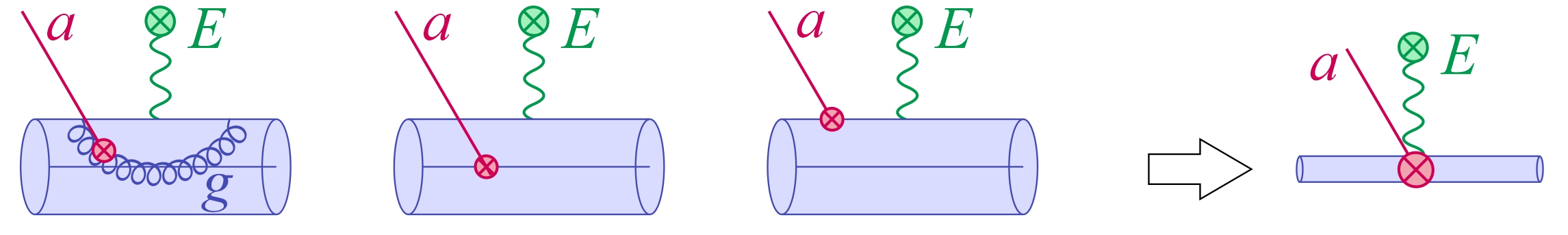}
\caption{Non-local partonic contributions to the nucleon local EDM operator of Eq.~(\ref{StrongCP}).}
\label{Fig1}
\end{figure}

\item[\textbf{3- Ward Identity}.] The pseudoscalar and/or derivative
couplings of the axion to the quarks also contribute to the nucleon operator
in Eq.~(\ref{StrongCP}) via similar non-local processes (see Fig.~\ref{Fig1}). To estimate their relative size compared to the gluonic contribution, a
crucial piece of information comes from the Ward identity of the anomalous
PQ symmetry (we assume for now that the axion couples only to a single quark 
$\psi _{q}$): 
\end{subequations}
\begin{equation}
\partial _{\mu }(\bar{\psi}_{q}\gamma ^{\mu }\gamma ^{5}\psi _{q})=2im_{q}%
\bar{\psi}_{q}\gamma ^{5}\psi _{q}+\frac{g^{2}}{16\pi ^{2}}G_{\mu \nu }%
\tilde{G}^{\mu \nu }\ .  \label{WardId}
\end{equation}%
This means that the Lagrangian interpolating between the derivative and
polar representations should read%
\begin{equation}
\mathcal{L}_{q,g}(\alpha )\supset \bar{\psi}_{q}\left( i\slashed D-m_{q}\exp
\left( 2(1-\alpha )i\gamma ^{5}\frac{a}{\Lambda }\right) +\alpha \frac{%
\gamma ^{\mu }\gamma _{5}\partial _{\mu }a}{\Lambda }\right) \psi
_{q}+\alpha \frac{a}{\Lambda }\frac{g^{2}}{16\pi ^{2}}G_{\mu \nu }\tilde{G}%
^{\mu \nu }\ ,  \label{LagrQuarks}
\end{equation}%
where one can recognize Eq.~(\ref{Lpol}) when $\alpha =0$, and Eq.~(\ref%
{Lder}) when $\alpha =1$, plus the anomalous term. Thus, the gluon-induced
and quark-induced contributions cannot truly be disentangled and must be
treated together. This fact is actually often used to estimate theoretically
the nucleon EDM for constant $\theta $, as it can be easier to deal with a
phase for the quark masses than with the anomalous $G\tilde{G}$ coupling~%
\cite{Baluni:1978rf,Crewther:1979pi}. Note, though, that if the $aG\tilde{G}$
coupling also receives contributions from other heavy states, either SM
quarks or new heavy fermions like in the KSVZ scenario, then the quark
Lagrangian should rather read%
\begin{eqnarray}
\mathcal{L}_{q,g}(\alpha ,\beta ) &\supset &\bar{\psi}_{q}\left( i\slashed %
D-m_{q}\exp \left( 2(g_{q}(1-\alpha )+g_{g}(1-\beta ))i\gamma ^{5}\frac{a}{%
\Lambda }\right) \right.   \notag \\
&&\left. \ \ \ +(g_{q}\alpha -g_{g}(1-\beta ))\frac{\gamma ^{\mu }\gamma
_{5}\partial _{\mu }a}{\Lambda }\right) \psi _{q}+(g_{q}\alpha +g_{g}\beta )%
\frac{a}{\Lambda }\frac{g^{2}}{16\pi ^{2}}G_{\mu \nu }\tilde{G}^{\mu \nu }\
,\ \ \ 
\end{eqnarray}%
where we have put back the coupling $g_{q}$ and $g_{g}$ to distinguish
axion-quark and axion-gluon couplings. Notice that in this case, it is
always possible to chose $\alpha $ (or $\beta $) such that one of the
coupling disappears, i.e., without the axial, the pseudoscalar, or the
anomalous coupling:
\begin{subequations}
\label{QuarkL}
\begin{eqnarray}
\alpha \overset{}{=}\frac{g_{g}}{g_{q}}(1-\beta ) &:&\bar{\psi}_{q}\left( i%
\slashed D-m_{q}\exp 2(g_{q}i\gamma ^{5}\frac{a}{\Lambda }\right) \psi
_{q}+g_{g}\frac{a}{\Lambda }\frac{g^{2}}{16\pi ^{2}}G_{\mu \nu }\tilde{G}%
^{\mu \nu }\ , \\
\alpha \overset{}{=}1+\frac{g_{g}}{g_{q}}(1-\beta ) &:&\bar{\psi}_{q}\left( i%
\slashed D-m_{q}+g_{q}\frac{\gamma ^{\mu }\gamma _{5}\partial _{\mu }a}{%
\Lambda }\right) \psi _{q}+(g_{q}+g_{g})\frac{a}{\Lambda }\frac{g^{2}}{16\pi
^{2}}G_{\mu \nu }\tilde{G}^{\mu \nu },\ \ \ \ \  \\
\alpha \overset{}{=}-\frac{g_{g}}{g_{q}}\beta  &:&\bar{\psi}_{q}\left( i%
\slashed D-m_{q}\exp \left( 2(g_{q}+g_{g})i\gamma ^{5}\frac{a}{\Lambda }%
\right) -g_{g}\frac{\gamma ^{\mu }\gamma _{5}\partial _{\mu }a}{\Lambda }%
\right) \psi _{q}\ .
\end{eqnarray}%
Also, notice that in the quark non-relativistic limit, $g_{g}$ never
contributes to the quark axion wind or the quark axionic EDM operator.

\item[\textbf{4- Sutherland-Veltman theorem}.] What the Ward identity Eq.~(%
\ref{WardId}) shows is that in the soft limit, i.e., $\partial _{\mu
}a\rightarrow 0$, the pseudoscalar axion-quark coupling is strictly
equivalent to the anomalous axion-gluon coupling, and thus that 
\end{subequations}
\begin{equation}
\left\langle N\gamma \right\vert -2i\frac{m_{q}}{\Lambda }a\bar{\psi}%
_{q}\gamma ^{5}\psi _{q}\left\vert N\right\rangle \overset{\partial _{\mu
}a\rightarrow 0}{\rightarrow }\left\langle N\gamma \right\vert \frac{a}{%
\Lambda }\frac{g^{2}}{16\pi ^{2}}G_{\mu \nu }\tilde{G}^{\mu \nu }\left\vert
N\right\rangle \ .  \label{SVTheo}
\end{equation}%
Actually, this is essentially the Sutherland-Veltman theorem~\cite%
{Sutherland:1967vf,Veltman67}, well-known in the context of $\pi
^{0}\rightarrow \gamma \gamma $. Here, it proves that the non-local
contributions of $a\bar{\psi}_{q}\gamma ^{5}\psi _{q}$ to the nucleon EDM
must reproduce that of $aG_{\mu \nu }\tilde{G}^{\mu \nu }$ quoted in Eq.~(%
\ref{GlueEDM}) in the soft limit. This serves as a baseline, and the
question now is whether the EDM can be enhanced compared to Eq.~(\ref%
{GlueEDM}) even slightly away from that limit. Note, for completeness, that
the Ward identity also relates the matrix elements of the axial current and
the anomalous term in the chiral limit. Indeed, for a massless quark, the
axion decouples entirely from Eq.~(\ref{LagrQuarks}), as can be seen taking $%
m_{q}=0$ and $\alpha =0$. Thus, the matrix elements of $\bar{\psi}_{q}\gamma
^{\mu }\gamma _{5}\psi _{q}\partial _{\mu }a$ and $aG_{\mu \nu }\tilde{G}%
^{\mu \nu }$ must match when $m_{q}\rightarrow 0$ since the two must cancel
each other~\cite{Quevillon:2019zrd}.

\item[\textbf{5- On the }$m_{a}\rightarrow 0$\textbf{\ limit. }] The soft
limit, $\partial _{\mu }a\rightarrow 0$, and the $m_{a}\rightarrow 0$ limit
are not entirely equivalent. Naively, if $a(t,\mathbf{x})=a_{0}\cos (m_{a}t-%
\mathbf{k}\cdot \mathbf{x}+\phi )$ for some constant $\phi $, $\partial
_{\mu }a\rightarrow 0$ is equivalent to $\dot{a}=0$ if $\mathbf{k}$ is
negligible, and $\dot{a}(t)=0$ can be attained for all time only with $%
m_{a}=0$. Now, simply setting $m_{a}=0$, the axion field becomes constant
and some of its couplings to quarks and gluon in Eq.~(\ref{LagrQuarks})
survive, and so is the nucleon EDM operator in Eq.~(\ref{StrongCP}). At the
same time, theoretically, $m_{a}\rightarrow 0$ requires $\Lambda \rightarrow
\infty $ since $m_{a}$ comes from the gluon coupling $aG_{\mu \nu }\tilde{G}%
^{\mu \nu }$. But if we send $\Lambda $ to infinity, the axion entirely
decouples, there will be no axionic EDM at all, and the strong CP puzzle is
back. Actually, there is probably a threshold for $m_{a}$ below which the
axion would take too much time to realign to compensate for some preexisting 
$\theta $ term, in which case all the axion contributions would be
overshadowed by the large constant EDM due to $\theta $. This goes beyond
what is discussed here, and we still assume that term is absent. Yet, if we
expand the coherent background axion field as $a(x^{\mu })=a_{0}+x^{\mu
}\partial _{\mu }a+...$, how to treat the constant term $a_{0}$ needs
caution. In effect, QCD with such an external field looks very much like QCD
with a non-zero $\theta $ term, which is CP-violating and quite different
from QCD with an axion and $\theta =0$, which is CP-conserving (see e.g.
Ref.~\cite{DiVecchia:2013swa} for a comparison). In particular, the $a_{0}$
term generates complex phases for the quark condensate, and changes how to
treat the chiral symmetry breaking terms. Specifically, when $a=a_{0}$ is
constant, Dashen theorem must be called in, and after the necessary
realignment of the chiral vacuum~\cite{Baluni:1978rf},%
\begin{equation}
\frac{m_{q}}{\Lambda }a_{0}\bar{\psi}_{q}(i\gamma _{5})\psi _{q}\rightarrow
\left( \frac{1}{m_{u}}+\frac{1}{m_{d}}+\frac{1}{m_{s}}\right) ^{-1}\frac{%
a_{0}}{\Lambda }\sum_{q=u,d,s}\bar{\psi}_{q}(i\gamma _{5})\psi _{q}\ .
\end{equation}%
This drastically changes the character of the axion-quark coupling, and
suppresses it significantly (it now vanishes if any of the quark masses
vanishes, as it should). This isospin singlet quark current accounts for the
isospin singlet anomalous term $a_{0}G_{\mu \nu }\tilde{G}^{\mu \nu }$. This
explains how Sutherland-Veltman theorem sets in: the $\mathcal{L}%
_{q,g}(\alpha =0)$ term contains both a suppressed term collapsing to the
anomalous one, and a term matching the axial interaction (i.e., proportional
to $\partial _{\mu }a$), so that any observable calculated from $\mathcal{L}%
_{q,g}(\alpha =0)$ or $\mathcal{L}_{q,g}(\alpha =1)$ are equal.

\item[\textbf{6- Axion couplings to nucleons}.] The three axion couplings in 
$\mathcal{L}_{q,g}(\alpha )$ of Eq.~(\ref{LagrQuarks}) generate the nucleon
EDM operator, but also simpler axion couplings to nucleons. With them, the
nucleon effective Lagrangian becomes:%
\begin{equation}
\mathcal{L}_{N}(\alpha ^{\prime })=\bar{\psi}_{N}\left( i\slashed %
D-m_{N}\exp \left( 2g_{N}(1-\alpha ^{\prime })i\gamma ^{5}\frac{a}{\Lambda }%
\right) +g_{N}\alpha ^{\prime }\frac{\gamma ^{\mu }\gamma _{5}\partial _{\mu
}a}{\Lambda }+i\frac{d_{N}}{2}\frac{a}{\Lambda }\sigma ^{\mu \nu }\gamma
^{5}F_{\mu \nu }\right) \psi _{N}\ ,  \label{LagrNucl}
\end{equation}%
for some prefactor $g_{N}$ a priori of $\mathcal{O}(1)$, and $D^{\mu
}=\partial ^{\mu }-ieQ_{N}A^{\mu }$ with $Q_{N}$ the nucleon electric
charge. From the previous points, all the $\mathcal{L}_{q,g}(\alpha )$
couplings appear to contribute to all the $\mathcal{L}_{N}(\alpha ^{\prime })
$ couplings. For instance, the axion-gluon coupling contributes to both $%
d_{N}$ and $g_{N}$, since $aG_{\mu \nu }\tilde{G}^{\mu \nu }$ does not need
photons to generate a CP-violating coupling. Similarly, the axion-quark
couplings in $\mathcal{L}_{q,g}(\alpha )$ naturally induce axion-nucleon
couplings, but also the local anomalous term $d_{N}$, as is obvious starting
from $\mathcal{L}_{q,g}(\alpha =0)$ or invoking Sutherland-Veltman theorem,
Eq.~(\ref{SVTheo}). Yet, there are subtelties at play here. First, even if
the the two axion couplings to nucleons are necessarily present and related
under the reparametrization $\psi _{N}\rightarrow \exp (ig_{N}\alpha \gamma
^{5}a/\Lambda )\psi _{N}$, $\alpha ^{\prime }$ is not necessarily equal to $%
\alpha $. We neglect the QED anomaly here, so the Ward identity underpinning
the Goldstone-boson reparametrization invariance at the nucleon level is
simply the classical one (and $\alpha ^{\prime }$ has to cancel from
observables). Second, for a constant axion background $a(t,\mathbf{x})=a_{0}$%
, the axion couplings to nucleons vanish. This is clear for $\alpha ^{\prime
}=1$, while it requires a chiral rotation $\psi _{N}\rightarrow \exp
(i\alpha g_{N}\gamma ^{5}a_{0}/\Lambda )\psi _{N}$ to make the mass term
real for $\alpha ^{\prime }=0$ (this can also be understood as a
nucleon-level Sutherland-Veltman theorem: the pseudoscalar coupling is
equivalent to the axial one, which vanishes in the $\partial _{\mu
}a\rightarrow 0$ limit). Third, the $d_{N}$ coupling does not represent the
whole axionic EDM of the nucleon. From an effective theory point of view, $%
d_{N}$ only represents the short-distance contribution, to which tree-level
(and loop-level once pions and other light mesons are included)
contributions from the leading couplings have to be added, see Figs.~\ref%
{Fig1} and~\ref{Fig2}.
\begin{figure}[t]
\centering\includegraphics[width=0.90\textwidth]{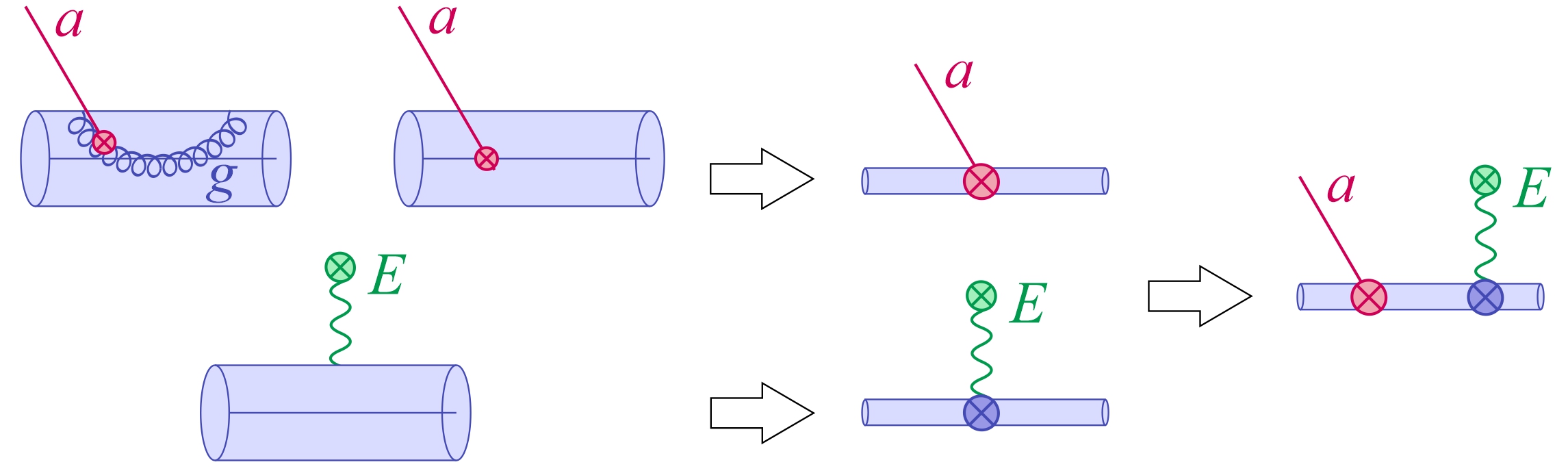}
\caption{Partonic contributions to the nucleon-axion and nucleon-photon
interaction generate a non-local, long-distance EDM effect for the proton,
at the hadronic level and in the non-relativistic limit. The axion-gluon and axion-quark contributions are related by a Ward identity, but their combination is imposed to ensure that the axion-nucleon interactions decouple entirely in the soft limit, $\partial_\mu a \rightarrow 0$. This is justified since in that limit, a non-decoupling axion-nucleon coupling corresponds to a complex mass term, so it has to be removed by a chiral rotation of the nucleon field.}
\label{Fig2}
\end{figure}

\item[\textbf{7- Nucleon magnetic moment.}] Deciding like in Sec.~\ref%
{SecEDM} to add the magnetic dipole operator to $\mathcal{L}_{N}(\alpha
^{\prime }=1)$, the nucleon Lagrangian becomes (see Eq.~(\ref{ShiftDipole})) 
\begin{eqnarray}
\mathcal{L}_{N}(\alpha ^{\prime }) &=&\bar{\psi}_{N}\left( i\slashed %
D-m_{N}\exp \left( 2g_{N}(1-\alpha ^{\prime })i\gamma ^{5}\frac{a}{\Lambda }%
\right) +g_{N}\alpha ^{\prime }\frac{\gamma ^{\mu }\gamma _{5}\partial _{\mu
}a}{\Lambda }\right.   \notag \\
&&\ \ \ \ \ \ \left. -\frac{\mu _{N}}{2}\sigma ^{\mu \nu }F_{\mu \nu }+i%
\frac{d_{N}-2g_{N}(1-\alpha ^{\prime })\mu _{N}}{2}\frac{a}{\Lambda }\sigma
^{\mu \nu }\gamma ^{5}F_{\mu \nu }\right) \psi _{N}\ ,  \label{LnuclMDM}
\end{eqnarray}%
up to dipole terms of $\mathcal{O}(a^{2})$. The magnetic dipole operator
accounts for the proton and neutron magnetic moments $2+\mu _{p}\approx 2.8$
and $\mu _{n}\approx -1.9$, respectively. This Lagrangian is invariant under
the Goldstone boson reparametrizations $\psi _{N}\rightarrow \exp
(ig_{N}\alpha ^{\prime }\gamma ^{5}a/\Lambda )\psi _{N}$, again up to terms
quadratic in the axion field. In the non-relativistic limit, a single
axionic EDM operator arises at leading order:%
\begin{equation}
\mathcal{L}_{N}(\alpha ^{\prime })\rightarrow -\left( d_{N}+\frac{eg_{N}}{%
m_{N}}Q_{N}\right) \dfrac{a}{\Lambda }\gamma ^{5}\boldsymbol{\gamma }\cdot 
\mathbf{E}\ .  \label{UniqEDM}
\end{equation}%
The contribution from $\mu _{N}$ drops out, independently of $\alpha
^{\prime }$ (as was already apparent for the neutron in Eq.~(\ref{Hfinal3}%
)). It is interesting to understand the mechanism of this cancellation (see
Ref.~\cite{Bigi:1990kz}): the $\mu _{N}$-dependent shift of $d_{N}$ is
compensated by the long-distance contributions arising from one-nucleon
reducible diagrams with the photon emitted via the $\mu _{N}\bar{\psi}%
_{N}\sigma ^{\mu \nu }F_{\mu \nu }\psi _{N}$ vertex. Now, this cancellation
crucially relies on the assumption that one should introduce the magnetic
dipole operator to $\mathcal{L}_{N}(\alpha ^{\prime }=1)$. In Sec.~\ref%
{SecEDM}, this was justified by the fact that quarks are PQ neutral in the
derivative representation. Nucleons, on the contrary, do not have definite
PQ charges if the PQ-breaking $aG\tilde{G}$ coupling contributes to the $a%
\bar{N}N$ vertex. To circumvent this, we require that only the local term $%
d_{N}$ should be present in the $\partial _{\mu }a\rightarrow 0$ limit.
Indeed, the reparametrization then becomes a chiral rotation, i.e., a change
of basis. What is to be called the magnetic moment and the EDM have to be
defined in the basis in which the nucleon mass is real (see e.g. Ref.~\cite%
{Abramczyk:2017oxr} for a detailed discussion). With the assumption that
this limit is smooth despite the fact that a change of basis for $\psi _{N}$
is implied, this restores a definite PQ charge for the nucleons. The
representation in Eq.~(\ref{LnuclMDM}) confines the contributions of the $aGG
$ coupling to the local term $d_{N}$, together with some axion-quark
contributions given Eqs.~(\ref{QuarkL}) and~(\ref{SVTheo}), leaving the
quark couplings to induce $g_{N}$. Thus, we expect $d_{N}(g_{q},g_{g})$, but 
$g_{N}(g_{q})$, with the $g_{N}(g_{q})$ contribution cancelling out when $%
\partial _{\mu }a=0$. Note, finally, that $\mathcal{L}_{N}(\alpha ^{\prime
}=1)$ corresponds to the usual form employed in the literature, see e.g.
Ref.~\cite{Graham:2013gfa}.

\item[\textbf{7- Proton EDM:} ] The important property of the first two
operators of $\mathcal{L}_{N}(\alpha ^{\prime })$ is that they combine with
the electromagnetic coupling to produce non-relativistic EDM operators for
the proton, but not for the neutron. The phenomenology of a precessing
proton is thus very similar to that discussed for charged leptons. In the
non-relativistic limit for the proton, using $\mathcal{L}_{N}(\alpha
^{\prime }=1)$, the axion EDM couplings are%
\begin{equation}
\mathcal{H}^{\mathrm{NR}}\supset g_{p}\dfrac{\gamma ^{5}\{\boldsymbol{\gamma 
}\cdot \mathbf{P},\dot{a}\}}{2m_{p}\Lambda }-d_{p}\dfrac{a}{\Lambda }\gamma
^{5}\boldsymbol{\gamma }\cdot \mathbf{E}\ .  \label{NRprot}
\end{equation}%
The covariant axioelectric operator cannot be rotated away, and does induce
spin precession, see Eq.~(\ref{PrecessAl}). We choose to write $\mathcal{H}^{%
\mathrm{NR}}$ starting from $\mathcal{L}_{N}(\alpha ^{\prime }=1)$ instead
of as in the equivalent form of Eq.~(\ref{UniqEDM}) to emphasize the
different nature of these two contributions. First, the $g_{p}$ is
explicitly vanishing if $\partial _{\mu }a=0$, but not the $d_{p}$ term.
Yet, both sum up to an EDM-like precession, and even more, both operators
becomes identical if $\left\vert \mathbf{E}\right\vert =E_{0}\sin (\omega t)$
since then $\dot{a}\gamma ^{5}\boldsymbol{\gamma }\cdot \mathbf{P}\supset 
\dot{a}\gamma ^{5}\boldsymbol{\gamma }\cdot \mathbf{A}=a\gamma ^{5}%
\boldsymbol{\gamma }\cdot \mathbf{E}$ provided $m_{a}=\omega \neq 0$.
Second, these two operators encode different physics: the $d_{p}$
contribution is local already at the hadronic level, but the axioelectric
contribution has an intrinsically non-local origin, see Fig.~\ref{Fig2}, and
becomes local in the non-relativistic limit only (as said earlier, we do not
consider nuclear systems here). From the discussion in the lepton case, we
do expect that the axion-induced proton EDM to increase to%
\begin{equation}
d_{p}(t)\approx g_{p}\frac{ea(t)}{m_{p}\Lambda }\approx 10^{-14}\frac{a(t)}{%
\Lambda }~e\text{ cm}\approx 10^{-14}\frac{\sqrt{2\rho _{DM}}}{m_{\pi }^{2}}%
\cos (m_{a}t)\approx 10^{-33}\cos (m_{a}t)~e\text{ cm\ ,}  \label{EDMp}
\end{equation}%
in the \textquotedblleft resonant\textquotedblright\ situation in which the
EM field matches the axion frequency, and assuming $g_{p}\sim \mathcal{O}(1)$%
. This represent an enhancement of the long-distance contribution by about
two orders of magnitude compared to the local contribution tuned by $d_{p}$,
Eq.~(\ref{GlueEDM}). Beware though that, obviously, the same provisions
about the implicit observation time constraints as in the lepton case do
apply, since the $g_{p}$ contribution does decouple if $\partial _{\mu }a=0$%
, as is manifest in Eq.~(\ref{NRprot}). Note, finally, that there is some
model dependence in comparing Eq.~(\ref{GlueEDM}) to Eq.~(\ref{EDMp}). For
instance, hadrophobic scenarios can be designed in which the axion couplings
to the $u$ and $d$ quarks conspire to suppress $g_{p}$ (see e.g. Ref.~\cite%
{Takahashi:2023vhv} and references there). Barring these possibilities
though, Eq.~(\ref{EDMp}) may represent our best window into the axion-light
quark couplings, even compared to axion wind operator that are suppressed by the local
galactic axion speed~\cite{Stadnik:2013raa}.

\item[\textbf{8- Neutron EDM: }] The purely long-distance enhancement
mechanism at play for the proton is not available for the neutron since it
is neutral, see Fig.~\ref{Fig2} (as said earlier, this crucially rely on how
the neutron magnetic dipole operator is introduced though). Instead, with
only the $a\bar{\psi}_{n}\sigma ^{\mu \nu }\gamma ^{5}F_{\mu \nu }\psi _{n}$
coupling, there will be an enhancement if the non-local quark-level matrix
elements 
\begin{equation}
\left\langle n\gamma \right\vert \frac{\bar{\psi}_{q}\gamma ^{\mu }\gamma
_{5}\partial _{\mu }a\psi _{q}}{\Lambda }\left\vert n\right\rangle \ ,
\label{SVboils}
\end{equation}%
can be significant away from the $m_{a}=0$ limit, so that the enhancement
identified at long-distance somehow spills over at short-distance. If that
is the case, this violation would show up in $d_{N}$, which should be
understood to be a form-factor:%
\begin{equation}
d_{N}(g_{q},g_{g})=d_{N}(g_{q},g_{g};q_{\gamma }^{2},q_{a}^{2},q_{a}\cdot
q_{\gamma })\ .
\end{equation}%
While we know that $d_{N}(g_{q},g_{g};q_{\gamma }^{2},q_{a}^{2},q_{a}\cdot
q_{\gamma })\rightarrow ~$Eq.~(\ref{GlueEDM}) when $\partial _{\mu }a=0$,
the behavior reaching that limit may not be that smooth if Eq.~(\ref{SVboils}%
) does not go to zero sufficiently fast as $\partial _{\mu }a\rightarrow 0$.
Let us imagine that the proton and the neutron are simply collections of
loosely bound non-relativistic constituent quarks. Then, the long-distance
hadronic mechanism at play for the proton would have a direct counterpart as
a non-local constituent quark mechanism (e.g. from the third diagram in Fig.~%
\ref{Fig1}). Both the proton and the neutron EDM would then be expected to
reach Eq.~(\ref{EDMp}) in the presence of \textquotedblleft
resonant\textquotedblright\ EM fields since, as explained in point 1 above,
the neutron is not neutral for spin-dependent electric interactions. In
practice, in this picture, one way to understand Eq.~(\ref{NRprot}) would be
from a term in $d_{N}$ scaling like $q_{a}^{2}/q_{a}\cdot q_{\gamma
}=m_{a}/\omega $, vanishing in the $m_{a}\rightarrow 0$ limit, but of $%
\mathcal{O}(1)$ when $\omega \approx m_{a}$ and $m_{a}$ is not too small. Of
course, this consituent quark picture is not particularly realistic, but in
our opinion, it nevertheless suggests that some level of enhancement of the
neutron EDM is possible. Indeed, the real world situation should lie
somewhere in between no enhancement, as expected looking at Fig.~\ref{Fig2}
with the neutron not interacting with photons, to a significant enhancement
thanks to residual interplays between the axion and photon couplings to the
quarks inside the neutron. Obviously, to get a definitive answer from first
principle is complicated and probably requires detailed lattice simulations
starting from the general Lagrangian of Eq.~(\ref{LagrQuarks}), away from
the $m_{a}^{2}=0$ and $q_{\gamma }^{2}=0$ limit.
\end{description}

To close this section, we stress once more that the above discussion does
not immediately apply to nuclear or atomic probes of the axion-induced
proton and neutron EDMs (assuming the axion-electron coupling is absent).
The non-relativistic limit appears crucial to collapse the axion couplings
to an EDM-like operator for the nucleons, which can then be
enhanced with suitable EM fields. Further, estimating how an oscillatory
external electric field can penetrate the nuclear, atomic, and/or even the
molecular system, accounting in addition for the presence of resonances, and
estimating the resulting observable EDM it would induce is beyond the scope
of the present work~\cite{Flambaum:2019emh,Flambaum:2018wkp,TranTan:2018nvu,Spevak:1996tu}.

\section{Summary}

\label{Ccl}

In this paper, the non-relativistic description of the axion interactions
with fermions was systematically analyzed. We relied on rather old and
well-established techniques like the Foldy-Wouthuysen transformation~\cite%
{Foldy:1949wa}, the unitary transformations of Ref.~\cite{Barnhill:1969ygg},
and Schiff theorem~\cite{Schiff:1963zz}. Yet, as these techniques had not
been fully combined and supplemented by the reparametrization invariance for
the axion field, to our knowledge, none of the final non-relativistic
expansions for the Hamiltonian presented here were derived before. Our
results can be summarized in three points:

\begin{itemize}
\item For a neutral fermion, we demonstrated by adapting Schiff theorem that
the axioelectric operator $\gamma ^{5}\{\boldsymbol{\gamma }\cdot \mathbf{p},%
\dot{a}\}$ is totally screened. As shown in the final Hamiltonian for this
scenario, Eq.~(\ref{HneutAxion}), there are only axion wind operators up to $%
\mathcal{O}(1/m^{3})$, except for a very suppressed $\dot{a}^{2}$ coupling.
Since there should be no finite-size effects, and because $\mathcal{O}%
(1/m^{3})$ relativistic corrections are of a different nature, this
screening should even hold to a much higher level than the usual Schiff
screening of charged fermion EDMs. Phenomenologically, this scenario is not
very relevant since normal matter is essentially made of charged particles,
but it provides the basis to understand the result in the charged case.

\item Specifically, for a charged fermion, the final Hamiltonian is in Eq.~(%
\ref{HfinalEM}). The covariant axioelectric operator $\gamma ^{5}\{%
\boldsymbol{\gamma }\cdot \mathbf{P},\dot{a}\}$ is found equivalent to an
axion-induced EDM operator $a\boldsymbol{\sigma }\cdot \mathbf{E}$, see Eq.~(%
\ref{AxioEquiv}) and Table~\ref{TableEquiv}. Both operators encode the same
physics, but the latter makes it manifest that this coupling disappears in
the absence of EM fields, or for a neutral fermion. Phenomenologically, the
usual axioelectric effect is recovered whatever the chosen form of the
operator, both having the same matrix elements for observables. Besides the
axioelectric effect, these operators can also induce EDMs for all charged
particles. The important points are first that these EDM operators are, in
some sense, tree-level. They are directly predicted by the Dirac equation
itself for all charged fermions, in a way totally analogous to the magnetic
moment factor of 2. Secondly, these EDMs are not constant in time, and
cannot be screened since Schiff transformation would simply change the
relative weight of $\gamma ^{5}\{\boldsymbol{\gamma }\cdot \mathbf{P},\dot{a}%
\}$ and $a\boldsymbol{\sigma }\cdot \mathbf{E}$, something irrelevant since
they lead to the same observables. Thus, though specific search strategies
have to be designed to tackle the oscillatory nature of these EDMs as well as their decoupling in the $m_a \rightarrow 0$ limit, their relatively large sizes, especially for the electron, makes them particularly promising.

\item Finally, concerning the proton and the neutron, the final Hamiltonians are in Eq.~(\ref{HfinalEM}) and Eq.~(\ref{Hfinal3}). The main issue here is whether the axion-induced quark EDMs, which are intrinsically non-relativistic, can manifest themselves at the hadronic level. We find that this is the case for the proton, whose axion-induced EDM is significantly enhanced by long-distance effects compared to current estimates based solely on a local EDM operator induced by the axion-gluon coupling (see Fig.~\ref{Fig2}). For the neutron, if taken as point-like in a first approximation, the axion-induced EDM coupling coming from the axion-quark couplings vanishes exactly since the purely long-distance hadronic contribution is absent. Beyond leading order, some effects are likely as the neutron is not transparent to quark EDM interactions, but further work is needed to estimate these finite-size effects and establish whether they can compete with the axionic EDM coming from the axion-gluon coupling.
\end{itemize}

All these results clarify the construction of non-relativistic expansions in
the presence of Goldstone bosons. Yet, to conclude, we would like to stress
again that this formalism, in itself, has some limitations. For instance,
our starting point was the Dirac equation for a single fermion in the
presence of external background fields, electromagnetic and axionic. In our
opinion, further work is urgently needed to obtain estimates for realistic
experimental settings, in particular in the atomic or nuclear contexts (or
even for the neutron-antineutron system~\cite{Arias-Aragon:2022byr}). Thus,
extending the formalism itself, or even grounding it within a fully
relativistic quantum field theory setting, would be very welcome, not least
to confirm the promising phenomenological opportunities we identified for
the detection of dark matter axions.

\subsection*{Acknowledgements}

The author acknowledges funding from the French Programme d'investissements
d'avenir through the Enigmass Labex, support from the IN2P3 Master project
\textquotedblleft Axions from Particle Physics to
Cosmology\textquotedblright, and from the French National Research Agency
(ANR) in the framework of the \textquotedblleft GrAHal\textquotedblright
project (ANR-22-CE31-0025).

\appendix

\section{Foldy-Wouthuysen transformation\label{AppFW}}

The Foldy-Wouthuysen (FW) procedure~\cite{Foldy:1949wa} is a systematic
order by order method to block-diagonalize the Dirac Hamiltonian via a
sequence of unitary transformations. Though it is well-known and can be
found in many textbooks on relativistic quantum mechanics, for completeness,
we here include a brief derivation up to $\mathcal{O}(1/m^{4})$. Also,
compared to the literature, we stick to the usual gamma matrices instead of
the original Dirac matrices. Though inessential, this permits to immediately
take advantage of computer packages, in particular FeynCalc~\cite%
{Shtabovenko:2020gxv}.

Being perturbative, the first step is to expand the impact of a specific
unitary rotation $\psi\rightarrow\psi^{\prime}=e^{iS}\psi$. If $i\partial
_{t}\left\vert \psi\right\rangle =\mathcal{H}\left\vert \psi\right\rangle $,
then $i\partial_{t}\left\vert \psi^{\prime}\right\rangle =\mathcal{H}%
^{\prime }\left\vert \psi^{\prime}\right\rangle $ with 
\begin{equation}
\mathcal{H}^{\prime}=e^{iS}\left( \mathcal{H}-i\partial_{t}\right) e^{-iS}\ .
\label{App1}
\end{equation}
Using the CBH formulas, 
\begin{equation}
e^{X}Ye^{-X}=\sum_{n=0}^{\infty}\frac{1}{n!}[[X]^{n},Y]\ ,\ \ \
e^{X}de^{-X}=\sum_{n=0}^{\infty}\frac{-1}{(n+1)!}[[X]^{n},dX]\ ,
\end{equation}
where $[[X]^{0},Y]=Y$, $[[X]^{1},Y]=[X,Y]$, $[[X]^{2},Y]=[X,[X,Y]]$, etc,
and $d$ is a differential operator acting only on $e^{-X}$, the expansion of
Eq.~(\ref{App1}) is%
\begin{equation}
\mathcal{H}^{\prime}=\mathcal{H}+\sum_{n=0}^{\infty}\frac{1}{(n+1)!}%
[[iS]^{n},[iS,\mathcal{H}]-\dot{S}]\ .
\end{equation}

In the first step, writing the Hamiltonian as $\mathcal{H}=\gamma ^{0}(m+%
\mathcal{O})+\mathcal{E}$ with $\mathcal{O}\gamma^{0}=-\gamma ^{0}\mathcal{O}
$ and $\mathcal{E}\gamma^{0}=\gamma^{0}\mathcal{E}$, we take $iS=\mathcal{O}%
/(2m)$. The various terms in the expansion are given by%
\begin{equation}
[[iS]^{n},\mathcal{H}]=\frac{\gamma^{0}(-\mathcal{O})^{n}}{m^{n-1}}+\frac{[[%
\mathcal{O}]^{n},\mathcal{E}]}{(2m)^{n}}-\frac{\gamma ^{0}(-\mathcal{O}%
)^{n+1}}{m^{n}}\ ,\ [[iS]^{n},-\dot{S}]=\frac {i[[\mathcal{O}]^{n},\mathcal{%
\dot{O}}]}{(2m)^{n+1}}\ .
\end{equation}
The new Hamiltonian is then $\mathcal{H}^{\prime}=\gamma^{0}(m+\mathcal{O}%
^{\prime})+\mathcal{E}^{\prime}$ with%
\begin{align}
\mathcal{E}^{\prime} & =\mathcal{E}-\frac{\gamma^{0}\mathcal{O}^{2}}{2m}+%
\frac{[\mathcal{O},\mathcal{V}_{1}]}{8m^{2}}-\frac{\gamma^{0}\mathcal{O}^{4}%
}{8m^{3}}+\frac{[\mathcal{O},[\mathcal{O},[\mathcal{O},\mathcal{V}_{1}]]]}{%
24(2m)^{4}}\ , \\
\mathcal{O}^{\prime} & =\frac{\gamma^{0}\mathcal{V}_{1}}{2m}+\frac {4%
\mathcal{O}^{3}}{3(2m)^{2}}+\frac{\gamma^{0}[\mathcal{O},[\mathcal{O},%
\mathcal{V}_{1}]]}{6(2m)^{3}}+\frac{8\mathcal{O}^{5}}{15(2m)^{4}}\ ,
\end{align}
with $\mathcal{V}_{1}\equiv\lbrack\mathcal{O},\mathcal{E}]+i\mathcal{\dot{O}}
$ an odd operator. The leading non-block diagonal term has disappeared, and
non-block diagonal terms in $\mathcal{O}^{\prime}$ start now at $\mathcal{O}%
(1/m)$. Those can be removed at that order by performing a second FW
transformation with $iS^{\prime}=\mathcal{O}^{\prime}/(2m)\sim\mathcal{O}%
(1/m^{2})$. Keeping terms up to $\mathcal{O}(1/m^{4})$ only, and using the
above formulas, we arrive at $\mathcal{H}^{\prime\prime}=\gamma^{0}(m+%
\mathcal{O}^{\prime\prime})+\mathcal{E}^{\prime\prime}$ with%
\begin{equation}
\mathcal{E}^{\prime\prime}=\mathcal{E}^{\prime}-\frac{\gamma^{0}\mathcal{O}%
^{\prime2}}{2m}+\frac{[\mathcal{O}^{\prime},[\mathcal{O}^{\prime },\mathcal{E%
}^{\prime}]+i\mathcal{\dot{O}}^{\prime}]}{2(2m)^{2}}\ ,\ \ \mathcal{O}%
^{\prime\prime}=\gamma^{0}\frac{[\mathcal{O}^{\prime },\mathcal{E}%
^{\prime}]+i\mathcal{\dot{O}}^{\prime}}{2m}\ .
\end{equation}
Proceeding further to eliminate $\mathcal{O}^{\prime\prime}$ with $%
iS^{\prime\prime}=\mathcal{O}^{\prime\prime}/(2m)\sim\mathcal{O}(1/m^{3})$
does not change the diagonal term anymore since $\mathcal{E}^{\prime
\prime\prime}-\mathcal{E}^{\prime\prime}\sim\mathcal{O}^{\prime\prime2}/(2m)$
is already $\mathcal{O}(1/m^{5})$. So, the final Hamiltonian can be read off
the result after only the $S$ and $S^{\prime}$ transformations, even though
a total of four FW transformations are actually necessary:%
\begin{align}
\mathcal{H}^{\mathrm{NR}} & =\gamma^{0}\left( m-\frac{\mathcal{O}^{2}}{2m}-%
\frac{\mathcal{O}^{4}}{8m^{3}}+\frac{\mathcal{V}_{1}^{2}}{8m^{3}}\right) +%
\mathcal{E}+\frac{[\mathcal{O},\mathcal{V}_{1}]}{8m^{2}}  \notag \\
& +3\frac{\{\mathcal{O}^{2},[\mathcal{O},\mathcal{V}_{1}]\}}{64m^{4}}+5\frac{%
\{\mathcal{O},[\mathcal{O}^{2},\mathcal{V}_{1}]\}}{128m^{4}}-\frac{[\mathcal{%
V}_{1},\mathcal{V}_{2}]}{32m^{4}}+\mathcal{O}(1/m^{5})\ ,
\end{align}
where all the higher order $\mathcal{E}$ and $\mathcal{\dot{O}}$ dependences
occur in the chain of odd operators $\mathcal{V}_{i+1}\equiv\lbrack \mathcal{%
V}_{i},\mathcal{E}]+i\mathcal{\dot{V}}_{i}$ (this remains true at higher
orders). In all the applications here, only the terms in the first line are
kept. Those are obtained by the sequence of transformations $\psi
\rightarrow e^{iS^{\prime\prime}}e^{iS^{\prime}}e^{iS}\psi$ with $iS=%
\mathcal{O}/(2m)$, $iS^{\prime}=\mathcal{O}^{\prime}/(2m)$ and $%
iS^{\prime\prime}=\mathcal{O}^{\prime\prime}/(2m)$.

\section{Non-relativistic electromagnetic interactions\label{AppEM}}

Let us start from the Hamiltonian after the Schiff transformations $S_{1}$
of Eq.~(\ref{SchiffTf1}) with $\alpha =md/e$ and $S_{2}$ of Eq.~(\ref{QEDS2}%
) with $\beta =1$, keeping terms at most linear in $a$ or $d$:%
\begin{align}
\mathcal{H}^{\mathrm{NR}}& =\gamma ^{0}\left( m+\frac{\mathbf{P}^{2}}{2m}-%
\frac{\mathbf{P}^{4}}{8m^{3}}-\frac{e\left( 1+a\right) \boldsymbol{\sigma }%
\cdot \mathbf{B}}{2m}+\frac{e\{\mathbf{P}^{2},\boldsymbol{\sigma }\cdot 
\mathbf{B}\}}{8m^{3}}\right) +e\phi   \notag \\
& \ \ \ \ +ie\frac{1+2a}{8m^{2}}[\boldsymbol{\gamma }\cdot \mathbf{P},%
\boldsymbol{\gamma }\cdot \mathbf{E}]+\frac{id(1+a)}{2m}[\boldsymbol{\gamma }%
\cdot \mathbf{P},\boldsymbol{\gamma }\cdot \mathbf{B}]  \notag \\
& \ \ \ \ +\gamma ^{0}\left( \frac{e^{2}(1+2a)}{8m^{3}}\mathbf{E}^{2}-\frac{%
e^{2}}{8m^{3}}\mathbf{B}^{2}-\frac{ed(1+a)}{4m^{2}}\{\boldsymbol{\gamma }%
\cdot \mathbf{E},\boldsymbol{\gamma }\cdot \mathbf{B}\}\right)   \notag \\
& \ \ \ \ +\gamma ^{5}\frac{[\boldsymbol{\gamma }\cdot \mathbf{P},[%
\boldsymbol{\gamma }\cdot \mathbf{P},(1+2a)d\boldsymbol{\gamma }\cdot 
\mathbf{E}]]+\{\boldsymbol{\gamma }\cdot \mathbf{P},\{\boldsymbol{\gamma }%
\cdot \mathbf{P},\boldsymbol{\gamma }\cdot (\delta _{\mu }\mathbf{B}-d%
\mathbf{E})\}\}}{8m^{2}}+\mathcal{O}(1/m^{4})\ .
\end{align}%
Further simplifications are possible. First, the redundancy induced by the
unitary transformation $S_{3}$ of Eq.~(\ref{QEDS3}) combined with the
identity $[A,[B,C]]+\{B,\{A,C\}\}=\{C,\{A,B\}\}$, i.e., 
\begin{equation}
\lbrack \boldsymbol{\gamma }\cdot \mathbf{P},[\boldsymbol{\gamma }\cdot 
\mathbf{P},\boldsymbol{X}]]+\{\boldsymbol{\gamma }\cdot \mathbf{P},\{%
\boldsymbol{\gamma }\cdot \mathbf{P},\boldsymbol{X}\}\}=\{\boldsymbol{X},\{%
\boldsymbol{\gamma }\cdot \mathbf{P},\boldsymbol{\gamma }\cdot \mathbf{P}%
\}\}\ ,  \label{IdAPPX}
\end{equation}%
with $\boldsymbol{X}=\boldsymbol{\gamma }\cdot \mathbf{E}$, permit to write 
\begin{align}
\gamma ^{5}\{\boldsymbol{\gamma }\cdot \mathbf{P},\{\boldsymbol{\gamma }%
\cdot \mathbf{P},\boldsymbol{\gamma }\cdot \mathbf{E}\}\}& =-\frac{1}{2}%
\gamma ^{5}\{\boldsymbol{\gamma }\cdot \mathbf{E},\{\boldsymbol{\gamma }%
\cdot \mathbf{P},\boldsymbol{\gamma }\cdot \mathbf{P}\}\}+\mathcal{O}(1/m)\ ,
\\
\gamma ^{5}[\boldsymbol{\gamma }\cdot \mathbf{P},[\boldsymbol{\gamma }\cdot 
\mathbf{P},\boldsymbol{\gamma }\cdot \mathbf{E}]]& =\frac{3}{2}\gamma ^{5}\{%
\boldsymbol{\gamma }\cdot \mathbf{E},\{\boldsymbol{\gamma }\cdot \mathbf{P},%
\boldsymbol{\gamma }\cdot \mathbf{P}\}\}+\mathcal{O}(1/m)\ .
\end{align}%
The anticommutator can then be simplified with $\{\boldsymbol{\gamma }\cdot 
\mathbf{P},\boldsymbol{\gamma }\cdot \mathbf{P}\}=-2\mathbf{P}^{2}-2e\gamma
^{0}\gamma ^{5}\boldsymbol{\gamma }\cdot \mathbf{B}$. For the magnetic
field, the identity Eq.~(\ref{IdAPPX}) with $\boldsymbol{X}=\boldsymbol{%
\gamma }\cdot \mathbf{B}$ becomes 
\begin{equation}
\gamma ^{5}[\boldsymbol{\gamma }\cdot \mathbf{P},[\boldsymbol{\gamma }\cdot 
\mathbf{P},\boldsymbol{\gamma }\cdot \mathbf{B}]]+\gamma ^{5}\{\boldsymbol{%
\gamma }\cdot \mathbf{P},\{\boldsymbol{\gamma }\cdot \mathbf{P},\boldsymbol{%
\gamma }\cdot \mathbf{B}\}\}=-2\{\mathbf{P}^{2},\gamma ^{5}\boldsymbol{%
\gamma }\cdot \mathbf{B}\}-4e\gamma ^{0}\mathbf{B}^{2}\ .  \label{Redund}
\end{equation}%
Notice that $\gamma ^{0}\{\mathbf{P}^{2},\boldsymbol{\sigma }\cdot \mathbf{B}%
\}=-\gamma ^{5}\{\mathbf{P}^{2},\boldsymbol{\gamma }\cdot \mathbf{B}\}$, so
this last relation introduces a redundancy between four of the operators
already present in the Hamiltonian.

Strictly speaking, there are not enough constraints to point us towards a
specific form for the Hamiltonian. To proceed, we therefore add the
requirement that the pure field-dependent terms should involve only the
electromagnetic invariants $\mathbf{E}^{2}-\mathbf{B}^{2}$ and $\mathbf{E}%
\cdot \mathbf{B}$. This matches the comments made in the text about higher
order operators, in particular $F_{\mu \nu }F^{\mu \nu }$ or $F_{\mu \nu }%
\tilde{F}^{\mu \nu }$, that could be added to the initial Hamiltonian and
would immediately contribute to these terms. With these requirements, we
obtain%
\begin{align}
\mathcal{H}^{\mathrm{NR}}& =\gamma ^{0}\left( m+\frac{\mathbf{P}^{2}}{2m}-%
\frac{\mathbf{P}^{4}}{8m^{3}}-\frac{e\left( 1+a\right) \boldsymbol{\sigma }%
\cdot \mathbf{B}}{2m}+\frac{e(1+a)\{\mathbf{P}^{2},\boldsymbol{\sigma }\cdot 
\mathbf{B}\}}{8m^{3}}\right) +e\phi   \notag \\
& \ \ \ \ +ie\frac{1+2a}{8m^{2}}[\boldsymbol{\gamma }\cdot \mathbf{P},%
\boldsymbol{\gamma }\cdot \mathbf{E}]+\frac{id(1+a)}{2m}[\boldsymbol{\gamma }%
\cdot \mathbf{P},\boldsymbol{\gamma }\cdot \mathbf{B}]  \notag \\
& \ \ \ \ +\gamma ^{0}\left( e^{2}\frac{1+2a}{8m^{3}}(\mathbf{E}^{2}-\mathbf{%
B}^{2})-ed\frac{1+2a}{4m^{2}}\{\boldsymbol{\gamma }\cdot \mathbf{E},%
\boldsymbol{\gamma }\cdot \mathbf{B}\}\right)   \notag \\
& \ \ \ \ -\frac{ea}{16m^{3}}\gamma ^{5}[\boldsymbol{\gamma }\cdot \mathbf{P}%
,[\boldsymbol{\gamma }\cdot \mathbf{P},\boldsymbol{\gamma }\cdot \mathbf{B}%
]]-d\frac{2+3a}{4m^{2}}\gamma ^{5}\{\mathbf{P}^{2},\boldsymbol{\gamma }\cdot 
\mathbf{E}\}+\mathcal{O}(1/m^{4})\ .
\end{align}%
up to terms of $\mathcal{O}(a^{2}/m^{3},d^{2}/m,1/m^{4})$. This form is
rather suggestive, with the $1+a$ factor occurring for both the Zeeman term
and the $\gamma ^{0}\{\mathbf{P}^{2},\boldsymbol{\sigma }\cdot \mathbf{B}\}$
operator once the $\mathbf{B}^{2}$ term is properly tuned to force the
appearance of the $\mathbf{E}^{2}-\mathbf{B}^{2}$ invariant. Remember though
that some redundancies remain in this Hamiltonian, as encoded in Eq.~(\ref%
{Redund}).

\end{document}